\documentclass[10pt]{article}

\usepackage{graphicx}
\usepackage{amsmath}
\usepackage{amsfonts}
\usepackage{setspace}
\usepackage{listings}
\usepackage{bm} % bold greek math symbols
\usepackage{multirow}
\usepackage[normalem]{ulem}
\usepackage[left=1.0in,top=1.0in,right=1.0in,bottom=1.0in]{geometry}
\usepackage{pslatex}
\usepackage{xspace}
\usepackage[usenames,dvipsnames]{color}

\usepackage{hyperref}
\hypersetup{
  colorlinks,
  urlcolor=blue,
  linkcolor=black,
  citecolor=black
}

\usepackage{xr}
\usepackage{cite}
\makeatletter
\renewcommand{\@biblabel}[1]{\quad#1.}
\makeatother

\newcommand{\togglefig}[1]{}

\newcommand{\comment}[1]{}

\newcommand{\V}[1]{{\bm #1}}

\newcommand{\hide}[1]{}

\newcommand{\wv}{{\em ARGweaver}\xspace}

\begin{document}

\begin{flushleft}
{\LARGE
\textbf{Genome-wide inference of ancestral recombination graphs}}
\\
Matthew D.\ Rasmussen$^{\ast,\dag}$, 
Melissa J.\ Hubisz,
Ilan Gronau,
Adam Siepel$^{\ast}$
\\[1ex]
Department of Biological Statistics and Computational Biology, 
Cornell University, Ithaca, New York 14853, USA
{ \singlespacing
$^\ast${\bf Corresponding authors:}\\[1ex]
Matthew D.\ Rasmussen, Adam Siepel \\
102 Weill Hall \\
Ithaca, NY 14853, USA \\
E-mail: rasmussen@cornell.edu, acs4@cornell.edu }
\\[2ex]

$^\dag${\bf Current address:} Counsyl, 180 Kimball Way, South San
Francisco, CA 94080, USA \\[1ex]

\\[1ex]

{\bf Running title:} Genome-wide ARG inference
\end{flushleft}

\thispagestyle{empty}

%=============================================================================
%\newpage

\section*{Abstract}
The complex correlation structure of a collection of orthologous DNA
sequences is uniquely captured by the ``ancestral recombination graph''
(ARG), a complete record of coalescence and recombination events in the
history of the sample.  However, existing methods for ARG inference are
computationally intensive, highly approximate, or limited to small numbers
of sequences, and, as a consequence, explicit ARG inference is rarely used
in applied population genomics.  Here, we introduce a new algorithm for ARG
inference that is efficient enough to apply to dozens of
complete mammalian genomes.  The key idea of our approach is to sample an ARG
of $n$ chromosomes conditional on an ARG of $n-1$ chromosomes, an operation
we call ``threading.''  Using techniques based on hidden Markov models, we
can perform this threading operation exactly, up to the assumptions
of the sequentially Markov coalescent and a discretization of time.  An
extension allows for threading of subtrees instead of individual sequences.
Repeated application of these threading operations results in highly
efficient Markov chain Monte Carlo samplers for ARGs. We have implemented
these methods in a computer program called \wv.  Experiments with simulated
data indicate that \wv converges rapidly to the true posterior distribution
and is effective in recovering various features of the ARG for dozens of
sequences generated under realistic parameters for human populations.  In
applications of \wv to 54 human genome sequences from Complete
Genomics, we find clear signatures of natural selection, including regions
of unusually ancient ancestry associated with balancing selection and
reductions in allele age in sites under directional selection.  The
patterns we observe near protein-coding genes are consistent with a primary
influence from background selection rather than hitchhiking.  Preliminary
results also indicate that our methods can be used to gain insight into
complex features of human population structure, even with a noninformative prior
distribution.

\clearpage

%=============================================================================

\section*{Author Summary}
\thispagestyle{empty}
The unusual and complex correlation structure of population samples of
genetic sequences presents a fundamental challenge for statistical analysis
that pervades nearly all areas of population genetics.  At a single genomic
position, the relationships among individual genotypes can be captured, in
a relatively straightforward manner, by a type of tree known as a
genealogy.  But recombination events in the history of the sample cause
these genealogies to change along the genome, leading to an intricate
network of intertwined genealogies.  This complex structure is ultimately
what makes many problems in population genetics difficult, including the
inference of ancestral population sizes, population divergence times, gene
flow between populations, loci under natural selection, recombination
rates, and genotype/phenotype associations.  It is possible, in principle,
to describe the genetic relationships among individuals precisely using a
generalized representation called the ancestral recombination graph (ARG),
which provides a complete description of both coalescence and recombination
events in the history of the sample.  However, previous methods for ARG
inference have not been fast and accurate enough for practical use with
large-scale genomic sequence data. In this article, we introduce a new
algorithm for ARG inference that has vastly improved scaling properties.
Our algorithm is implemented in a computer program called \wv, which is
fast enough to be applied to sequences megabases in length.  With the aid
of a large computer cluster, \wv can be used to sample full ARGs for entire
mammalian genome sequences.  We show that \wv performs well in simulation
experiments and demonstrate that it can be used to provide new insights
about both demographic processes and natural selection when applied to 
real human genome sequence data.

\clearpage

%=============================================================================
\setcounter{page}{1}
\section*{Introduction}

At each genomic position, orthologous DNA sequences drawn from one or more
populations are related by a branching structure known as a genealogy
\cite{HEINETAL05,WAKE09}.  Historical recombination events lead to changes
in these genealogies from one genomic position to the next, resulting in a
correlation structure that is complex, analytically intractable, and
poorly approximated by standard representations of high-dimensional data.
Over a period of many decades, these unique features of genetic data have
inspired numerous innovative techniques for probabilistic modeling and
statistical inference
\cite{FISH30,WRIG31,KIMU62,FELS73,FELS81,MENO78,KING82A}, and, more
recently, they have led to a variety of creative approaches that achieve
computational tractability by operating on various summaries of the data
\cite{SAWYHART92,VOIGETAL05,KEIGEYRE07,BOYKETAL08,LAWSETAL12,PALAETAL12,RALPCOOP13,HARRNIEL13}.
Nevertheless, none of these approaches fully captures the correlation
structure of collections of DNA sequences, which inevitably leads to
limitations in power, accuracy, and generality in genetic analysis.

In principle, the correlation structure of a collection of colinear
orthologous sequences can be fully described by a network known as an {\em
ancestral recombination graph} (ARG)
\cite{HUDS91,Griffiths1996,GRIFMARJ97}.  An ARG provides a record of all coalescence and recombination
events since the divergence of the sequences under study and specifies a
complete genealogy at each genomic position (Figure
\ref{fig:arg}A).  In many senses, the ARG
is the ideal data structure for population genomic analysis.  Indeed, if an
accurate ARG could be obtained, many problems of interest today---such as
the estimation of recombination rates or ancestral effective population
sizes---would become trivial, while many other problems---such as the
estimation of population divergence times, rates of gene flow between
populations, or the detection of selective sweeps---would be greatly
simplified.  Various data representations in wide use today, including the
site frequency spectrum, principle components, haplotype maps, and identity
by descent spectra, can be thought of as low-dimensional summaries of the
ARG and are strictly less informative.

An extension of the widely used coalescent framework
\cite{KING82A,HEINETAL05,WAKE09} that 
includes recombination
\cite{Hudson1983}
%It is relatively straightforward to extend the widely used coalescent
%framework for describing genealogies \cite{KING82A} to include
%recombination, and the resulting {\em coalescent-with-recombination} 
%\cite{Hudson1983} 
is regarded as an adequately rich generative process for
ARGs in most settings of interest.  While simulating an ARG under this model
is fairly straightforward, however, using it to 
reconstruct an ARG from sequence data is notoriously difficult.  Furthermore, the 
data are generally only weakly informative about the ARG, so it is often
desirable to regard it as a ``nuisance'' variable to be integrated out
during statistical inference (e.g., \cite{Fearnhead2001}).  During the past
two decades, various attempts have been made to perform explicit
inference of ARGs using techniques such as importance sampling~\cite{Griffiths1996,Fearnhead2001} and Markov chain Monte Carlo sampling
\cite{KUHNETAL00,NIEL00,Kuhner2006a,OFAL13}.  There is also a considerable
literature on heuristic or approximate methods for ARG
reconstruction in a parsimony framework
\cite{HEIN90,HEIN93,KECEGUSF98,WANGETAL01,Song2005a,SONGETAL05,Minichiello2006,Wu2009}.
Several of these approaches have shown promise, but they are
generally highly computationally intensive and/or limited in accuracy, and
they are not suitable for application to large-scale data sets.  As a
result, explicit ARG inference is rarely used in applied population
genomics.

The coalescent-with-recombination is conventionally described as a
stochastic process in time 
\cite{Hudson1983}, but Wiuf and Hein \cite{Wiuf1999a} showed that it could
be reformulated as a mathematically equivalent process along the genome
sequence.  Unlike the process in time, this ``sequential'' process
is not 
Markovian
%that is, the genealogies for genomic positions $1, \dots, i-1$
%cannot be ``forgotten'' when generating the genealogy for position $i+1$
%based on the genealogy for position $i$.  The reason is that 
because long-range
dependencies 
are induced 
by so-called ``trapped'' sequences (genetic material nonancestral to the
sample flanked by ancestral segments).  As a result, the full sequential
process is complex and computationally expensive to manipulate.
%The
%process can 
%be precisely described
%at considerable cost in complexity and computational efficiency, 
%by keeping track not just of the genealogy at each
%position, but of an expanded graph that includes nonancestral lineages.
Interestingly, however, simulation
processes that simply disregard the non-Markovian features of the
sequential process produce collections of sequences that are
remarkably consistent in most respects with those generated by the full
coalescent-with-recombination \cite{McVean2005,Marjoram2006}.  In other
words, the 
coalescent-with-recombination is almost Markovian, in the sense that the 
long-range correlations induced by trapped material are fairly weak and
have a minimal impact on the data.  The original Markovian approximation to
the full process \cite{McVean2005}
is known as the {\em sequentially Markov coalescent} (SMC), and an extension
that allows for an additional class of recombinations
\cite{Marjoram2006} is known as the SMC$'$.

In recent years, the SMC has become favorite starting point for approximate
methods for ARG inference
\cite{Hobolth2007,MAILETAL11,MAILETAL12,Li2011a}.  The key insight behind
these methods is that, if the continuous state space for the Markov chain
(consisting of all possible genealogies) is approximated by a moderately
sized finite set---typically by enumerating tree topologies and/or
discretizing time---then inference can be performed efficiently using
well-known algorithms for hidden Markov models (HMMs).  Perhaps the
simplest and most elegant example of this approach is the pairwise
sequentially Markov coalescent (PSMC) \cite{Li2011a}, 
which applies to pairs of homologous chromosomes (typically the two
chromosomes 
in a diploid individual) and is used to reconstruct a profile of effective
population sizes over time.  In this case, there is only one possible tree
topology and one coalescence event to consider at each genomic position, so
it is sufficient to discretize time and allow for coalescence within any of
$k$ possible time slices.  Using the resulting $k$-state HMM, it is
possible to perform inference integrating over all possible ARGs.  A
similar 
HMM-based approach has been used to estimate ancestral effective population
sizes and divergence times from individual representatives of a few closely
related species \cite{Hobolth2007,MAILETAL11,MAILETAL12}. 
Because of their dependency on a complete characterization of the SMC state
space, however, these methods can
only be applied to small numbers of samples.  This limits their utility
with newly emerging population genomic datasets and leads to reduced power
for certain features of interest, such as recent effective population
sizes, recombination rates, or local signatures of natural selection.

An alternative modeling approach, with better scaling properties, is the
product of approximate conditionals (PAC) or ``copying'' model of Li and
Stephens \cite{Li2003a} (see also \cite{Stephens2000,Fearnhead2001}).  The
PAC model is motivated primarily by computational tractability and is not
based on an explicit evolutionary model.  The model generates the $n$th
sequence in a collection by concatenating (noisy) copies of fragments of
the previous $n-1$ sequences.  The source of each copied fragment
represents the ``closest'' (most recently diverged) genome for that
segment, and the noise process allows for mutations since the source and
destination copies diverged.  The PAC framework has been widely used in
many applications in statistical genetics, including recombination rate
estimation, local ancestry inference, haplotype phasing, and genotype
imputation (e.g.,
\cite{STEPSCHE05,MARCETAL07,HOWIETAL09,PRICETAL09,LIETAL10}), and it
generally offers good performance at minimal computational cost.  Recently,
Song and colleagues have generalized this framework to make use of
conditional sampling distributions (CSDs) based on models closely related
to, and in some cases equivalent to, the SMC
\cite{Paul2010,Paul2011,SHEEETAL13,STEIETAL12}.  They have demonstrated
improved accuracy in conditional likelihood calculations
\cite{Paul2010,Paul2011} and have shown that their methods can be effective
in demographic inference \cite{SHEEETAL13,STEIETAL12}.  However, their
approach avoids explicit ARG inference and therefore can only be used to 
characterize properties of the ARG that are directly determined by model
parameters (see Discussion).

In this paper, we introduce a new algorithm for ARG inference that combines
many of the benefits of the small-sample SMC-based approaches and the
large-sample CSD-based methods.  Like the PSMC, our algorithm requires no
approximations beyond those of the SMC and a discretization of time, but it
improves on the PSMC by allowing multiple genome sequences to be considered
simultaneously.  The key idea of our approach is to sample an ARG of $n$
sequences conditional on an ARG of $n-1$ sequences, an operation we call
``threading.''  Using HMM-based methods, we can efficiently sample new
threadings from the exact conditional distribution of interest.  By
repeatedly removing and re-threading individual sequences, we obtain an
efficient Gibbs sampler for ARGs.  This basic Gibbs sampler can be improved
by including operations that rethread entire subtrees rather than
individual sequences.  Our implementation of these methods, called \wv, is
efficient enough to sample full ARGs on a genome-wide scale for dozens of
diploid individuals.  Simulation experiments indicate that \wv converges
rapidly and is able to recover many properties of the true ARG with
good accuracy.  In addition, our explicit characterization of the ARG
enables us to examine many features not directly described by model
parameters, such as local times to most recent common ancestry, allele
ages, and gene tree topologies.  These quantities, in turn, shed light on
both demographic processes and the influence of natural selection across
the genome.  For example, we demonstrate, by applying \wv to 54 individual
human sequences from Complete Genomics, that it provides insight into the
sources of reduced nucleotide diversity near functional elements, the
contribution of balancing selection to regions containing very old
polymorphisms, and the relative influences of direct and indirect selection
on allele age.  The method also show promises in addressing questions
related to human population structure.  Our \wv software
(\url{https://github.com/mdrasmus/argweaver}), our sampled ARGs
(\url{http://compgen.bscb.cornell.edu/ARGweaver/CG_results}), 
and genome-browser tracks summarizing these ARGs
(\url{http://genome-mirror.bscb.cornell.edu}; assembly hg19) are all
freely available.

%=============================================================================

\section*{Results}

%\subsection{Generative Model}

\subsection{The Sequentially Markov Coalescent}

The starting point for our model is the Sequentially Markov Coalescent
(SMC) introduced by McVean and Cardin \cite{McVean2005}.  We begin by
briefly reviewing the SMC and introducing notation that will be useful
below in 
describing a general discretized version of this model.  

The SMC is a stochastic process for generating a sequence of local trees,
$\V{T}^n = T^n_1, ..., T^n_m$ and corresponding genomic breakpoints $\V{b}
= b_1, \dots, b_{m+1}$, such that each $T^n_i$ ($1 \leq i \leq m$)
describes the ancestry of a collection of $n$ sequences in a nonrecombining
genomic interval $[b_i, b_{i+1})$, and each breakpoint $b_{i}$ between
intervals $T^n_{i-1}$ and $T^n_{i}$ corresponds to a recombination event
(Figure \ref{fig:arg}B).  The model is continuous in both space and time,
with each node $v$ in each $T^n_i$ having a real-valued age $t(v) \geq 0$
in generations ago, and each breakpoint $b_i$ falling in the continuous
interval $[0, L]$, where $L$ is the total length of the genomic segment of
interest in nucleotide sites.  The intervals are exhaustive and
nonverlapping, with $b_1=0$, $b_{m+1}=L$, and $b_i < b_{i+1}$ for all $i$.
Each $T^n_i$ is a binary tree with $t(v) = 0$, for all leaf nodes $v$, and
has a marginal distribution given by the standard coalescent.  We will use
the convention of indexing branches in the trees by their descendant nodes;
that is, branch $v$ is the branch between node $v$ and its parent.

As shown by Wiuf and Hein \cite{Wiuf1999a}, the correlation structure of
the local trees and recombinations under the full
coalescent-with-recombination is complex.  The SMC 
approximates this 
distribution by assuming
that $T^n_i$ is conditionally independent of
$T^n_1,
\dots, T^n_{i-2}$ given $T^n_{i-1}$, and, similarly, that $b_i$ depends only on
$b_{i-1}$ and $T_{i-1}$,  
% $(b_i, T^n_i)$ is conditionally independent of
%$(b_1, T^n_1),
%\dots, (b_{i-2}, T^n_{i-2})$ given $(b_{i-1}, T^n_{i-1})$, 
so that,
\begin{equation}
P(\V{T}^n, \V{b} \;|\; N, \rho) = P(T^n_1 \;|\; N) \left[
\prod_{i=2}^m P(b_i \;|\; b_{i-1}, T^n_{i-1})
\; P(T^n_i \; | \;
T^n_{i-1}, N, \rho)
\right]  P(b_{m+1}=L \;|\; b_m, T^n_{m}), 
\end{equation}
where $N$ is the effective population size, $\rho$ is the recombination
rate, and it is understood that $b_1=0$.  Thus, the SMC
can be viewed as generating a sequence of local trees and corresponding
breakpoints
by a first-order Markov process.  The key to the model is to define the
conditional distributions $P(b_i \;|\; b_{i-1}, T^n_{i-1})$ and $P(T^n_i \; | \;
T^n_{i-1}, N, \rho)$ such that this 
Markov process closely approximates the coalescent-with-recombination.
Briefly, this is accomplished by first sampling the initial tree $T^n_1$
from the 
standard coalescent and setting $b_1=0$,
and then iteratively
(i) determining the next breakpoint, $b_i$, by incrementing $b_{i-1}$ by an
exponential random
variate with rate $\rho |T^n_{i-1}|$, where $|T^n_i|$ denotes the
total branch length of $T^n_i$; 
(ii) sampling a recombination point
$R_i=(w_i,u_i)$ uniformly along the branches beneath the root of $T^n_{i-1}$,
where $w_i$ is a branch and $u_i$ is a time along that branch; (iii)
dissolving the branch $w_i$ above point $u_i$; and (iv) allowing $w_i$ to
rejoin the remainder of tree $T^n_{i-1}$ above time $u_i$ by the standard
coalescent process, creating a new tree $T^n_i$ (Figure \ref{fig:arg}B).
%The initial tree $T^n_1$ is
%sampled from the standard coalescent.  
%The genomic breakpoint between
%$T^n_{i-1}$ and $T^n_i$, $b_i$, is obtained by drawing a random variate from an
%exponential distribution with rate $\rho |T^n_{i-1}|$ and adding this value to
%$b_{i-1}$.  
As a generative process for an arbitrary number of genomic segments, 
the SMC can be implemented 
by simply repeating the iterative process until $b_i \geq L$, then setting
$m$ equal to $i-1$ and $b_{m+1}$ equal to $L$.

Notice that, if the sampled recombination points $R_i$ are retained, this process
generates not only a 
sequence of local trees but a complete ARG.  In addition, a sampled
sequence of local trees, $\V{T}^n$, is sufficient for generation of $n$
aligned DNA sequences corresponding to the leaves of the trees (Figure \ref{fig:arg}C).
% (assuming real-valued interval lengths are rounded to the nearest
% integer).  
%We
%denote the alignment $\V{D}^n = (D^n_1, \dots, D^n_L)$, where each $D^n_j$
%represents an alignment column of height $n$.  
%If a finite sites model is
%assumed (as will be true throughout this paper), each alignment column
%$D^n_j$ can be generated by sampling an ancestral allele from an
%appropriate background distribution, and then allowing this allele to
%mutate stochastically along the branches of the corresponding local tree,
%in a branch-length-dependent manner.  We denote the induced conditional probability
%distribution over alignments by $P(D^n_j \;|\; T^n_i, \mu)$,
%where $\mu$ is the mutation rate.
Augmented in this way, the SMC can be
considered a full generative model for ARGs and sequence data.

%=============================================================================
\subsection{The Discretized Sequentially Markov Coalescent}

We now define an approximation of the SMC that is discrete in both space
and time, which we call the Discretized Sequentially Markov Coalescent
(DSMC).  The DSMC can be viewed as a generalization to multiple genomes of
the discretized pairwise 
sequentially Markov coalescent (PSMC) used by Li and Durbin \cite{Li2011a}.
It is also closely 
related to several other recently described discretized Markovian
coalescent models 
\cite{Hobolth2007,MAILETAL11,Paul2011}. 

The DSMC assumes that time is partitioned into $K$ intervals,
whose boundaries are given by a sequence of time points ${\cal P} =
(s_0,..., s_K)$, with $s_0=0$, $s_{j+1} > s_j$ for all $j$ ($0 \leq j <
K$), and $s_K$ equal to a user-specified maximum value.  (See Table \ref{tab:notation}
for a key to the notation used in this paper.) Every coalescence
or recombination event is assumed to occur precisely at one of these $K+1$
time points.  Various strategies can be used to determine these time points
(see, e.g., \cite{Paul2011}).
In this paper, we simply 
distribute them uniformly on a logarithmic scale, so
that the resolution of the discretization scheme is finest near the leaves
of the ARG, where the density of events is expected to be greatest (see
Methods).  Each local block is assumed to have an integral
length measured in base pairs, with all recombinations occurring between
adjacent nucleotides.  The DSMC approaches the SMC as the number of
intervals $K$ and the sequence length $L$ grow large, for fixed $N$ and
$\rho$.

Like the SMC, the DSMC generates an ARG $\V{G}^n$ for $n$ (haploid)
sequences, each  
containing $L$ nucleotides (Figure~\ref{fig:arg}B).  In the discrete
setting, it is convenient to define local trees and recombination events at
the level of individual nucleotide positions.  Assuming that $R^n_i$
denotes a recombination between $T^n_{i-1}$ and $T^n_i$, we write
$\V{G}^n = (\V{T}^n, \V{R}^n)$, with
$\V{T}^n = (T_1^n, ..., T_L^n)$ for positions $1, \dots, L$, and $\V{R}^n =
(R_2^n, ..., R_L^n)$.   Notice that it is possible in this setting
that $R^n_i = \emptyset$ and $T^n_i = T^n_{i-1}$.  
Where a recombination occurs ($R^n_i \ne \emptyset$), we write 
$R_i^n =
(w_i, u_i)$ where $w_i$ is the branch in $T^n_{i-1}$ and $u_i \in {\cal
  P}$ is the time point
of the recombination.  For simplicity and computational efficiency, we
assume that at most one recombination occurs between each 
pair of adjacent sites.  Given the
sparsity of variant sites in most data sets, this simplification is likely
to have, at most, a
minor effect during inference (see Discussion).

Like the SMC, the DSMC can additionally be used to generated an alignment
of DNA sequences (Figure \ref{fig:arg}C).   
We
denote such an alignment by $\V{D}^n = (D^n_1, \dots, D^n_L)$, where each $D^n_i$
represents an alignment column of height $n$.  
Each 
$D^n_i$ can be generated, in the ordinary way, by sampling an ancestral
allele from an 
appropriate background distribution, and then allowing this allele to
mutate stochastically along the branches of the corresponding local tree,
in a branch-length-dependent manner.  We denote the induced conditional
probability 
distribution over alignment columns by $P(D^n_i \;|\; T^n_i, \mu)$,
where $\mu$ is the mutation rate.
In 
this work, we assume a Jukes-Cantor model \cite{JUKECANT69} for nucleotide
mutations along the branches of the tree, but another mutation model
can easily be used instead.  Notice that, while the recombinations
$\V{R}^n$ are required to define the ARG completely, the probability of the
sequence data given the ARG depends only on the local trees $\V{T}^n$.

%\subsection{Inference Strategy}
%=============================================================================
\subsection{The Threading Problem}
\label{sec:arghmm}

In the case of an observed alignment, $\V{D}^n$, and an unobserved ARG,
$\V{G}^n = (\V{T}^n, \V{R}^n)$, the DSMC can be viewed as a hidden
Markov model (HMM) with a state space given by all possible local trees,
transition probabilities given by 
expressions of the form
$P(R_i^{n} \;|\; T_{i-1}^n, \rho)$
$P(T_i^n \;|\; R_i^{n}, T_{i-1}^n, N)$,
%the conditional distributions $P(T^n_i \;|\; T^n_{i-1}, \rho, N)$, 
and emission probabilities given by the conditional distributions for
alignment columns, $P(D^n_i \;|\; T^n_i, \mu)$.  The complete data
likelihood function of this model---that is, the joint probability of an ARG
$\V{G}^n=(\V{T}^n, \V{R}^n)$ and a sequence alignment $\V{D}^n$ given model
parameters $\Theta = (\mu, \rho, \V{N})$---can be expressed as a product of
these terms over 
alignment positions (see Methods for further details):
\begin{equation}
P(\V{T}^n, \V{R}^n, \V{D}^n \;|\; \Theta) = P(T_1^n \;|\; N) \; P(D_1^n
\;|\; T_1^n, \mu) \;\prod_{i=2}^L  
P(R_i^{n} \;|\; T_{i-1}^n, \rho) \;
P(T_i^n \;|\; R_i^{n}, T_{i-1}^n, N) \;
P(D_i^n \;|\; T_i^n, \mu).
\label{eqn:dsmc-lik}
\end{equation}
This HMM
formulation is impractical as a framework for direct inference, however, because
the set of possible local trees---and hence the state space---grows
super-exponentially with $n$.  Even with additional 
assumptions, similar approaches have only been able to accommodate small
numbers of sequences \cite{HUSMWRIG01,Song2005a,Wu2009}.

\begin{figure}[h!]
\begin{center}
\includegraphics[width=5.6in]{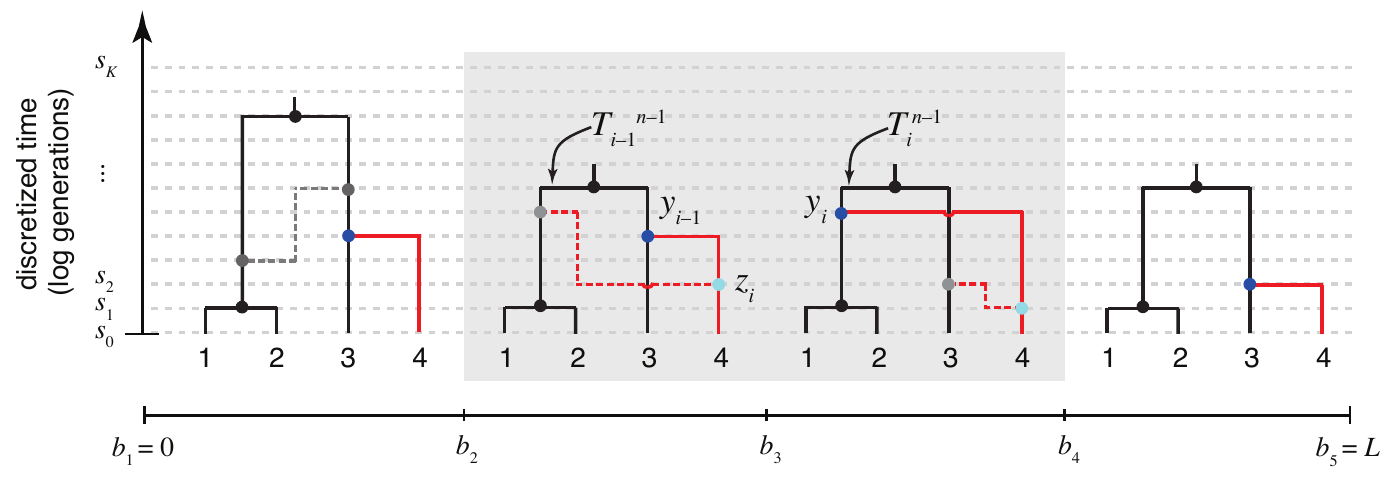}
\caption{\textbf{The ``threading'' operation.}  
\label{fig:arghmm}
}
\end{center}
\end{figure}

Instead, we use an alternative strategy with better scaling properties.  The key
idea of our approach is to sample the ancestry of only one sequence at a
time, while conditioning on the ancestry of the other $n-1$ sequences.
Repeated applications of this ``threading'' operation form 
the basis of a Markov chain Monte Carlo sampler that explores the posterior
distribution of ARGs.  In essence, the threading operation adds one branch
to each local tree in a manner that is consistent with the assumed
recombination process and the observed data (Figure \ref{fig:arghmm}).  While
conditioning on a given set of local trees introduces a number of technical
challenges, the Markovian properties of the DSMC are retained in the
threading problem, and it can be solved using standard dynamic programming
algorithms for HMMs.

The threading problem can be precisely described as follows.  Assume we are
given an ARG for $n-1$ sequences, $\V{G}^{n-1}$, a corresponding data set
$\V{D}^{n-1}$, and a set of model parameters $\Theta = (\mu,
\rho, \V{N})$.  Assume further that $\V{G}^{n-1}$ is consistent with the
assumptions of the DSMC (for example, all of its recombination and
coalescent events occur at time points in ${\cal P}$ and it contains at most
one recombination per position).  Finally, assume that we are given an
$n$th sequence $d$, of the same length of the others, and let $\V{D}^{n} =
(\V{D}^{n-1}, 
d)$.  The threading problem is to sample a new ARG $\V{G}^n$ from the
conditional distribution $P(\V{G}^n\; | \; \V{G}^{n-1},\V{D}^n,\Theta)$
under the DSMC.

The problem is simplified by recognizing that $\V{G}^n$ can be defined by
augmenting $\V{G}^{n-1}$ with the additional recombination and coalescence
events required for the $n$th sequence.  First, let $\V{G}^{n-1}$ be represented in
terms of its local trees and recombination points: $\V{G}^{n-1} =
(\V{T^{n-1}}, \V{R^{n-1}})$.  Now, observe 
that specifying the new coalescence events in $\V{G}^{n-1}$ is equivalent
to adding one 
branch to each local tree, $T^{n-1}_i$ for $i \in \{1, \dots, L\}$, to
obtain a new tree $T^n_{i}$ (Figure \ref{fig:arghmm}).  Let us denote the
point at which each of these new branches attaches to the smaller subtree
at each genomic position $i$ by $y_i = (x_i, t_i)$, where $x_i$
indicates a branch in $T^{n-1}_i$ and $t_i \in {\cal P}$ indicates the
coalescence time  
along that branch.  Thus, the {\em coalescence threading} of the $n$th
sequence is given by the sequence $\V{Y} = (y_1, ..., y_L)$.

To complete the definition of $\V{G}^n$, we must also specify the precise
locations of the additional recombinations associated with the
threading---that is, the specific time point at which each branch in a
local tree $T_{i-1}$ was broken before the branch was allowed to
re-coalesce in a new location in tree $T_i$.
%Notice that these points are needed to
%define a full ARG but they do not influence the local trees and hence they
%have no impact on the probability of the observed data given the ARG.
Here it is useful to partition the recombinations into those
that are given by $\V{G}^{n-1}$, denoted $\V{R^{n-1}}$, and those new
to $\V{G}^n$, which we denote $\V{Z} = (z_1,\dots,z_L)$ (Figure
\ref{fig:graphmodels}A\&B).  Each $z_i$ is 
either null ($z_i = \emptyset$), meaning that there is no new recombination
between $T^n_{i-1}$ and $T^n_i$, or defined by
$z_i = (w_i,
u_i)$, where $w_i$ is a branch in $T^n_{i-1}$ and $u_i \in {\cal P}$ is the time along
that branch at which the recombination occurred.  We call $\V{Z}$ the {\em
  recombination threading} of the $n$th sequence.
For reasons of efficiency, we take a two-step approach to threading: first, 
%Our approach is first to 
we sample the
coalescence threading $\V{Y}$, and second, we sample the recombination threading
$\V{Z}$ conditional on $\V{Y}$.  This separation into two
steps allows for a substantially reduced state space during the coalescence
threading operation, leading to significant savings in computation.  When
sampling the coalescence threading (step one), we integrate over 
the locations of the new recombinations $\V{Z}$, as in previous work
\cite{Li2011a,Paul2011}.  Sampling the recombination threading (step two)
can be accomplished in a 
straightforward manner independently for each recombination event, by
taking advantage of the conditional independence structure of the DSMC
model (see 
Methods for details).

The core problem, then, is to accomplish step one by
sampling the coalescence threading $\V{Y}$ from the distribution,
\begin{align}
&P(\V{Y} \;|\; \bar{\V{T}}^{n-1}, \bar{\V{R}}^{n-1}, \bar{\V{D}}^n, \Theta) \notag 
\propto P(\V{Y}, \bar{\V{T}}^{n-1},\bar{\V{R}}^{n-1}, \bar{\V{D}}^n
\;|\; \Theta) \notag \\
%&\qquad = \sum_{\V{Z}} P(\bar{\V{T}}^{n-1},\V{Y}, \bar{\V{R}}^{n-1}, \V{Z}, \bar{\V{D}}^n \;|\; \Theta) \notag \\
& \qquad = P(\bar{T}_1^{n-1},y_1
\;|\; N) \; P(\bar{D}_1 \;|\; \bar{T}_1^{n-1}, y_1, \mu) \; \prod_{i=2}^L 
P(\bar{R}_i^{n-1}, \bar{T}_i^{n-1}, y_i  \;|\; \bar{T}_{i-1}^{n-1}, y_{i-1}, \rho, N) \;
P(\bar{D}_i \;|\; \bar{T_i}^{n-1}, y_i, \mu),
\label{eqn:reduced-abbrev}
\end{align}
where the notation $\bar{A}$ indicates that random variable $A$ is held
fixed (``clamped'') at a particular value throughout the procedure.  This
equation defines a hidden 
Markov model with a state space given by the possible values of each $y_i$,
transition probabilities given by $a^i_{l,m} = P(\bar{R}_i^{n-1},
\bar{T}_i^{n-1}, y_i=m \;|\; \bar{T}_{i-1}^{n-1}, y_{i-1}=l, \rho, N)$ and
emission probabilities given by $b_l^i(D_i^{n}) = P(D_i \;|\; \bar{T}_i^{n-1},
y_i=l, \mu)$ (Figure \ref{fig:graphmodels}C).  Notice that the location of
each new
recombination, $z_i$, is implicitly integrated out in the definition of
$a^i_{l,m}$.
Despite some unusual features of this model---for example, it has a
heterogenous state space and normalization structure along the
sequence---its Markovian dependency structure is retained, and the
problem of drawing a coalescent threading $\V{Y}$ from the
desired conditional distribution can be solved exactly by dynamic programming
using the stochastic traceback algorithm for HMMs (see Methods for details).

%=============================================================================

\subsection{Markov chain Monte Carlo sampling}

The main value of the threading operation is in its usefulness as a
building block for Markov chain Monte Carlo methods for sampling from an
approximate posterior distribution over ARGs given the data.  We employ
three main types of sampling algorithms based on threading, as described below.

% sequential sampling
\paragraph{Sequential sampling.}
First, the threading operation can be applied iteratively to a series of
orthologous sequences to obtain an ARG of size $n$ from sequence data
alone.  This method works by randomly choosing one sequence and
constructing for it a trivial ARG $\V{G}^1$ (i.e. every local tree is a single
branch).  Additional sequences are then threaded into the ARG, one at a
time, until an ARG $\V{G}^n$ of $n$ sequences has been obtained.  Notice that
an ARG derived in this manner is not a valid sample from the posterior
distribution, because each successive $\V{G}^k$ (for $k \in \{2, \dots, n-1\}$)
is sampled conditional on only $\V{D}^{1:k}$ (the first $k$ sequences). Nevertheless, the sequential sampling algorithm is an efficient
heuristic method for obtaining an initial ARG, which can subsequently be
improved by other methods.  If desired, this operation can be applied
multiple times, possibly with various permutations of the sequences, to
obtain multiple initializations of an MCMC sampler.  Heuristic methods can
also be used to choose a ``smart'' initial ordering of sequences.  For
example, one might begin with one representative of each of several
populations, to first approximate the overall ARG
structure, and subsequently add
more representatives of each population.

% Gibbs sampler
\paragraph{Gibbs sampling for single sequences.}
Second, the threading operation can serve as the basis of a Gibbs sampler
for full ARGs.  Starting with an initial ARG of $n$ sequences, individual
sequences can be removed, randomly or in round-robin fashion, and
rethreaded.  Since the threading procedure samples from the conditional
distribution $P(\V{G}^n \;|\; \V{G}^{n-1}, \V{D}^n, \Theta)$, this produces a valid
Gibbs sampler for the ARG up to the assumptions of the DSMC.
The ergodicity of the Markov chain follows, essentially, from the fact that
any tree is reachable from any other by a finite sequence of branch
removals and additions (see
Text S1 for details).

The main limitation of this method is that it leads to poor mixing when the
number of sequences grows large.  The essential problem is that rethreading a
single sequence is equivalent to resampling the placement of external
branches in the local trees, so this method is highly inefficient at
rearranging the ``deep structure'' (internal branches) of the ARG.
Furthermore, this mixing 
problem becomes progressively worse as $n$ grows.  
As a result, an alternative strategy is needed for large numbers of sequences.
%We have found
%that this strategy is sufficient for no more than 6--8 sequences (Supplemental
%Figure XXX), and other strategies become necessary with larger data sets.

% MCMC sampling
\paragraph{Subtree sampling.}
The third sampling strategy addresses the mixing limitations of the Gibbs
sampler by generalizing the threading operation to accommodate not only
individual sequences but subtrees with arbitrary numbers of leaves.  As a
result, internal branches in the local trees can be resampled and the deep
structure of the ARG can be perturbed.  The subtree threading problem is
considerably more difficult than the single-sequence threading problem,
because, in general, the subtrees change in composition and structure along
the sequence.
% maybe this should be first?
In principle, one could arbitrarily select a subtree for each
nonrecombining segment and resample its attachment point to the remainder
of the tree, but because the attachment points at both ends of a segment
would be constrained by the flanking local trees, there would be a strong
tendency to resample the original attachment points, resulting in poor
mixing of the sampler.  Instead, we use an approach that allows us to
select a sequence of subtrees guaranteed to have good continuity
properties in order to enable long-range rethreading of internal branches.  To
select these sequences of subtrees, we use a data structure called a {\em
  branch graph}, which traces the parent/child relationships among branches
across genomic positions.  Using dynamic programming, it is possible to
identify paths through the branch graph that correspond to sequences of
internal branches with good continuity properties, resulting in efficient
sampling of subtree threadings (see Text S1 for details).

After a sequence of internal branches is identified, 
the selected branch is removed from each local tree, splitting it into a
main tree and a subtree.  A new branch is then added above the root of
every subtree and allowed to re-coalesce with the corresponding main tree
in a manner consistent with the DSMC.  As with the single-sequence threading
operation, it is possible to sample directly from the desired conditional
distribution under the DSMC.  However, since the number of ways of removing
internal branches depends on the current structure of the ARG, the Hastings
ratio is not equal to one in this case, and a more general
Metropolis-Hastings algorithm (with rejection of some proposed threadings)
is required (see Text S1 for details).  In practice, the
acceptance rates for proposed threadings are fairly high ($\sim$40\% for
typical human data) and this strategy
substantially improves the mixing properties of the Gibbs sampler.  This
generalized sampling strategy allows the number of sequences to be
increased substantially (see below).

\subsection{{\em ARGweaver} Program and Visualization}

We implemented these sampling strategies in a computer program called \wv,
that ``weaves'' together an ARG by repeated applications of the threading
operation.  The program has subroutines for threading of both individual
sequences and subtrees.  Options allow it to be run as a Gibbs sampler with
single-sequence threading or a general Metropolis-Hastings sampler with
subtree threading.  In either case, sequential sampling is used to obtain
an initial ARG.  Options to the program specify the number of sampling
iterations and the frequency with which samples are recorded.  The program
is written in a combination of C++ and Python and is reasonably well
optimized.  For example, it 
requires about 1 second to sample a threading of a single 1 Mb sequence in
an ARG of 20 sequences with 20 time steps. Our source code is freely available
via GitHub (\url{https://github.com/mdrasmus/argweaver}).

To summarize and visualize samples from the posterior distribution over
ARGs, we use two main strategies.  First, we summarize the sampled ARGs
in terms of the time to most recent common ancestor (TMRCA) and total
branch length at each position along the genome.  We also consider the
estimated age of the derived alleles at polymorphic sites, which we obtain
by mapping the mutation to a branch in the local tree and calculating the
average time for that branch (see Methods).  We compute posterior mean and
95\% credible intervals for each of these statistics per genomic position,
and create genome browser tracks that allow these values to be visualized
together with other genomic annotations.

Second, we developed a novel visualization device for ARGs called a
``leaf trace.''  A leaf trace contains a line for each haploid sequence in
an analyzed data set.  These lines are ordered according to the local
genealogy at each position in the genome, and the spacing between adjacent
lines is proportional to their TMRCAs
(Figure \ref{fig:leaf-trace}).  The lines are parallel in nonrecombining
segments of the genome, and change 
in order or spacing where recombinations occur.
As a result, several features of interest are immediately evident
from a leaf trace.  For example, recombination hot spots show up as regions
with dense clusters of vertical lines, whereas recombination cold spots are
indicated by long blocks of parallel lines.  

%=============================================================================

\subsection{Simulation Study}

\subsubsection{Effects of Discretization and Convergence of Sampler}
Before turning to inference, we performed a series of preliminary
experiments to verify that our discretization strategy allowed for an
adequate fit to the data,
and to ensure that \wv
was capable of converging to a close approximation of the true
posterior distribution for realistic simulated data sets.
Briefly, 
%following McVean and Cardin \cite{McVean2005}, 
we found 
that the DSMC 
produces similar numbers of recombination counts and segregating sites as
the coalescent-with-recombination and SMC,
when generating data under various recombination rates 
and effective population sizes (see Text S1 and Supplementary Figure
\ref{fig:sim-dsmc}). 
With small numbers of sequences, the Gibbs sampler based on the
single-sequence threading operation appeared to converge rapidly, according to
both the log likelihood of the sampled ARG
and the inferred numbers of recombination events.   
% put this in the caption
%(Notice that
%the complete data log likelihood---the logarithm of equation
%\ref{eqn:dsmc-lik}---is a measure that captures both recombination and
%mutation events.)
When the number of sequences grew larger than about 6--8
(depending on the specific details of the simulation), the Gibbs sampling
strategy was no longer adequate.
However, the subtree threading operation and 
Metropolis-Hastings sampler 
appeared to address this problem effectively, allowing the number of
sequences to be pushed to 20 or more.  With 20 
sequences 1 Mb in length, the sampler converges to true values within about
500 sampling iterations, which takes about 20 minutes on a typical desktop
computer (Supplementary Figure \ref{fig:sim-converge}).
% FIXME: say which simulated data set used here
% perhaps give processor details in fig caption?
% mention limitations of Gibbs strategy in caption
% give the details on the simulated data sets in the caption

% may need to do a bit more with convergence diagnostics.  See Fearnhead
% and Donnelly.  (after CSHL)

\subsubsection{Recovery of Global ARG Features}

Next, we systematically assessed the ability of \wv to recover several
features of interest from simulated ARGs over a range of plausible ratios
of mutation to recombination rates (see Methods for simulation parameters).
In these experiments, we considered three ``global'' features of the ARG:
(i) the log joint probability of the ARG and the data (log of equation
\ref{eqn:dsmc-lik}), 
(ii) the total number of recombinations, and (iii) the total branch length
of the ARG.  We define the total branch length of the ARG to be the sum of
the total branch lengths of the local trees at all sites, a quantity
proportional to the expected number of mutations in the history of the
sample.
%and, if the population-scaled mutation rate is low (as it is here),
%one approximately proportional to the expected number of segregating sites.
%We examined these features in
%data sets simulated with six different
%recombination-to-mutation rate ratios ($\mu/\rho$), ranging from 1--6 (data
%set \#3, Table \ref{tab:sim-params}).
We applied \wv to each simulated data set with 500 burn-in iterations,
followed by 
1000 sampling iterations, with every tenth sample retained (100 samples total).

We found that \wv was able to recover the features of interest with fairly high
accuracy at all parameter settings (Figure
\ref{fig:sim-eval-multipanel}A and Supplementary Figure
\ref{fig:sim-stats-eval}).  In addition, the variance of our estimates is
generally fairly low, but does show a clear reduction
as $\mu/\rho$ increases from 1 to 6.
Most current estimates of average
rates would place the true value of $\mu/\rho$ for
human populations between 1 and 2 \cite{KONGETAL02,KONGETAL12,SUNETAL12},
%suggesting that we may expect substantial variance in our estimates for
%real data.
but the concentration of recombination events in hot spots implies that the
ratio should be considerably more favorable for our methods across most of
the genome.  Notably, we do observe a slight tendency to under-estimate the
number 
of recombinations, particularly at low values of $\mu/\rho$.  This underestimation is paired with an over-estimation of the joint
probability (left column), suggesting that it reflects model
misspecification of the DSMC.  It is possible that this bias could be improved
by the use of the SMC$'$ rather than the SMC, or by a finer-grained
discretization scheme (see Discussion).

\subsubsection{Recovery of Local ARG Features}

% FIXME: this section needs to be reworked with expanded results from
% Matt's new experiments.  Try to include more comparisons across values of
% mu/rho, show improvement in estimation of mutation rage relative to
% frequency alone, etc.

An advantage of explicitly sampling full ARGs is that it enables inferences
about local features of the ARG that are not directly determined by model
parameters.  Using the same simulated data and
inference procedure as in the previous section, we evaluated the
performance of \wv in estimating three representative quantities along the
genome sequence: (i) time to most recent common ancestry (TMRCA), (ii)
recombination rate, and (iii) allele age.  We estimated each quantity using
an approximate posterior expected value, computed by averaging across
sampled ARGs.  
%The TMRCA was computed directly per genomic position, while
%the recombination rate was computed in a sliding window of 1 kb.  The
%mutation age was computed only at polymorphic sites.  
With 20 sequences, we
found that \wv was able to recover the TMRCA with fairly high accuracy
and resolution (Figure \ref{fig:sim-eval-multipanel}B).  The quality
of the estimates degrades somewhat at lower values of the ratio $\mu/\rho$ but
remains quite good even with $\mu/\rho=1$ (Supplementary Figure
\ref{fig:tmrca-full}). 
%To study recombination rates, we generated simulated data by a slightly
%modified  procedure, allowing for intermittent recombination hot
%spots designed to mirror those in real human sequences.  
We found that our power for recombination rates
was weak with only 20 sequences, but with 
100 sequences the reconstructed ARGs clearly displayed elevated rates of
recombination in simulated hotspots compared with the flanking regions
(Supplementary Figure 
\ref{fig:sim-recomb-map}).
% FIXME: need more info: how high was the true rate in the hotspots.  how
% big were they?  how realistic are these parameter settings?
Estimates of allele ages appeared to be unbiased, with good
concordance between true and estimated values, although the variance in the
estimates was fairly high (Supplementary Figure \ref{fig:alleleAgeSim},
left column).
Notably, the ARG-based estimates of allele age appear to be considerably
better than estimates based on allele-frequency alone  (Supplementary
Figure \ref{fig:alleleAgeSim}, right column).
Together, these results suggest that, even with modest numbers of
sequences, the distributions of ARGs inferred by our methods may be
informative about loci under natural selection, local recombination rates,
and other local features of evolutionary history.

%informative about selective sweeps, background selection, and other local
%influences from natural selection.

\subsubsection{Accuracy of Local Tree Topologies}

In our next experiment, we evaluated the accuracy of \wv in inferring
the topology of the local trees, again using the same simulated data.  The
local trees are a more complex 
feature of the ARG but are of particular interest for applications
such as genotype imputation and association
mapping.  
%running the MCMC
%sampler each time for 1000 iterations (200 burn-in).
% be clear done for four values of mu/rho
% something about burn-in?  
% considering multiple samples from posterior or just one here?
For comparison, we also inferred local trees using the heuristic {\em Margarita} program
\cite{Minichiello2006}, which is, to our
knowledge, the only other available
ARG-inference method that can be applied at
this scale.  To compare the two programs, we identified
100 evenly spaced locations in our simulated data sets, and extracted
the local trees reconstructed by both methods at these positions.  
%We
%compared the trees inferred by \wv and {\em Margarita} with the true
%local trees generated during simulations with various values of $\mu/\rho$.
%by two measures of
%topological similarity: branch correctness (one minus the
%normalized Robinson-Foulds (RF) distance \cite{Robinson1981}) and Maximum
%Agreement Subtree (MAST) percentage (the size of the largest leaf-set
%such that induced subtrees are topologically equivalent,
%expressed as 
%a percentage of the total number of leaves).  
% FIXME: perhaps should move this to methods and include more detail
We found that \wv produced substantially more 
accurate local tree topologies than {\em Margarita} 
(Supplementary Figure \ref{fig:sim-top-eval}).
% FIXME: have we done an actual statistical test?  Should be clear about this
The improvements were most pronounced at high $\mu/\rho$ values (where
topological information is greatest) but were evident across all ratios
considered.  
%They were also most pronounced when measured
%by MAST percentage, which is less sensitive to minor differences between
%trees than the branch-wise measure.  
In addition, the absolute accuracy of
the trees inferred by \wv was fairly high, given the sparseness of
informative sites in these data sets.  
%For example, at $\mu/\rho =6$, more
%than four of five predicted branches were correct and MAST percentages
%approached 75\%, and even in the challenging case of $\mu/\rho =1$, roughly
%two of three branches were correct and MAST percentages exceeded 50\%.
These results
indicate that the sampler is effectively pooling information from
many sites across the multiple alignment in making inferences about 
local tree topologies.

% mention absolute measures first?

%\subsubsection{Accuracy of Branch Uncertainty}

Finally, we evaluated the accuracy of
%In principle, a fully Bayesian method for ARG inference should also provide 
%not only accurate estimates of ARG features of interest, but also 
the \wv's assessment of the uncertainty in the local trees given the data.
%the performance of \wv in measuring uncertainty of this kind, 
We grouped
individual branches into bins according to their 
estimated posterior probabilities (i.e., the
fraction of sampled local trees in which each branch is found), and
compared these values with 
the relative frequencies with which the same branches were observed in
the true trees.  
We found that 
the predicted and actual probabilities of correctness were closely
correlated, indicating that \wv is accurately measuring the uncertainty
associated with the local trees (Supplementary Figure
\ref{fig:sim-uncertainty}).  By 
contrast, the heuristic {\em Margarita} sampler shows a clear tendency to
overestimate the confidence associated with branches in the local trees,
often by 10--20\%.  This comparison is is not entirely fair, because the
authors of {\em Margarita} do not claim that it samples from the posterior
distribution, but it nevertheless highlights an important advantages of the
Bayesian approach.  
%The ability to accurately measure uncertainty in
%features of the ARG has many potential benefits, including improved
%accuracy of posterior expected values that depend on highly uncertain
%aspects of the ARG (such as mutation age estimates) and improved tests of
%statistical significance.

%=============================================================================

\subsection{Analysis of Real Data}

Having demonstrated that \wv was able to recover many features of simulated
ARGs with reasonable accuracy, we turned to an analysis of real human
genome sequences.  For this analysis we chose to focus on sequences for 54
unrelated individuals from the ``69 genomes'' data set from Complete
Genomics (\url{http://www.completegenomics.com/public-data/69-Genomes})
\cite{Drmanac2010}.  
% maybe could mention (i) and (ii) in passing
%This data set is well suited for an initial analysis
%with \wv because (i) the sequences are of high coverage and high quality,
%minimizing any potential impact from genotyping error; (ii) they include
%representation from multiple, diverse human populations; and (iii) the
%number of sequences is of a scale that can be accommodated 
%with our current implementation.  
% update below based on re-analysis.  Check with Melissa
The 54 genome sequences were computationally phased using SHAPEIT v2
\cite{DELAETAL13} and were 
filtered in 
various ways to minimize the influence from alignment and genotype-calling
errors. 
They were partitioned into $\sim$2-Mb blocks and \wv was applied to these
blocks in parallel using the Extreme Science and
Engineering Discovery Environment 
(XSEDE).  For this analysis, we assumed
$K=20$, $s_K=$ 1,000,000 generations, and $N=$ 11,534.
We allowed for
variation across loci in mutation and recombination rates.  For each $\sim$2-Mb
block, we collected samples for 2,000 iterations of the sampler and retained
every tenth sample, after an appropriate burn-in (see Methods for complete
details).   The entire 
procedure took $\sim$36 hours for each of the 1,376 2-Mb blocks, or 5.7
CPU-years of total compute time.  
The sampled ARGs were summarized 
by UCSC Genome Browser tracks
describing site-specific times to most recent common ancestry (TMRCA),
total branch length, allele ages, leaf traces,
and other features across the human genome.
These tracks are publicly available from our
local mirror 
of the UCSC Genome Browser (\url{http://genome-mirror.bscb.cornell.edu},
assembly hg19).  

\subsubsection{Distortions in the ARG due to Natural Selection}

While our prior distribution over ARGs 
is based on the neutral coalescent, 
we were interested in exploring whether 
natural selection produces a sufficiently strong signal in the
data to create detectable distortions in the ARG near functional elements.
We began by examining the estimated posterior expected values of the TMRCA
%and total branch length 
around known protein-coding genes, focusing on fourfold degenerate (4d)
sites within coding exons and noncoding sites flanking exons.
For comparison with our ARG-based measures, we also computed a simple
measure of nucleotide 
diversity, $\pi$.  Both $\pi$ and the ARG-based TMRCA behave in a
qualitatively similar 
manner near genes, achieving minimal values in coding exons and
gradually 
increasing with distance from exon boundaries
(Figure~\ref{fig:metagene-sweeps}A). 
These observations are consistent with several recent studies indicating
reduced neutral diversity near both coding and noncoding functional
elements, which has been attributed to indirect effects from
selection at linked sites
\cite{MCVIETAL09,CAIETAL09,HERNETAL11,GOTTETAL11,LOHMETAL11}.  However, it has been
difficult to distinguish between two alternative modes of selection 
both predicted to have similar influences on patterns of neutral diversity: 
``background selection'' (BGS) associated with negative or purifying
selection at linked sites \cite{CHARETAL93,HUDSKAPL95,NORDETAL96,CHAR12},
and ``hitchhiking'' (HH) (selective sweeps) associated with linked
mutations under 
positive selection \cite{MAYNHAIG74}.  In principle, explicit ARG inference
could help to resolve this controversy, because BGS and HH lead to
different predictions for the structure of genealogies (e.g.,
\cite{BART98,WALCETAL12}).

% do we need to include more explicit argument for selection from linked sites?
% Don't know *which* elements but has to be BGS -- effect will be same if
% direct selection on subset of closely linked sites or on all sites.  But
% allow could be BGS or HH
% something about this in caption?

To examine these questions further, 
%We attempted to shed light on these questions by comparing patterns of
%diversity at protein-coding genes and
%at regions predicted to have undergone recent selective
%sweeps.  In particular, 
we computed the same three statistics for 255
putative selective sweeps identified in CEU populations and 271 sweeps
identified in YRI populations based on the integrated extended haplotype
homozygosity statistic (iHS) \cite{VOIGETAL06}.
%(We excluded sweeps identified in East Asian [ASN]
%populations because the Complete Genomics data contains relatively few
%individuals from comparable populations.)
As expected, the sweep regions were broadly similar to the
protein-coding genes in terms of nucleotide diversity $\pi$ (Figure
\ref{fig:metagene-sweeps}B). 
However, unlike the protein-coding genes, the sweep regions displayed no
clear depression in TMRCA.
One possible way of understanding this observation is that, while 
sweeps tend to be enriched overall for recent coalescence events 
(as indicated by the reductions in $\pi$),
the oldest coalescence events are relatively unaffected by selective
sweeps, perhaps because 
some lineages tend to 
 ``escape'' each sweep, leading to near-neutral patterns of coalescence
 near the roots of genealogies (where the contribution to the TMRCA is
 greatest).  This may be particularly true for the partial sweeps
 identified by the iHS method, but a similar phenomenon should occur in
 flanking regions of the causal mutations for complete sweeps.
%In other words, unless a sweep results in complete fixation of the
%advantageous allele, it may not lead to a pronounced depression
%in the TMRCA.  
BGS, by contrast, is expected to affect both the total branch
length and TMRCA approximately equally, by effectively reducing the time
scale of the coalescence process, 
but to have a minimal influence on the
relative intervals between coalescence events.

% comment on difficulty of distinguishing BGS and HH.  Look at sweeps for
%  contrast.  Notice simility in total branch length but difference in
%  TMRCA.  One possible way of understanding this is....
% suggest possibility that lineages escape, distribution of coalescences
%  is different even if total branch length is affected in a similar way.
%  Maybe start with distribution idea, then go to escape.

% propose new summary statistic, RTH, as a way of getting at "bursts" of
%  coalescence that are incomplete.  Look at RTH for genes and sweeps.  Note
%  stark contrast.
% suggest consistent with BGS rather than HH dominating in genes.  Caveats:
%  recurrent sweeps, sweeps of different ages, complete sweeps, etc.
%  (sweeps could occur in a different way)

In an attempt to distinguish further between BGS and HH, we introduced a
statistic called the {\em relative TMRCA halflife} (RTH), 
% or should it be relative half-TMRCA?
%which
%is easily computed from a fully defined genealogy and can therefore be
%calculated for every genomic position based on an ARG.  The RTH is 
defined
as the ratio between the time to most recent common ancestry for the first
50\% of chromosomes and the full TMRCA.  The RTH captures the degree to
which coalescence events are skewed toward the recent past, in a manner that
does not depend on the
overall rate of coalescence.  Thus, the RTH should be relatively insensitive to
BGS, but sensitive to HH if, as proposed above, sweeps tend to
affect many but not all lineages (see Supplementary Figure
\ref{fig:rth-conceptual}). 
%Because the identified sweeps tend to be population-specific, we computed the
%RTH separately for the European (CEU and TSI) and African (YRI, LWK, and
%MKK) individuals in our data set. 
%
In the European populations, the statistic showed a pronounced valley near
selective sweeps (Figure \ref{fig:metagene-sweeps}B), as expected, but it
was much more constant across genic regions (Figure
\ref{fig:metagene-sweeps}A).  Its behavior was similar in the African
populations, except that it showed somewhat more variability near genes, yet
in an opposite pattern from the sweeps (Supplementary
Figure \ref{fig:metagene-sweeps-afr}).  
%Particular examples of known selective sweeps, such as the
%lactase ({\em LCT}) gene, support the general trend by exhibiting 
%striking depressions in RTH but TMRCAs similar to those in flanking regions
%(Figure \ref{}).
Overall, these results suggest that, while the total rate
of coalescence differs substantially across genic regions, the relative
depths of middle and extreme coalescence events do not, consistent with the
predictions of a model in which BGS dominates in genes.  By contrast, the
sharp decrease in the RTH within the sweeps is striking, especially
considering that these sweeps were identified using rather different
methods and data from ours.  These observations do not rule out the
possibility that alternative modes of hitchhiking---such as recurrent hard
selective sweeps---might make a non-negligible contribution to patterns of
variation near human protein-coding genes, but they generally 
support the emerging view that BGS likely plays a dominant role in
determining these patterns \cite{MCVIETAL09,HERNETAL11,LOHMETAL11}.

\subsubsection{Genomic Regions with Extremely Ancient Most Recent Common
  Ancestry} 

%Another possible use of the ARG is in identifying loci under selection.
%In general, natural selection on a particular mutation, or multiple
%closely linked mutations, is expected to distort the distribution of
%coalescence times in the surrounding region.  In turn, several properties
%readily estimatable from the ARG---such as the TMRCA, the lengths of 
%haplotypes, the ages of alleles, or the density along the genome in
%recombination events---should be affected by these differences in
%coalescence time.
% too much foreshadowing?
%For example, unusually large TMRCAs might reflect selection to maintain
%nucleotide diversity (balancing selection), for example, due to
%heterozygote advantage or frequency-dependent selection.  Similarly,
%unusually small TMRCAs might reflect the elimination of diversity in a
%region surrounding an advantageous allele due to a (hard) selective sweep.

The previous section showed that genomic regions with reduced TMRCAs are
often associated with purifying selection.  To see whether the opposite
signal was also of interest, we computed the posterior expected TMRCA in
10-kb blocks across the human genome and examined the regions
displaying the oldest shared ancestry.  Not surprisingly, four of the top
twenty 10-kb blocks by TMRCA fall in the human leukocyte antigen (HLA)
region on chromosome 6 (see Table \ref{tab:highTmrca}).  It has been known
for decades that that the HLA region exhibits extraordinary levels of
genetic diversity, which is believed to be maintained by some type of
balancing selection (overdominance or frequency-dependent selection)
associated with the immunity-related functions of the HLA system
\cite{HUGHNEI88,APANETAL97,HUGHYEAG98}.  The four HLA-related high-TMRCA
blocks include three regions near {\em HLA-F} and one region between {\em
  HLA-A} and {\em HLA-J} (Supplementary Figure \ref{fig:hla}).
All four high-TMRCA regions exhibit more than 12 polymorphisms per kilobase
of unfiltered sequence, 8--10 times the expected neutral rate after
normalizing for local mutation rates.  The estimated TMRCAs for these
regions range from $\sim$340,000--380,000 generations, or $\sim$8.5--9.5
My (assuming 25-year generations).

Among these high-TMRCA blocks were two additional regions that displayed
extraordinary levels of mutation-rate-normalized nucleotide
diversity.  The first of these, in a gene desert near the
telomere of the long arm of chromosome 4, exhibits the deepest expected
TMRCA in the genome, at $>$600,000 generations (15 My), and has $>$30 times the neutral polymorphism rate (Table
\ref{tab:highTmrca}).  The second region is the {\em PRIM2} gene on
chromosome 6, which contributes the 4th and 7th highest TMRCA blocks in the
genome, exhibiting polymorphism rates 28.0 and 12.8 times the neutral
expectation, respectively.  Both of these regions were identified as extreme
outliers in a recent study of coincident SNPs in humans and chimpanzees,
and it was argued that the {\em PRIM2} gene was a likely target of
balancing selection \cite{HODGEYRE10}.
%Hodgkinson
%and Eyre-Walker identified these same two regions as extreme outliers in a
%scan for coincident SNPs in human and chimpanzee.  They were skeptical
%about the biological significance of the region on chromosome 4, but argued
%that {\em PRIM2} may be under balancing selection, based on both the strong
%enrichment for coincident SNPs and a significant positive skew in this
%region in
%Tajima's $D$ in human populations \cite{HODGEYRE10}.
On closer inspection, however, we found that both regions were
flagged by Complete Genomics as having ``hypervariable'' or ``invariant''
read depth across individuals,
suggesting that the elevated SNP rates in our data 
are likely artifacts of copy number 
variation (CNV)
at loci unduplicated in the reference genome.  (Leffler et al.\
recently reached a similar conclusion about {\em PRIM2} \cite{LEFFETAL13}.)
Despite that these flags were associated with only $\sim$5\% of genomic
positions, they indicated that
five of our top six regions were likely CNVs (Table \ref{tab:highTmrca}).  
%These five regions included
%the block at the telomere of chromosome 4 and both {\em PRIM2} blocks, the
%only three non-{\em HLA} regions to display normalized polymorphism rates
%in excess of about five times the neutral expectation.  Manual inspection
%of these loci suggests that it is likely that the SNP rates in these
%regions are indeed spuriously elevated owing to mapping ambiguity from
%CNVs.  
%This finding demonstrates the importance of the use of rigorous data
%quality filters in interpreting the inferred ARGs, particularly when
%extreme outliers are identified.  
%They also suggest that reported
%predictions of balancing selection should perhaps be viewed with skepticism
%until the influence of CNVs and other sources of elevated SNP rates is
%fully understood.  
Thus, for all subsequent analyses reported in this paper and
for our publicly available browser tracks, we filtered out all regions
labeled as invariant or hypervariable.

% careful with the Eyre-Walker thing.  Don't be too heavy handed.  Are we
% sure CG is right?  maybe should look at it more closely

%%%%%

Once these extreme outliers were excluded, 
several loci of interest remained.  In addition to the four {\em HLA} loci,
these included
(\#5 in Table \ref{tab:highTmrca}) an apparent {\em cis}-regulatory region downstream of the {\em KCNE4} gene, which
encodes a potassium voltage-gated channel (Supplementary Figure \ref{fig:kcne4});
(\#9)~an intronic interval in {\em BCAR3}, a gene involved in the
development of 
anti-estrogen 
resistance in breast cancer (Supplementary Figure \ref{fig:bcar3});
%(\#13) an intronic interval in {\em CEP112}, which encodes a centrosomal
%protein; 
(\#16)~an apparent regulatory region upstream of {\em TULP4}, a tubby-like
protein that may be involved in 
ubiquitination and proteasomal degradation with a possible
association with cleft lip (Supplementary Figure \ref{fig:tulp4}); 
and (\#18)~an intronic region in {\em CRHR1}, 
which encodes a GPCR that binds corticotropin
releasing hormones, has roles in  
in stress, reproduction, immunity, and obesity, and is
associated with alcohol abuse, asthma, and depression.
Notably, all of these are predominantly noncoding regions
that include multiple ChIP-seq-supported transcription factor binding
sites.  
%All of them are in regions of conserved synteny among the great
%apes with no evidence of recent human duplication and high-quality SNP calls.
The estimated TMRCAs of these regions range from 335,000--450,000
generations, (8.4--11.3 My), suggesting that genetic variation in these
loci considerably predates the human/chimpanzee divergence.

\subsubsection{Segregating Haplotypes Shared Between Humans and Chimpanzees}

%The inclusion of the HLA locus and other regions predicted to be under
%balancing selection among the highest-TMRCA regions in the genome suggests
%that extreme values of the \wv-estimated expected TMRCA may generally be
%useful as a predictor of balancing selection.  
To explore the connection between extreme TMRCAs and balancing selection
further, we examined 125 loci recently identified as having segregating
haplotypes that are shared between humans and chimpanzees
\cite{LEFFETAL13}.  These loci are expected to be enriched for ancient
polymorphisms maintained by balancing selection, although some may reflect
independent occurrences of the same mutation in both species.  We compared
these putative balancing selection loci with neutral sequences having the
same length distribution (see Methods), and found that their \wv-estimated
TMRCAs were clearly shifted toward higher values, with a mean value nearly
twice as large as that of the neutral sequences (Supplementary Figure
\ref{fig:przeworski}).  In addition, the putative balancing selection loci
that do not contain polymorphisms in CpG dinucleotides---which are less
likely to have experienced parallel mutations---had slightly higher TMRCAs
than the group as a whole.

If these loci are sorted by their estimated TMRCAs, several loci that were
highlighted by Leffler et al.\ \cite{LEFFETAL13} for having more than two
pairs of shared SNPs in high LD appear near the top of the list (Table
\ref{tab:lefflerTmrca}).  For example, the haplotype between the {\em
  FREM3} and {\em GYPE} genes (\#11 in Table \ref{tab:lefflerTmrca};
Supplementary Figure \ref{fig:frem3}) contains shared SNPs in
almost perfect LD with several expression quantitative trait loci (eQTLs)
for {\em GYPE}, a close paralog of a gene ({\em GYPA}) that encodes a
receptor for {\em Plasmodium falciparum} and may be under balancing
selection.  Another haplotype (\#3) contains shared SNPs
in significant LD with an eQTL for {\em MTRR}, a gene implicated in the
regulation of folate metabolism, including one SNP that is also segregating
in gorillas.  In a third case (\#18), the shared SNPs occur in a likely
enhancer in 
an intron of {\em IGFBP7}, a gene that plays a role in innate immunity,
among other functions.  Another example is a locus near the {\em ST3GAL1}
gene (\#7) that contains only one pair of shared SNPs but was suggested by
a phylogenetic analysis to have an ancient origin \cite{LEFFETAL13}.
Notably, all of these shared haplotypes fall outside of coding regions and
several show signs of regulatory activity based on functional genomic data
\cite{LEFFETAL13}.  Their expected TMRCAs range from roughly 150,000 to
250,000 generations, or 3.8--6.3 My.  Thus, the ARGweaver estimates of age
are reasonably consistent with the hypothesis that these hapolotypes
predate the human/chimpanzee divergence (estimated at 3.7--6.6 Mya
\cite{SUNETAL12}), an observation that is especially notable given that our
analysis does not make direct use of data from chimpanzees.  
%In addition,
%the regions with very high TMRCAs are significantly depleted for CpG SNPs
%($p=0.04$, one-sided Fisher's exact test), suggesting that they are less
%likely to have undergone parallel mutations.
% this is based on 5 Cpgs and 15 non CpGs among top 20, compared with 51
% and 54 in the remainder of the list 
%Notably, Leffler et al.\ were careful to filter out duplicated regions from
%their analysis and 
%only one of the top 20 loci by TMRCA overlaps likely CNVs according to the
%Complete 
%Genomics tracks (Table \ref{tab:lefflerTmrca}).

By contrast, the loci near the bottom of the list (with the shortest
TMRCAs) appear to be much less convincing.  For example, the bottom 20 have
expected ages of only 25,000--50,000 generations (0.65--1.3 My), suggesting
that they actually post-date the human/chimpanzee divergence by millions of
years.  In addition, many of these regions appear hundreds of kilobases
from the nearest gene, and they typically do not overlap regions with
strong functional or comparative genomic evidence of regulatory potential.
Indeed, if our ARG-based estimates of the TMRCA are interpreted literally,
a majority of the 125 segregating haplotypes may post-date the
human/chimpanzee divergence, which current estimates would place at
$\geq$150,000 generations ago (see Supplementary Figure
\ref{fig:przeworski}).  This observation is in general agreement with rough
calculations by Leffler et al.\ suggesting that the false discovery rate
for ancient balancing selection in this set could be as high as 75\%
\cite{LEFFETAL13}.  Thus, it appears that our ARG-based methods may be
useful in distinguishing true ancestral polymorphisms from shared
haplotypes that occur by chance due to homoplasy.

%possibly more useful
% than would measures of
%total diversity or total branch length, which have been used in some cases
%(cite DeGiorgio, others), because balancing selection is primarily expected
%to maintain two alleles in the population, and to have a minimal impact on
%patterns of coalescence within each allele class.  
% not sure want to go here

\subsubsection{Natural Selection and Allele Age}

Next we examined the ARG-based expected ages of derived alleles at
polymorphic sites in various annotation classes.
%As shown from our simulation study (UPDATE), another statistic that is
%potentially informative about natural selection is the ARG-based expected
%age of the derived allele at each polymorphic site.  
Classical theory
predicts that both deleterious and advantageous alleles will not only
have skewed population frequencies but will also tend to be younger than
neutral alleles at the same frequency, because directional selection will tend to 
accelerate a new mutation's path to fixation or loss \cite{MARU74}.  This 
idea has recently been used to characterize selection in the human genome
based on a haplotype-based summary statistic that serves as a proxy for allele
age \cite{KIEZETAL13}.
% should some of this background go in simulation section instead?
%To see whether our more direct, ARG-based predictions of age were
%informative about natural selection, 
We computed ARG-based estimates of allele age in
putatively neutral regions (Neut), fourfold degenerate sites in coding
regions (4d), conserved noncoding sequences (CNS), missense coding
mutations predicted by PolyPhen-2 to be ``benign'' (PPh:Benign), ``possibly
damaging'' (PPh:PosDam), or ``probably damaging'' (PPh:ProbDam), and coding
or noncoding mutations classified by the ClinVar database
(\url{http://www.ncbi.nlm.nih.gov/clinvar}) as ``nonpathogenic''
(categories 1--3; CV:NonPath) or ``pathogenic'' (categories 4 \& 5;
CV:Path) based on direct supporting evidence of phenotypic effects.  
% FIXME: put something about ClinVar in caption
%The ClinVar sets contain both coding and noncoding mutations but
%they are currently dominated by missense mutations in coding regions.
% be clear that this is based on direct evidence of pathogenicity (CHECK)
% rather than computational predictions
We found, indeed, that the Neut mutations were significantly older, on
average, than 
all other classes (Figure \ref{fig:alleleAge}A).  In addition, among the
missense coding mutations, PPh:Benign mutations were the oldest, PPh-PosDam
were significantly younger, and PPh-ProbDam mutations were the youngest.
Similarly, mutations in the CV:NonPath class were significantly older than
those in the CV:Path class.  Interestingly, the 4d mutations showed
substantially lower
average ages (by $>$30\%) than the Neut mutations.  We attribute this
reduction primarily to
the effects of selection from linked sites (see \cite{MCVIETAL09}),
although direct selection from mRNA secondary structure and exonic
regulatory elements may also contribute to it.

% better to leave this for the later discussion
%Interestingly, 4d mutations were significantly younger than CNS mutations,
%suggesting that their increased influence from selection at
%linked sites more than offsets a reduced influence from direct selection.
%PPh:Benign, and CV:NonPath mutations.  The reduction in age of 4d compared
%with CNS mutations suggests that the increased influence from selection at
%linked sites more than offsets the reduced influence from direct selection
%at 4d sites.  The slight reduction in age of 4d compared with the
%PPh:Benign and CV:NonPath mutations may reflect constraints in 4d sites due
%to mRNA secondary structure, exonic splicing enhancers, or other features.

In part, these differences in age simply reflect differences in the
site frequency spectrum (SFS) across classes of mutations.  For example,
missense mutations are well known to be enriched for low-frequency derived
alleles, which will tend to be younger, on average, than higher-frequency
derived alleles.  To account for the influence of allele frequency, we further
grouped the sites in each annotation class by derived allele frequency and
compared the average allele ages within each group (Figure
\ref{fig:alleleAge}B).  As expected, the estimated ages increase with the
derived allele frequency across all annotation classes.  In addition, within each
class we continue to observe approximately the expected rank-order in
allele ages, with Neutral mutations being the oldest, 4d, PPh:Benign, CNS,
and CV:NonPath mutations coming next, followed by PPh:PosDam, PPh:ProbDam,
and CV:Path mutations.  This analysis demonstrates that \wv is able to obtain
information about natural selection from allele ages beyond what can be
obtained from the SFS alone.

% check Slatkin and Rannala paper  1997.  Worth citing?

Another way of viewing these results is to consider the reduction in allele
age relative to the neutral expectation within each frequency group, across
annotation classes (Supplementary Figure \ref{fig:alleleAgeDiff}).  As
expected, these reductions are larger at higher allele frequencies, where
sojourn times will tend to be longer.  However, from this representation it
is also clear that the reductions in age increase with frequency much more
rapidly for the mutations under strong, direct selection
than for the mutations at
which selection from linked sites is expected to dominate.  For example, at
very low derived 
allele frequencies (singletons), the reduction in age of 4d mutations is
roughly equal to that at PPh:PosDam mutations, whereas at higher derived
allele frequencies the damaging mutations exhibit reductions in age 2--3
times larger.  
% ask Shamil for feedback on this?
The reason for this observation is probably that the reduction in age for
the nearly neutral sites is largely a consequence of reductions in drift,
while the reductions at selected sites are more directly driven by the
influence of directional selection on sojourn times (see 
Supplementary Figure \ref{fig:alleleAgeDiff}).  Consistent with this
interpretation, CNS mutations show less reduction in age than 4d and
PPh:Benign mutations at low frequencies, and more reduction at high
frequencies, suggesting that CNS mutations are influenced less by selection
at linked sites and more by direct selection.

%It is worth noting that, as with more indirect measures of allele age
%\cite{KIEZETAL13}, the per site variance in allele age is high across all
%categories of sites (even after normalizing by allele frequency), and our
%ability to characterize selective effects at individual sites using allele
%age remains weak.  Nevertheless, it is clear that there is information
%about natural selection in bulk distributions of allele ages, and direct
%ARG inference is a powerful means for accessing this information.  
%This
%approach integrates information from mutations over an entire region,
%considering full genealogies and specific recombination events, which
%results in improved accuracy and reduced variance compared with methods
%that consider allele frequencies only or summary statistics based on
%patterns of LD.  Together with other sources of information, allele age may
%turn out to be a valuable tool for predicting functional consequences of
%mutations.

\subsubsection{Information in the ARG about Population Phylogenies}

\begin{figure}[h!]
\begin{center}
\includegraphics[width=6.6in]{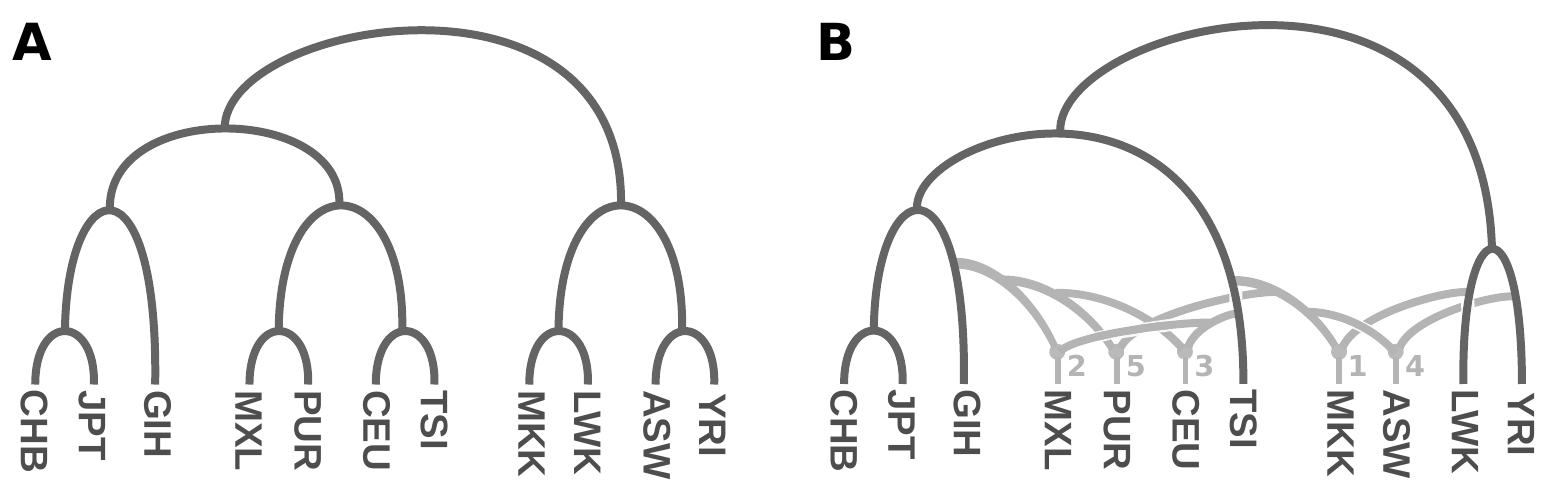}
\caption{\textbf{Human population phylogenies inferred from sampled
    ancestral recombination graphs.}}
\label{fig:poptrees}
\end{center}
\end{figure}

%In this section, we turn to the use of our ARG inference framework in studying
%human demographic history.
%More work will be needed to develop our ARG-sampling methods into a full
%framework for demography inference, similar to coalescent-based methods
%that apply to small numbers of sequences
%\cite{Hobolth2007,MAILETAL12,Li2011a,HARRNIEL13} or ignore intralocus
%recombination \cite{RANNYANG03,BURGYANG08,Gronau2011} (see Discussion).
%Nevertheless, even when a demographically naive prior distribution is
%assumed (based on a single panmictic population of fixed size), we expect
%samples drawn from the posterior distribution of ARGs to contain useful
%information about demographic history.  
To explore the usefulness of \wv in demographic analysis, 
we attempted to infer a
population phylogeny with admixture edges for the 11 human populations
represented in the Complete Genomics data set (see Figure
\ref{fig:poptrees} for a list of populations).
%we
%extracted local trees from our sampled ARGs and analyzed them with methods
%that attempt to infer a species or population network from a collection
%of (possibly discordant) gene trees based on parsimony criteria.
%We focused in this initial analysis on the question of inferring a
%population phylogeny with admixture edges for the 11 human populations
%represented in the Complete Genomics data set (see Figure
%\ref{fig:poptrees} for the list of populations).
%the Yoruba (YRI), Luhya
%(LWK), Maasai (MKK), African American (ASW), Tuscan (TSI), Centre d'Etude
%de Polymorphisme Humain samples (CEU; Utah residence of Northern and
%Western European ancestry), Mexican (MXL), Puerto Rican (PUR), East Indian
%(GIH), Han Chinese (CHB), and Japanese (JPT) populations.  
%
%We identified one locus at random for approximately each megabase of
%analyzed sequence (excluding regions with missing data or eliminated by our
%filters), obtained a consensus of the local trees sampled at each locus,
%then extracted a subtree containing one randomly selected chromosome for
%each of the 11 distinct populations (see Methods).  This resulted in 2,304
%approximately unlinked local trees with 11 leaves each.  
We extracted 2,304 widely spaced loci from our inferred ARGs, obtained a
consensus tree at each locus,  
and reduced this tree to a subtree with
one randomly selected chromosome for each of the 11 distinct populations
(see Methods).
We then analyzed
these 11-leaf trees with the PhyloNet program.  PhyloNet finds a population tree
that minimizes the number of ``deep coalescences'' required for
reconciliation with a given set of local trees, allowing both for
phylogenetic discordance from incomplete lineage sorting (see, e.g.,
\cite{SIEP09}) and for a specified number of hybridization (admixture)
events between groups \cite{THANNAKH09,YUETAL13}.  We ran the program six
times, allowing for 0--5 hybridization nodes.

In the absence of hybridization (0 nodes), PhyloNet recovers
the expected phylogeny for these populations, with the deepest divergence
event between African and Eurasian populations, and successively more
recent events separating the European from the Asian populations, the South
Asian Indian population from the East Asian Han Chinese and Japanese, and
the West African Yoruba from the East African Luhya and Maasai (Figure
\ref{fig:poptrees}A).  The most recent events separate the relatively
geographically and ethnically similar Han Chinese and Japanese populations,
Luhya and Massai populations, and Tuscan and CEU populations (Figure
\ref{fig:poptrees}A).  The Mexican and Puerto Rican individuals cluster
with the Europeans, and the African American individual clusters with the
West African Yorubans, consistent with recent analyses of these admixed
populations \cite{KIDDETAL12}.
% read through the Kidd paper, see if other relevant points to include
% cite another paper on african americans?

When admixture nodes are permitted, PhyloNet uses them to explain gene flow
in several populations identified as admixed in other analyses
\cite{HAPMCONS10,KIDDETAL12}, including the Maasai (MKK), African Americans
(ASW), Mexicans (MXL), and Puerto Ricans (PUR) (Figure
\ref{fig:poptrees}B).  In most cases, the inferred source populations are
consistent with previous studies, but there are two major anomalies in the
inferred networks.  The first anomaly is the use of GIH as a source
population for the admixed MXL and PUR populations.  This may be a
consequence of the absence in this data set of a better surrogate for
Native American source populations for the MXL and PUR or it may reflect
European admixture in India \cite{HAPMCONS10}.  The second anomaly is the
inference of admixture from the MXL and TSI individuals in the CEU sample.
Overall, it appears that the program correctly identifies a complex pattern
of gene flow among the Latino, European, and African populations but is
unable to reconstruct the precise topology of this subnetwork.  These
experiments suggest that additional work will be needed to fully exploit
the use of ARG inference in demographic analysis, but that
the posterior 
distribution of ARGs does appear to contain useful information about
population structure even when an uninformative prior distribution is used.

%\subsubsection{High TMRCA regions placeholder}
%See Table \ref{tab:highTmrca}.

%\subsubsection{Balancing selection placeholder}
%See Figure \ref{fig:przeworski}, Table \ref{tab:przeworskiTmrca}. Regions of ancient balancing selection came from this paper: \cite{LEFFETAL13}.

%\subsubsection{Allele age analysis}
%See Figure \ref{fig:allelAgeSimpleBarplot}, Figure \ref{fig:alleleAgeAbsolute} and Figure \ref{fig:alleleAgeDiff}.

%=============================================================================

\section{Discussion}
Several decades have passed since investigators first worked out the
general statistical 
characteristics of 
population samples of 
genetic markers in the presence of recombination
\cite{HILLROBE66,KARLMCGR68,STROMORG78,GRIF81,Hudson1983}.  Nevertheless, 
solutions to the problem of explicitly
characterizing this structure in the general case of multiple markers and
multiple sequences---that is, of making direct inferences about the
ancestral recombination graph (ARG) \cite{Griffiths1996,GRIFMARJ97}---have
been elusive.  Recent investigations have led to important progress on this
problem based on the Sequentially Markov Coalescent (SMC)
\cite{McVean2005,Marjoram2006,Hobolth2007,MAILETAL11,MAILETAL12,Li2011a,HARRNIEL13}, but existing methods are still
either restricted to small numbers of sequences or require severe
approximations.  In this paper, we  
introduce a method that is faithful to the SMC yet has much better scaling
properties than previous methods.  These properties depend on a novel
``threading'' operation that can be performed in a highly efficient manner
using hidden Markov modeling techniques.  Inference does require the use of
Markov chain Monte Carlo (MCMC) sampling, which has certain costs, but we
have shown that the sampler mixes fairly well and converges rapidly,
particularly if the threading operation is generalized from single
sequences to
subtrees.  Our methods allow explicit
statistical inference of ARGs on the scale of complete mammalian genomes
for the first time.
Furthermore, the sampling of ARGs from their posterior distribution has the
important advantage of allowing estimation 
of any ARG-derived quantity, such as times to most recent
common ancestry, allele ages, or regions of identity by descent.

Despite our different starting point, our methods are similar in several
respects to the conditional sampling 
distribution (CSD)-based methods of Song and colleagues
\cite{Paul2010,Paul2011,STEIETAL12,SHEEETAL13}.
%---by generalizing
%small-sample SMC-based methods for explicit ARG inference rather than the
%product of approximate conditionals (PAC) framework
%\cite{Stephens2000,Fearnhead2001,Li2003a}.  
Both approaches
consider a conditional distribution for the $n$th sequence given the
previous $n-1$ sequences, and in both cases a discretized SMC is exploited
for efficiency of inference.  However, 
%there is a crucial difference between the two strategies: 
the CSD-based methods consider the marginal
distribution of the $n$th sequence only, given the other $n-1$ sequences,
while ours considers the joint distribution of an ARG of size $n$ and the
$n$th sequence, given an ARG of size $n-1$ and the previous $n-1$ sequences.
In this sense, we have employed a ``data augmentation'' strategy
by explicitly representing full ARGs in our inference procedure.
%with consideration of full genealogies.  
The cost of this
strategy is that it requires 
Markov chain Monte Carlo methods for inference, rather than allowing direct
likelihood calculations and maximum-likelihood parameter estimation.  The
benefit is that it
provides an approximate posterior distribution over complete ARGs and
derived quantities.
%, including times to most recent common ancestry, 
%distributions of coalescence times, gene flow between populations, and
%similar quantities.  
By contrast, the CSD-based methods provide information
about only those properties of the ARG that are directly described by the model
parameters.
%and therefore would require reparameterization (with potentially
%large sets of 
%free parameters) for many applications of interest.  
We view these two
approaches as complementary and expect that they will have somewhat
different strengths and 
weaknesses, depending on the application in question.

% check Song's likelihood funciton
% issue with coalescence rate

Our explicit characterization of genealogies 
can be exploited to characterize the influence of natural selection
across the genome, as shown in our analysis of the Complete Genomics
data set.  In particular, we see clear evidence of an enrichment for
ancient TMRCAs in regions of known and predicted balancing selection,
reduced TMRCAs near protein-coding genes and selective sweeps, and reduced
allele ages in sites experiencing both direct selection and selection at
closely linked sites.  Interestingly, the genealogical view appears to have
the potential to shed light on the difficult problem of distinguishing
between background selection and hitchhiking.  Our initial attempt at
addressing this problem relies on a genealogy-based summary statistics, the
relative TMRCA halflife (RTH), that does appear to distinguish effectively
between protein-coding genes and partial selective sweeps.  However, more
work will be needed to determine how well this approach generalizes to other
types of hitchhiking (e.g., complete sweeps, soft sweeps, recurrent sweeps)
and whether additional genealogical information can be used to characterize the
mode of selection more precisely.  Additional work is also needed to
determine whether our ARG-based allele-age estimator---which is highly informative in
bulk statistical comparisons but has high variance at individual
sites---can be used to improve functional and evolutionary characterizations
of particular genomic loci.  A related challenge is to see whether our
genome-wide ARG samples can be used to improve methods for 
 association/LD
mapping
(see \cite{RANNREEV01,LARRETAL02,ZOLLPRIT05,Minichiello2006,Wu2008,BESEETAL09}).  

In addition to natural selection, our methods for ARG inference have the
potential to shed light on historical demographic processes, 
an area of particular interest in the
recent literature
\cite{RALPCOOP13,HARRNIEL13,SHEEETAL13,PRADETAL13,STEIETAL12}.  In this
paper, we have taken the simple approach of sampling ARGs under a noninformative
prior distribution (reflecting the assumption of a panmictic population of
constant size) and then attempting to make demographic inferences based
on the sampled genealogies.  This approach shows promise for the
reconstruction of population phylogenies but may be of limited value in the
estimation of features such as divergence times, ancestral effective
population sizes, and rates of gene flow.  An alternative strategy would be
to extend our methods to incorporate a full phylogenetic demographic model,
such as the one used by G-PhoCS \cite{Gronau2011},
thereby generalizing this fully Bayesian method to a setting in which
recombination is 
allowed and complete genome sequences are considered.  Importantly, the use
of the complete ARG would allow information about demographic history from
both patterns of mutation and patterns of linkage disequilibrium to be
naturally integrated (see \cite{Gronau2011}).  However, as with CSD-based
methods \cite{SHEEETAL13,STEIETAL12}, an extension to a full, parametric
multi-population model for application on a genome-wide scale would be
technically challenging.  In our case, it would require the ability to
sample ``threadings'' consistent with the constraints of a population model
(e.g., with no coalescent events between genetically isolated populations)
and exploration of a full collection of population parameters, which would
likely lead to slow convergence and long running times.  Nevertheless, a
version of this joint inference strategy may be feasible with appropriate
heuristics and approximations.  Our methods may also be useful for a wide
variety of related applications, including local ancestry inference
\cite{TANGETAL06,SANKETAL08,PRICETAL09}, haplotype phasing / genotype
imputation \cite{SCHESTEP06,BROWBROW07,HOWIETAL09,LIETAL10}, and
recombination rate estimation \cite{Fearnhead2001,MCVEETAL04}.

Our initial implementation of \wv relies on several simplying assumptions
that appear to have minimal impact on performance with (real or simulated)
human sequence data, but may produce limitations in other settings.
Following Li and Durbin \cite{Li2011a}, we compute probabilities of
recombination between discrete genomic positions under the assumptions of
the continuous-space SMC \cite{McVean2005}.  When recombination rates are
low, the discrete and continuous models are nearly identical, but the
differences between them can become significant when recombination rates
are higher (A.\ Hobolth and J.\ L.\ Jensen, ``Markovian approximation to
the finite loci coalescent with recombination along multiple sequences,''
under revision.).  Similarly, our assumption of at most one recombination
event per site and our use of the SMC rather than the improved SMC$'$
\cite{Marjoram2006} may lead to biases in cases of higher recombination
rates, larger numbers of sequences, or more divergent sequences.  In
addition, our heuristic approach of accommodating zero-length branches by
randomly sampling among ``active'' branches for coalescence and
recombination events (see Methods) may lead to biases when the
discretization scheme is coarse relative to evolutionary events of
interest.  Finally, we currently assume haploid genome sequences as input,
which, in most cases of current interest, requires computational phasing as
a pre-processing step.  Phasing errors may lead to over-estimation of
recombination and mutation rates and associated biases, because the sampler
will tend to compensate for them with additional recombination and/or
mutation events.  In principle, most of these limitations can be addressed
within our framework.  For example, it should be fairly straightforward to
extend \wv to use the SMC$'$ and Hobolth and Jensen's finite-loci transition
density.  In addition, we believe it is possible to enable the program to
work directly with unphased data and integrate over all possible phasings
(see, e.g., \cite{WUGUSF07,Gronau2011}).

% do we need this?  shorten?
The ability to perform explicit ARG inference on the scale of complete
genomes opens up a wide range of possible applications, but the long
running times required for these analyses and the unwieldy data structures
they produce (numerous samples of ARGs) are potential barriers to practical
usefulness.  In our initial work, we have attempted to address this problem
by precomputing ARGs for a highly informative public data set and releasing
both our complete ARGs and various summary statistics (as browser tracks)
for use by other groups.  In future work, it may be possible to improve
data access by providing more sophisticated tools for data retrieval and
visualization.  For example, 
%one 
%could carefully analyze a particularly rich public data set, such as the
%high-coverage genome sequences currently being produced by the 1000 Genomes
%Project \cite{1KGCONS10}, and extract a modest number of samples (say,
%1000) from a lengthy MCMC run.  
sampled ARGs could be stored in a database
in a manner that allowed researchers to construct queries to 
efficiently 
extract various
features of interest, such as marginal genealogies, recombination events,
regions of IBD, or times to most recent common ancestry for designated
subsets of samples.  A related
possibility would be to support on-the-fly 
threading of user-specified query sequences into precomputed ARGs.
This operation would be analogous to local ancestry inference
\cite{TANGETAL06,SANKETAL08,PRICETAL09}, but would reveal not only the
population sources of query sequence segments, but also additional information
about recombination events, coalescence times, approximate mutation ages,
and other features.  The same operation could be used to allow our sampling
methods 
to scale to thousands of genomes: one could infer ARGs for, say, 100
genomes, then simply thread in hundreds more, without full MCMC sampling.
%Finally, precomputed ARG samples could also be used as
%the basis for various visualization tracks, perhaps including the tracks
%like the ones introduced in this paper, as well as complementary tracks
%describe features of the ARG such as population divergence times, migration
%rates, or mutation ages.  
In general, we believe that posterior samples of ARGs
will be a rich resource for genetic analysis, but
careful work will be needed on
efficient and effective strategies for making these samples
practically useful to the genomics community.

%=============================================================================
% Methods

\section{Methods}

\subsection{Discretized Sequentially Markov Coalescent}
%=============================================================================
\subsubsection{Discretization Scheme and Notation}

The Discretized Sequentially Markov Coalescent (DSMC)
assumes that all coalescence and
recombination events occur at $K+1$ discrete time points,
${\cal P} = \{s_0, s_1, s_2, ..., s_K\}$, with $s_0 = 0$ (the present time)
and $s_K$ 
equal to a user-specified maximum value.  These time points are defined
in units of generations before the present time.
We evenly distribute these time points on a logarithmic scale, 
so that the discretization scheme has finer resolution
near the leaves of the ARG, where more events are expected to occur.
Specifically, we define $s_j$ (for $0 \leq j \leq K$) to be $s_j = g(j)$, where
\begin{equation}
g(j) = \frac{1}{\delta} 
  \left\{ \exp\left[\frac{j}{K} \log(1 + \delta s_K)\right] - 1\right\}.
\end{equation}
Here, $s_K$ is the maximum time and $\delta$ is a tuning parameter that,
when increased, causes the time points to become more densely clustered
near the leaves of the ARG.  Notice that $g(0) = 0$ and $g(K) = s_K$.
In this work, we have assumed $s_K =$ 200,000 generations and
$\delta=10$.  
We denote the length of time interval $j$ as
$\Delta s_j = s_{j+1} - s_j$.  
The DSMC process is defined such that it approaches the
continuous SMC as a limit as $K\rightarrow \infty$ and each $\Delta
s_j\rightarrow 0$, with $s_K$ sufficiently large that the probability of a
coalescence event older than $s_K$ is close to zero.
%[CHECK -- SEE COAL].

It is useful to specify ``midpoints'' between time points (on a log
scale), to facilitate rounding of continuous-valued times to the nearest
discrete time point.  We define the midpoint between times $s_j$ and
$s_{j+1}$ (for $0 \leq j < K$) as $s_{j+\frac12} = g(j+\frac12)$.  We can
alternatively refer to the midpoint between times $s_{j-1}$ and $s_j$ as
$s_{j-\frac12} = g(j-\frac12)$ (for $1\leq j\leq K$), noting that
$s_{j-\frac{1}{2}} = g(j-\frac12) = g((j-1)+\frac12) =
s_{(j-1)+\frac{1}{2}}$.  Coalescence events that occur between
$s_{j-\frac{1}{2}}$ and $s_{j+\frac{1}{2}}$ are ``rounded'' to time point
$s_j$.  We found that it was less critical to round recombination events to
the nearest time point, so they are simply rounded to the next most recent
time point (see below).  We denote the lengths of the half intervals
between $j-\frac12$ and $j$, and between $j$ and $j + \frac12$, as $\Delta
s_{j-\frac12,j}$ and $\Delta s_{j,j+\frac12}$, respectively.

% branch counts
Because all coalescence events must occur at the designated time points, the
collection of branches is fixed for each interval $j$ between time points
$s_j$ and $s_{j+1}$.  Given a local tree $T^n_i$ that is consistent with the
DSMC, we denote the set of branches in time interval $j$ as $B(T^n_i,j)$.
The size of this set, $|B(T^n_i, j)|$, is of particular interest, and is
abbreviated $B_j$ (with $T^n_i$ clear from context).  In addition, it is often of
interest to consider the branch sets for a tree $T^n_i$ from which a branch
$w$ has been removed.  We denote such a tree by $T_i^{n,(-w)}$ and abbreviate
the number of branches in interval $j$ as $B_j^{(-w)}$ (again, with $T^n_i$
clear from context). 

% discretization rationale
One consequence of discretizing time is that the DSMC will tend to generate
ARGs that contain many branches of length zero (corresponding to polytomies
in the local trees), which will have zero probability of recombination,
coalesce, or 
mutation events.  In effect, the rounding procedure will tend to shrink
short branches to zero, which may lead to distortions in data generation
and inference.  We address this problem heuristically, by defining the DSMC
to first sample
the times of recombination and coalescence events, and then randomly
select a branch from all of those ``active'' at the sampled time point.
We define the set of active branches at a time point $s_j$, for a local
tree $T^n_i$, to be those branches in $T^n_i$ that start, end, or pass through
$s_j$.  This set is denoted $A(T_i, j)$ and its size is abbreviated as 
$A_j$.  As above, we use $A_j^{(-w)}$ to indicate the active branches at
$s_j$ excluding branch $w$. 
Simulations indicate that this heuristic solution to the problem of
zero-length branches works fairly well in practice (see
Figure \ref{fig:sim-dsmc}).

%=============================================================================
\subsubsection{Recombination Process}
\label{sec:dsmc-recomb-generation}

As in the standard SMC, recombinations are assumed to occur according to a
Poisson process with 
rate $\rho |T^n_{i-1}|$, where $|T^n_{i-1}|$ is the total branch length 
of local tree $T^n_{i-1}$ and $\rho$ is the average number of
recombinations/generation/site. 
Once a recombination occurs, the ordinary SMC process places the
recombination uniformly along the branches of $T^n_{i-1}$.  The analogous
operation of sampling a recombination branch and time point, $R^n_i=(w, s_k)$,
in the DSMC is accomplished 
by first sampling a
time point $s_k$ in proportion to the total branch length present
during time interval $k$, then randomly selecting one of the $A_k$ branches
active at that time point.  Consistent with the assumptions of the SMC,
the recombination point cannot occur above the time point associated with
the root $r$ of tree $T^n_{i-1}$, which we denote $s_r$.
Thus, the sampling distribution for a
recombination point $R^n_i$ on a local tree $T^n_{i-1}$ is given by,
\begin{equation}
P(R^n_i \;|\; T^n_{i-1},\, \Theta) 
 = 
\begin{cases}
\exp(- \rho |T^n_{i-1}|) & \text{if } R^n_i = \emptyset \\
\frac1{A_k} \cdot \frac{B_k \, \Delta s_k}{C} \cdot \left[1 - \exp(- \rho
  |T^n_{i-1}|)\right] & \text{if } R^n_i= (w, s_k),\,\, w \in A(T^n_{i-1}, k),\, 0 \leq s_k < s_r \\
\frac12 \cdot \frac{\Delta s_k}{C} \cdot \left[1 - \exp(- \rho
  |T^n_{i-1}|)\right] & \text{if } R^n_i= (w, s_k),\, w \in A(T^n_{i-1}, r)
\setminus \{r\},\, s_k=s_r \\
0 & \text{otherwise},
\end{cases}
\label{eqn:recomb}
\end{equation}
where $C = \sum_{j=0}^{r} B_j \, \Delta s_j$ is a constant that explicitly
normalizes the distribution over time points $s_0, \dots, s_r$.  The
special case for the time point at
the root of the tree, $s_k=s_r$, is required because the
SMC does not allow recombinations to occur beyond this point, so the
effective number of active branches is only two at this time point, despite
that $A_r$ will have a value of three.  The number of branches in the
interval above the root, $B_r$, is necessarily one, so this term can be
omitted in this case.  

This sampling distribution effectively rounds the times of recombination
events downward to the next most recent time point.  However, a strict policy of
downward rounding, together with a prohibition again recombination events
above the root node, would make it impossible to sample recombination
events at time point $s_r$, which turns out to have undesirable effects in
inference (it makes some trees unreachable by the threading operation).
Therefore, when sampling time points, we use the heuristic approach of
imagining that recombinations can also occur in the time interval
immediately above the
root and assigning these events to the time point $s_r$.  This has the
effect of redistributing some of the probability mass from later time
points to the root, without altering the overall rate at which
recombinations occur ($\rho|T^n_{i-1}|$).  For this reason, the normalizing
constant $C$ differs slightly from the total branch length $|T^n_{i-1}|$; in
particular, $C = |T^n_{i-1}| + \Delta s_r$.  It would be slightly more
elegant to allow upward as well as downward rounding of times for
recombinations, as we do with coalescence events (see below), but as long
as the time discretization is not too coarse these differences are of minor
importance, and the approach we have used seems to be adequate.

%=============================================================================
\subsubsection{Re-coalescence Process}
\label{sec:dsmc-recoal-generation}

Once a recombination point $R^n_i = (w, s_k)$ is sampled,
the selected branch $w$ is removed from
time points $s_k$ and older, and allowed to
re-coalesce to the remainder of the tree, in a manner analogous to the SMC.
Because we explicitly prohibit multiple
recombinations between adjacent positions, 
the local tree $T^n_i$ must be reachable from $T^n_{i-1}$ by a single
``subtree pruning and regrafting'' (SPR) operation corresponding to the
recombination,
i.e., an operation
that cuts a branch of the tree at the recombination point and re-attaches it
(and any descendant nodes) to 
the remainder of the tree.  Thus, we can write,
\begin{equation}
P(T^n_i \;|\; R^n_i,\, T^n_{i-1},\, \Theta)
 = \begin{cases} 
   1 & \text{if } R^n_i=\emptyset,\, T^n_i = T^n_{i-1} \\
   P(x,\, s_j \;|\; w,\, s_k,\, T^n_{i-1},\, \Theta) & \text{if } R^n_i=(w, s_k),\;
   (x, s_j) \text{ s.t. } T^n_i = SPR(T^n_{i-1},\, w,\, s_k,\, x,\, s_j), \;
   s_j \geq s_k \\
   0 & \text{otherwise},
\end{cases}
\label{eqn:new-tree}
\end{equation}
where $SPR(T^n_{i-1},\, w,\, s_k,\, x,\, s_j)$ is a 
function that returns the new tree produced by
an SPR operation on $T^n_{i-1}$ that cuts branch $w$ at time $s_k$ and
re-attaches it 
to branch $x$ at time $s_j$, and $P(x,\, s_j \;|\; w,\, s_k,\,
T^n_{i-1},\, \Theta)$ is a joint conditional
distribution over re-coalescence branches and time points.

The main challenge is therefore to define the discrete re-coalescence
distribution,
$P(x,\, s_j \;|\; w,\, s_k,\, T^n_{i-1},\, \Theta)$, for $s_j \geq s_k$ (as required by
the SMC). 
There are two distinct cases to consider: $s_j > s_k$ and $s_j = s_k$.  When $s_j >
s_k$, the unattached branch $w$ must first fail to re-coalesce during the
interval between $s_k$ and $s_{j-\frac12}$, and then must re-coalesce
between $s_{j-\frac12}$ and $s_{j+\frac12}$ (because all such
re-coalescence events will be rounded to $s_j$).  By contrast, when $s_j =
s_k$, the branch $w$ must simply re-coalesce between $s_j$ ($=s_k$) and 
$s_{j+\frac12}$ (because the re-coalescence time is strictly bounded by the
recombination time).  

In all cases, the instantaneous rate of
re-coalescence in each 
interval $l$ ($k \leq l \leq j)$ is given by $B^{(-w)}_l/(2N_l)$, in the standard
manner for the coalescent.  (Note that we use $B^{(-w)}_l$ rather than $B_l$,
because we are concerned with the coalescence rate to the remainder of the
tree, excluding branch $w$.  We also assume a diploid species throughout,
so the total number of chromosomes per locus is $2N$.)
The
probability that a lineage starting at a time $s_l$ coalesces before
$s_{l+1}$ is given by the cumulative distribution function for 
exponentially distributed waiting times,
\begin{equation}
W(l, l+1) = 1 - \exp\left(-\frac{B^{(-w)}_l\, \Delta s_l}{2N_l}\right),
\end{equation}
and
the probability of coalescence during a sequences of intervals, $m,
m+1, \dots, n-1$ is given by,
\begin{equation}
W(m, n) = 1 - \exp\left(-\sum_{l=m}^{n-1} \frac{B^{(-w)}_l\, \Delta s_l}{2N_l}  \right).
\end{equation}
Similarly, the probabilities of coalescence during the half intervals before
and after 
time point $s_l$ are given, respectively, by,
\begin{equation}
W\left(l-\frac12, l\right) = 1 - \exp\left(-\frac{B^{(-w)}_{l-1}\, \Delta
    s_{l-\frac12, l}}{2N_{l-1}} \right),\qquad
W\left(l, l+\frac12\right) = 1 - \exp\left(-\frac{B^{(-w)}_l\, \Delta s_{l,l+\frac12}}{2N_l} \right).
\end{equation}
Thus, the distribution of re-coalescence times
for the case of $s_j>s_k$ is given by,
\begin{align}
P(s_j \;|\; w,\, s_k,\, T^n_{i-1}, \, \Theta) &= 
\left[1-W\left(k, j-\frac12\right)\right] \times W\left(j-\frac12, j+\frac12\right)
\notag \\
& = \exp\left[ -\left(\sum_{l=k}^{j-2} \frac{B_l^{(-w)}\, \Delta s_l}{2N_l}\right) -
  \frac{B_{j-1}^{(-w)}\, \Delta s_{j-1,j-\frac12}}{2N_{j-1}} \right] \times \left[ 1- \exp
  \left( -\frac{B^{(-w)}_{j-1}\, \Delta s_{j-\frac12,j}}{2N_{j-1}} -
    \frac{B_{j}^{(-w)}\, \Delta s_{j,j+\frac12}}{2N_{j}}\right) \right].
\label{eqn:old-recomb}
\end{align}
The probability of re-coalescence for the case of $s_j=s_k$ is simply,
\begin{align}
P(s_j=s_k \;|\; w,\, s_k,\, T^n_{i-1}, \, \Theta) &= 
W\left(k, k+\frac12\right)
 = \left[ 1- \exp
  \left( -\frac{B_{k}^{(-w)}\, \Delta s_{k,k+\frac12}}{2N_{k}}\right) \right].
\label{eqn:young-recomb}
\end{align}
Finally, the requirement for re-coalescence by the maximum time, $s_K$,
is enforced by explicitly normalizing the distribution:
\begin{equation}
P(s_j = s_K \;|\; w,\, s_k,\, T^n_{i-1}, \, \Theta) = 1 - \sum_{l=k}^{K-1} P(s_l \,|\, w,\,
s_k,\, T^n_{i-1}, \, \Theta).
\label{eqn:recomb-norm}
\end{equation}

Once 
the coalescence time point $s_j$ is chosen, a lineage $x$ is uniformly chosen
from the $A_j^{(-w)}$ active lineages in $T_i$ at that time point, similar to
the process for recombination 
events.  Thus, $P(x,\, s_j \;|\; w,\, s_k,\, T_{i-1}, \, \Theta) = 
\frac1{A_j^{(-w)}} P(s_j \;|\; w,\, s_k,\, T_{i-1}, \, \Theta)$,
%for $x \in A(T_i^{(-w)}, j),\, s_j \geq s_k$, 
and equation \ref{eqn:new-tree} can be rewritten as,
\begin{equation}
P(T^n_i \;|\; R^n_i,\, T^n_{i-1}, \, \Theta)
 = \begin{cases} 
   1 & \text{if } R^n_i=\emptyset,\, T^n_i = T^n_{i-1} \\
   \frac1{A_j^{(-w)}} P(s_j \;|\; w,\, s_k,\, T^n_{i-1}, \, \Theta) & \text{if } R^n_i=(w,
   s_k),\; 
   (x, s_j) \text{ s.t. } T^n_i = SPR(T^n_{i-1},\, w,\, s_k,\, x,\, s_j), \;
   s_k \leq s_j \leq s_K \\
   0 & \text{otherwise},
\end{cases}
\label{eqn:recoal-master}
\end{equation}
where $P(s_j \;|\; w,\, s_k,\, T^n_{i-1}, \, \Theta)$ is given by equations
\ref{eqn:old-recomb}--\ref{eqn:recomb-norm}. 

%the joint 
%sampling distribution over times and branches is given by,
%\begin{align}
%P(x,\, s_j \;|\; w,\, s_k,\, T_{i-1}) = 
%\begin{cases}
%\frac1{A_j^{(-w)}} P(s_j \;|\; w,\, s_k,\, T_{i-1}) & \text{if } x \in
%A(T_i^{(-w)}, j),\, s_j \geq s_k \\
%0 & \text{otherwise}.
%\end{cases}
%\label{eqn:recomb-joint}
%\end{align}

%=============================================================================
\subsubsection{Initial Local Tree}
\label{sec:dsmc-init-tree-generation}

The DSMC begins by generating an initial local tree, $T^n_1$, using a
discretized version of the coalescent process.  This 
process can be decomposed into two steps: (1) the generation of a sequence of
branch counts, $\V{C} = (C_0, C_1, \dots, C_K)$ for time points $s_0, s_1,
\dots, s_K$, and (2) sampling of a topology ${\cal T}^n_1$ consistent with 
these branch counts.  The probability of an observed initial tree $T^n_1$ can
therefore be calculated as,
\begin{equation}
P(T^n_1 \;|\; \Theta) = P({\cal T}^n_1, \V{C} \;|\; \Theta) 
= P(\V{C} \;|\; \V{N}) \; P({\cal T}^n_1 \;|\; \V{C}),
\label{eqn:init-tree}
\end{equation}
where $\V{N}$ is a vector of effective population sizes, $\V{N} = (N_0,
\dots, N_K)$.  The branch count for time 0 is constrained to be equal to
the number of samples, $C_0=n$, and the branch count for time $K$ is
required to be one, $C_K=1$ (see below).

Since the coalescent
process is Markovian in time, the distribution for the vector of branch counts
can be factored by time intervals,
\begin{align}
P(\V{C} \;|\; \V{N}) =& P(C_0) \, \prod_{l=1}^{K} P(C_l \;|\; C_{l-1}, \Delta
s_{l-1}, N_{l-1}),
\label{eqn:branch-counts}
\end{align}
with degenerate first and last terms, 
$P(C_0) = I[C_0=n]$ and $P(C_K \;|\; C_{K-1},
N_{K-1}) = I[C_K=1]$. 

The conditional distributions of the form $P(C_l \;|\; C_{l-1}, \Delta
s_{l-1}, N_{l-1})$, for $1 \leq l < K$, have been derived previously as
\cite{Tavare1984},
\begin{align}
\label{eqn:coal_counts}
P(C_l = b \;|\; C_{l-1} = a, \Delta
s_{l-1} = t, N_{l-1}) = 
\sum_{k=b}^a \exp\left(\frac{-k(k-1)}{4N_{l-1}} t\right)
  \frac{(2k-1)(-1)^{k-b}}{b! (k-b)! (k+b-1)}
  \prod_{y=0}^{k-1} \frac{(b+y)(a-y)}{a+y}.
\end{align}

%The conditional distribution for the topology given the branch counts, 
%$P({\cal T}^n_1, \V{C} \;|\; \Theta)$, is determined by .... [MATT?]

% Matt says same as used for phylogenetic 

% need to specific order in which lineages coalesce
% P({\cat T}_1 \;|\; \V{k})

%=============================================================================
\subsection{Hidden Markov Model}

\subsubsection{Hidden Markov Model for Full Threading Problem}

\begin{figure}[h!]
\begin{center}
\includegraphics[width=3in]{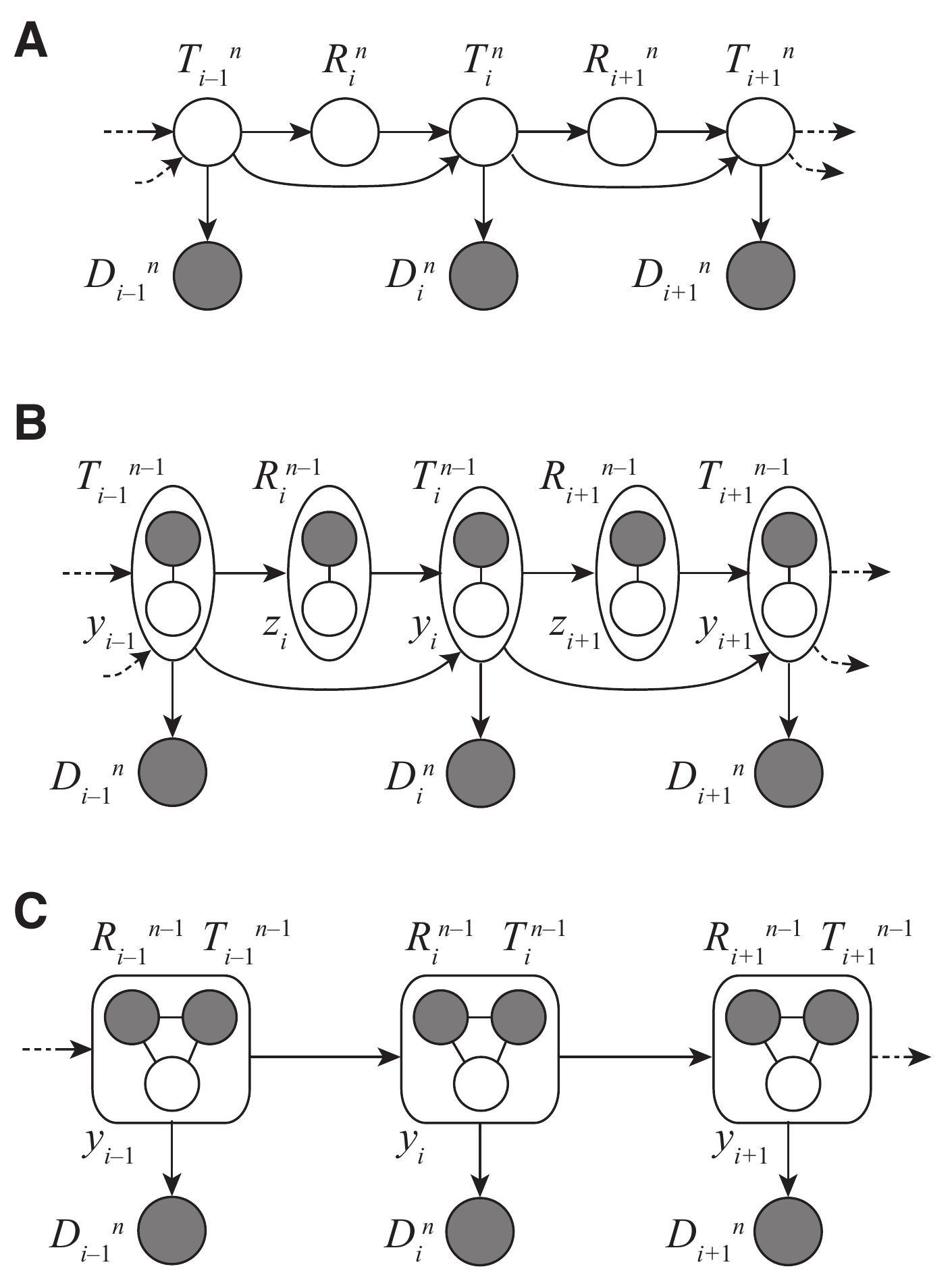}
\caption{\textbf{Graphical models for Discretized Sequentially Markov
    Coalescent (DSMC) models}.  
\label{fig:graphmodels}
}
\end{center}
\end{figure}

As noted in the Results section, 
the complete data likelihood function under the DSMC is given by equation
\ref{eqn:dsmc-lik}.  
%that is, the probability of an ARG $\V{G}^n=(\V{T}^n, \V{R}^n)$ and a sequence
%alignment $\V{D}^n$ given the model parameters.  
%In our notation, this
%function is given by,
%\begin{equation}
%P(\V{T}^n, \V{R}^n, \V{D}^n \;|\; \Theta) = P(T_1^n \;|\; N) \; P(D_1^n
%\;|\; T_1^n, \mu) \;\prod_{i=2}^L  
%P(R_i^{n} \;|\; T_{i-1}^n, \rho) \;
%P(T_i^n \;|\; R_i^{n}, T_{i-1}^n, N) \;
%P(D_i^n \;|\; T_i^n, \mu).
%\label{eqn:dsmc-lik}
%\end{equation}
If the full ARG $\V{G}^n=(\V{T}^n, \V{R}^n)$ is regarded as a latent
variable, this equation defines a hidden Markov model with a state space
given by all possible pairs $(T^n_i, R^n_i)$, transition probabilities
given by expressions of the form $P(R_i^{n} \;|\; T_{i-1}^n, \rho)$
$P(T_i^n \;|\; R_i^{n}, T_{i-1}^n, N)$ and emission probabilities given by
$P(D_i^n \;|\; T_i^n, \mu)$ (see Figure \ref{fig:graphmodels}A).  The
transition probabilities can be computed 
using equations \ref{eqn:recomb} and \ref{eqn:recoal-master}, and the emission
probabilities can be computed using Felsenstein's pruning algorithm.
This
model can be viewed as an instance of the ``phylo-HMMs''
that have been widely used in comparative genomics \cite{SIEPHAUS05}.
As discussed in the Results section, however, unless the number of 
sequences $n$ is very small, the state space of this HMM will be too large to
allow it to be used directly for inference.

Instead, we constrain the inference problem by fixing the ARG for the first
$n-1$ sequences, $\V{G}^{n-1}$, and sampling from the conditional
distribution $P(\V{G}^n \;|\; \V{G}^{n-1}, \V{D}, \Theta)$.  Using the notation
$\V{G}^n=(\V{T}^n, \V{R}^n)$ and $\V{G}^{n-1}=(\V{T}^{n-1}, \V{R}^{n-1})$,
we define $\V{T}^n = (\V{T}^{n-1}, \V{Y})$, where $\V{Y} = (y_1, \dots,
y_L)$ is a vector of coalescence points such that $y_i = (x_i, t_i)$
indicates a coalescence of the $n$th sequence to branch $x_i$ and time
point $y_i$ of local tree $T^{n-1}_i$, and $\V{R}^n = (\V{R}^{n-1},
\V{Z})$, where $\V{Z} = (z_2, \dots, z_L)$ is a vector of recombination
points such that $z_i = (w_i, u_i)$ indicates a recombination at branch
$w_i$ and time point $u_i$ of local tree $T^{n-1}_{i-1}$ between positions
$i-1$ and $i$. (Note that $z_1$ is undefined.)  Thus, we can sample from the
desired conditional distribution $P(\V{G}^n \;|\; \V{G}^{n-1}, \V{D},
\Theta)$ by sampling from $P(\V{Y}, \V{Z} \;|\;
\V{T}^{n-1}, \V{R}^{n-1}, \V{D}^n, \Theta)$.  We refer to a sample
$(\V{Y}, \V{Z})$ from this distribution as a {\em threading} of the $n$th
sequence through the ARG (see Figure \ref{fig:graphmodels}B).  For now,
we will consider a complete threading $(\V{Y}, \V{Z})$, but in later
sections we will describe our two-step process for sampling, first, the
coalescent 
threading 
$\V{Y}$, and second, the recombination threading $\V{Z}$ given $\V{Y}$.

Note that the restriction to one
recombination event per position implies that $z_i = \emptyset$
wherever $R^{n-1}_i \ne \emptyset$, and that $T_{i-1}^{n-1} =
T_i^{n-1}$ wherever $z_i \ne \emptyset$.  This restriction is not strictly
required but it simplifies the description of new recombination events $z_i$,
and in the setting of interest here it comes with little cost
(see Discussion). 

It turns out to be more convenient to work with the joint
distribution 
$P(\V{T}^{n-1}, \V{Y}, \V{R}^{n-1}, \V{Z}, \V{D}^n \;|\; \Theta)$ (the
complete data likelihood) than with
the conditional distribution $P(\V{Y}, \V{Z} \;|\;
\V{T}^{n-1}, \V{R}^{n-1}, \V{D}^n, \Theta)$.  However, to emphasize
that the variables $\V{T}^{n-1}$ and $\V{R}^{n-1}$ are held fixed
(``clamped'') at pre-specified values throughout the threading operation,
we denote them as $\bar{\V{T}}^{n-1}$ and $\bar{\V{R}}^{n-1}$, and
refer to the distribution of interest as $P(\bar{\V{T}}^{n-1}, \V{Y},
\bar{\V{R}}^{n-1}, \V{Z}, \bar{\V{D}}^n \;|\; \Theta)$.  (Notice that
the data $\V{D}^n$ are also clamped, as usual for HMMs.)  When
$\V{T}^{n-1}$, $\V{R}^{n-1}$, and $\V{D}^{n}$ are clamped, 
\begin{equation}
P(\bar{\V{T}}^{n-1}, \V{Y}, \bar{\V{R}}^{n-1}, \V{Z}, \bar{\V{D}}^n
\;|\; \Theta) \propto 
P(\V{Y}, \V{Z} \;|\;
\bar{\V{T}}^{n-1}, \bar{\V{R}}^{n-1}, \bar{\V{D}}^n, \Theta).
\end{equation}
Thus, samples of $(\V{Y}, \V{Z})$ drawn in proportion to the unnormalized
density $P(\bar{\V{T}}^{n-1}, \V{Y}, \bar{\V{R}}^{n-1}, \V{Z},
\bar{\V{D}}^n \;|\; \Theta)$ 
will be valid samples from the desired conditional distribution.

We can now write the density function for the (unnormalized) sampling
distribution for a threading  $(\V{Y}, \V{Z})$ as,
\begin{align}
&P(\bar{\V{T}}^{n-1},\V{Y}, \bar{\V{R}}^{n-1}, \V{Z},
\bar{\V{D}}^n 
\;|\; \Theta) = \notag \\
& \qquad P(\bar{T}_1^{n-1},y_1
\;|\; N) \; P(\bar{D}_1 \;|\; \bar{T}_1^{n-1}, y_1, \mu) \; \prod_{i=2}^L 
P(\bar{R}_i^{n-1}, z_i \;|\; \bar{T}_{i-1}^{n-1}, y_{i-1}, \rho) \;  P(\bar{T}_i^{n-1}, y_i \;|\; \bar{R}_i^{n-1}, z_i,
\bar{T}_{i-1}^{n-1}, y_{i-1}, N) \notag \\
& \qquad \qquad \qquad \qquad \qquad \qquad \qquad \qquad \qquad \times 
P(\bar{D}_i \;|\; \bar{T}_i^{n-1}, y_i, \mu),
\end{align}
where all terms are computable using previously described expressions, as
for equation \ref{eqn:dsmc-lik}.

Notice that this threading HMM has the same conditional 
independence structure as the HMM for the full DSMC (equation
\ref{eqn:dsmc-lik}, Figure \ref{fig:graphmodels}), but its state space is
now defined by sets of possible 
$(y_i, z_i)$ pairs rather than the set of possible $(T^n_i, R^n_i)$ pairs,
making it far more tractable for inference.

\subsubsection{Reduced Model for Coalescent Threading}
The state space can be reduced further by proceeding in two steps.  First, we
sample a {\em coalescent threading} $\V{Y}$ from the marginal distribution 
$P(\bar{\V{T}}^{n-1},\V{Y}, \bar{\V{R}}^{n-1}, \bar{\V{D}}^n \;|\;
\Theta) \propto P(\V{Y} \;|\; \bar{\V{T}}^{n-1}, \bar{\V{R}}^{n-1},
\bar{\V{D}}^n, \Theta)$.  Then we sample a {\em recombination threading}, $\V{Z}$, from the
conditional distribution $P(\V{Z} \;|\; \V{Y}, \bar{\V{T}}^{n-1},
\bar{\V{R}}^{n-1}, \Theta)$.  Notice that the data need not be considered
when sampling the recombination threading, because $\V{Z}$ is conditionally
independent of 
$\V{D}^n$ given  $\V{Y}$, $\V{T}^{n-1}$, and
$\V{R}^{n-1}$.

The marginal distribution $P(\bar{\V{T}}^{n-1},\V{Y},
\bar{\V{R}}^{n-1}, \bar{\V{D}}^n \;|\; \Theta)$ can be computed
efficiently by changing the order of products and sums in the usual way
for HMMs:
%(e.g., \cite{FELSCHUR96}):  
\begin{align}
&P(\bar{\V{T}}^{n-1},\V{Y}, \bar{\V{R}}^{n-1}, \bar{\V{D}}^n \;|\; \Theta) 
= \sum_{\V{Z}} P(\bar{\V{T}}^{n-1},\V{Y}, \bar{\V{R}}^{n-1}, \V{Z}, \bar{\V{D}}^n \;|\; \Theta) \notag \\
& \qquad = P(\bar{T}_1^{n-1},y_1
\;|\; N) \; P(\bar{D}_1 \;|\; \bar{T}_1^{n-1}, y_1, \mu) \; \prod_{i=2}^L 
\left[\sum_{z_i}
P(\bar{R}_i^{n-1}, z_i \;|\; \bar{T}_{i-1}^{n-1}, y_{i-1}, \rho) \;
P(\bar{T}_i^{n-1}, y_i \;|\; \bar{R}_i^{n-1}, z_i,
\bar{T}_{i-1}^{n-1}, y_{i-1}, N) \right]\; \notag \\
& \qquad \qquad \qquad \qquad \qquad \qquad \qquad \qquad \qquad \qquad \qquad \times P(\bar{D}_i \;|\; \bar{T}_i^{n-1}, y_i, \mu) \notag \\
& \qquad = P(\bar{T}_1^{n-1},y_1
\;|\; N) \; P(\bar{D}_1 \;|\; \bar{T}_1^{n-1}, y_1, \mu) \; \prod_{i=2}^L 
P(\bar{R}_i^{n-1}, \bar{T}_i^{n-1}, y_i  \;|\; \bar{T}_{i-1}^{n-1}, y_{i-1}, \rho, N) \;
P(\bar{D}_i \;|\; \bar{T_i}^{n-1}, y_i, \mu).
\label{eqn:reduced}
\end{align}
This equation defines an HMM with a state space given by the possible
values of $y_i$ only, the size of which is bounded by $nK$, where $n$ is
the number of sequences and $K$ is the number of time intervals  (see
Figure \ref{fig:graphmodels}C).   

While this model has the conditional independence structure of a standard HMM,
the state space is heterogeneous along the 
sequence, 
because the set of possible coalescent points at each position $i$
depends on the local tree, $T_i^{n-1}$.  (The full
threading HMM described above also has this property.) If we denote the state space at
position $i$ as ${\cal S}_i$, the transition probabilities between states
in position $i-1$ and states in position $i$ are defined by a $|{\cal
  S}_{i-1}| \times |{\cal S}_{i}|$ transition matrix $\V{A}_{i-1} = \{a_{l,m}^{i-1}\}$
where $l$ and $m$ index the states of $|{\cal S}_{i-1}|$ and $|{\cal
  S}_{i}|$, respectively, and $a_{l,m}^{i-1}$ can be computed as,
\begin{align}
a^{i-1}_{l,m} &= P(\bar{R}_{i}^{n-1}, \bar{T}_{i}^{n-1}, y_{i}=m  \;|\; \bar{T}_{i-1}^{n-1}, y_{i-1}=l,
\rho, N) \notag \\
&= \sum_{z_{i}}
P(\bar{R}_{i}^{n-1}, z_{i} \;|\; \bar{T}_{i-1}^{n-1}, y_{i-1}=l, \rho) \;
P(\bar{T}_{i}^{n-1}, y_{i}=m \;|\; \bar{R}_{i}^{n-1}, z_{i},
\bar{T}_{i-1}^{n-1}, y_{i-1}=l, N)
\label{eqn:transition-prob}
\end{align}
using equations \ref{eqn:recomb} and \ref{eqn:recoal-master}.  
The emission probability for
alignment column $D_i^{n}$ in state $l$ in ${\cal S}_i$
is denoted $b_l^i(D_i^{n}) = P(D^n_i \;|\; \bar{T}_i^{n-1}, y_i=l, \mu)$ and can be
computed using Felsenstein's pruning algorithm, as in all cases above.  The
initial state probabilities for the HMM are given by $\pi_l = 
P(\bar{T}_1^{n-1}, y_1=l \;|\; N)$ for $1 \leq l \leq |{\cal S}_1|$ and
can be computed using equations \ref{eqn:init-tree}--\ref{eqn:coal_counts}.

Notice that, unlike with a standard, locally normalized HMM, it is not true
in this model that $\sum_m a^i_{l,m} = 1$.  Furthermore, for two positions
$i$ and $j$, it 
is not true in general that $\sum_m a^i_{l,m} = \sum_m a^j_{l,m}$, because of
differences across positions in the local trees $\bar{T}_i^{n-1}$ and
recombination 
points $\bar{R}_i^{n-1}$.  Similarly, 
% ACS: this is not correct as written!  maybe just omit
%it is not true that the sum of the
%emission probabilities $b_l^i(D_i^n)$ over all possible values of $D_i^n$
%(all alignment columns of size $n$) is equal to one, nor that
%this sum must be the same at any two positions $i$ and $j$.  Finally, 
it is not
true that $\sum_l \pi_l = 1$.  Thus, this model is not only globally
unnormalized, but it also has a heterogenous local
normalization structure across positions.  It is these unusual features
of the threading HMM that make it more convenient to work with the
clamped joint distribution than with the conditional distributions of
direct interest.

\subsubsection{Stochastic Traceback}

Despite the unusual features of the HMM described in the previous section,
it still permits the use of standard dynamic programming algorithms to
integrate over all coalescent threadings $\V{Y}$ (the forward or backward
algorithms), obtain a most likely threadings $\hat{\V{Y}}$ (the Viterbi
algorithm), compute marginal posterior distributions for each $y_i$
(forward-backward algorithm), and sample threadings in proportion to their
conditional probability \cite{RABI89,DURBETAL98}.  These algorithms depend
only on the linear conditional independence structure of the model (and,
equivalently, on its factorization into local transition and emission
probabilities) and on the use of nonnegative potential functions,
% (CHECK!),
both properties that are maintained in this model.

We are primarily interested in a dynamic programming algorithm for
sampling from the posterior distribution over HMM paths that is
sometimes referred to as the {\em stochastic traceback} algorithm
\cite{CAWLPACH03,ZHUETAL98,DURBETAL98}.  In our case, each application of
this algorithm is guaranteed to sample a coalescent threading $\V{Y}$ in
proportion to the density $P(\bar{\V{T}}^{n-1},\V{Y},
\bar{\V{R}}^{n-1}, \bar{\V{D}}^n \;|\; \Theta)$, and equivalently, in
proportion to the desired conditional distribution.

The stochastic traceback algorithm consists of a deterministic forward pass
and a stochastic backward pass.  The forward pass is identical to the
forward algorithm.  In our notation, the algorithm recursively fills out a
matrix $\V{F} = \{f_{i,m}\}$, $1 \leq i \leq L$, $1 \leq m \leq
\max_i(|{\cal S}_i|)$.  Each $f_{i,m}$ represents the probability of
a prefix of the data joint with a constraint on the state path at position
$i$.  Here, $f_{i,m} = P(\bar{\V{T}}_{1:i}^{n-1},
\bar{\V{R}}_{1:i}^{n-1}, \bar{\V{D}}_{1:i}, y_i = m \;|\; \Theta)$, where
the notation $\V{X}_{i:j}$ indicates the subsequence $(X_i, \dots, X_j)$.
After an initialization of $f_{1,m} = \pi_m b_m^1(D_1^{n})$, for
$1 \leq m \leq |{\cal S}_1|$, the algorithm proceeds iteratively for $i$
from 2 to $L$ and sets each value $f_{i,m}$ (for $1 \leq l \leq |{\cal
  S}_i|$) equal to,
\begin{equation}
f_{i,m} = b_m^i(D_i^{n-1}) \; \sum_{l=1}^{|{\cal S}_{i-1}|} f_{i-1, l} \; a^{i-1}_{l, m}.
\end{equation}
Note that the heterogeneity of the state space along the sequence implies
that portions of the 
matrix are left undefined.

In the backward pass, the algorithm samples a sequence $\V{Y}$ one element
at a time, 
starting with 
$y_L$ and working backward to $y_1$.  
First, $y_L=l$ is simply sampled in proportion to $f_{L,l}$.
Then, for $i$ from $L-1$ down to 1, each $y_i$ is
sampled conditional on $y_{i+1}$ in proportion to,
\begin{equation}
q_i(y_i = l \;|\; y_{i+1}=m) \propto f_{i,l} \; a^{i}_{l,m}.
\label{eqn:backward}
\end{equation}
The limiting step of the algorithm is the forward pass, which in general
requires $O(C^2L)$ time, where $C$ is the size of the state space.
However, in our case the structure of the $\V{A}_i$ matrices can be
exploited to reduce the running time to $O(nK^2L)$ (see Text S1).

It can be shown by induction on suffixes of $\V{Y}$ that this procedure
will correctly sample from the target distribution, $P(\V{Y} \;|\;
\bar{\V{T}}^{n-1}, \bar{\V{R}}^{n-1}, \bar{\V{D}}^n, \Theta)$.
Briefly, in the base case, the suffix $y_L=l$ is by construction sampled from
the density $f_{L,l} = P(\bar{\V{T}}^{n-1}, \bar{\V{R}}^{n-1},
\bar{\V{D}}^n, y_L = l \;|\; \Theta)$, which is proportional to the
desired conditional distribution, $P(y_L=l \;|\; \bar{\V{T}}^{n-1},
\bar{\V{R}}^{n-1}, \bar{\V{D}}^n, \Theta)$.  For the inductive case,
assume $\V{Y}_{i+1:L}$ has been sampled from $P(\V{Y}_{i+1:L} \;|\;
\bar{\V{T}}^{n-1}, \bar{\V{R}}^{n-1}, \bar{\V{D}}^n, \Theta)$.  The
procedure of sampling $y_i$ from $q_i$ given $y_{i+1}$ is equivalent to
sampling from, 
\begin{align}
q_i(y_i = l \;|\; y_{i+1}=m) &\propto f_{i,l} \; a^i_{l,m}
= P(\bar{\V{T}}_{1:i}^{n-1}, \bar{\V{R}}_{1:i}^{n-1},
\bar{\V{D}}_{1:i}^n, y_i = l \;|\; \Theta) \;
P(\bar{R}_{i+1}^{n-1}, \bar{T}_{i+1}^{n-1}, y_{i+1}=m  \;|\; \bar{T}_{i}^{n-1},
y_{i}=l, \rho, N) \notag \\
&= P(\bar{\V{T}}_{1:i+1}^{n-1}, \bar{\V{R}}_{1:i+1}^{n-1},
\bar{\V{D}}_{1:i}^n, y_i = l, y_{i+1}=m \;|\; \Theta) \notag \\
&\propto P(y_i=l \;|\; \V{Y}_{i+1:L}, \bar{\V{T}}^{n-1},
\bar{\V{R}}^{n-1}, \bar{\V{D}}^n, \Theta),
\end{align}
where the last step is possible because $y_i$ is conditionally independent
of $\V{Y}_{i+2:L}$, $\V{T}^n_{i+2:L}$, $\V{R}^n_{i+2,L}$, and $D^n_{i+1:L}$
given $y_{i+1}$.  Thus, the algorithm will correctly sample
from $P(\V{Y}_{i:L} \;|\;
\bar{\V{T}}^{n-1}, \bar{\V{R}}^{n-1}, \bar{\V{D}}^n, \Theta)$ for all
$i$ such that $1 \leq i \leq L$.

%=============================================================================
\subsubsection{Sampling a Recombination Threading}
\label{sec:sample-recomb}

The final step in the threading operation is to sample a
recombination threading $\V{Z}$ conditional on a coalescent threading $\V{Y}$ and
the clamped parameters.  This step is greatly simplified by the fact that
the individual $z_i$ values are conditionally independent of one another
given the $y_i$ 
variables and the clamped $\V{T}^{n-1}_i$ and $\V{R}^{n-1}_i$ variables
(see Figure \ref{fig:graphmodels}B). 
Consequently, each $z_i$ can be sampled separately from the distribution,
\begin{equation}
P(z_i \;|\; \bar{R}_i^{n-1}, \bar{T}_i^{n-1}, \bar{y}_i, \bar{T}_{i-1}^{n-1},
\bar{y}_{i-1}, \Theta) \propto 
P(\bar{R}_i^{n-1}, z_i \;|\; \bar{T}_{i-1}^{n-1}, \bar{y}_{i-1}, \rho) \;
P(\bar{T}_i^{n-1}, \bar{y}_i \;|\; \bar{R}_i^{n-1}, z_i,
\bar{T}_{i-1}^{n-1}, \bar{y}_{i-1}, N),
\end{equation}
where the $y_i$ variables are now clamped along with the $\V{T}^{n-1}_i$
and $\V{R}^{n-1}_i$ variables.  Notice that the distribution on the RHS is
the same one considered in equations \ref{eqn:reduced} \&
\ref{eqn:transition-prob}.  The normalizing constant for this
distribution, for clamped values $\bar{y}_{i-1}=l$ and $\bar{y}_i=m$,
is given by the transition probability $a^{i-1}_{l,m}$.

Notice that this distribution is implicitly degenerate in the case in which
$\bar{R}_i^{n-1} \ne \emptyset$, owing to the limitation of at most one
recombination event per position.
In particular, if $\bar{R}_i^{n-1} \ne \emptyset$, then
$P(\bar{R}_i^{n-1}, z_i \;|\; \bar{T}_{i-1}^{n-1}, \bar{y}_{i-1},
\rho) = I[z_i=\emptyset]$, hence $P(z_i \;|\; \bar{R}_i^{n-1}, \bar{T}_i^{n-1}, \bar{y}_i, \bar{T}_{i-1}^{n-1},
\bar{y}_{i-1}, \Theta) = I[z_i = \emptyset]$. 
At the same time, notice that, if $\bar{R}_i^{n-1} = \emptyset$, 
 a new recombination is still possible ($z_i \ne \emptyset$) even if
 $\bar{\V{T}}_{i-1}^{n-1} = \bar{\V{T}}_i^{n-1}$ and $\bar{y}_{i-1} =
 \bar{y}_{i}$, because a branch could be broken by a recombination event
 but then re-coalesce at precisely its original position in the local tree.

When $\bar{R}^{n-1}_i = \emptyset$, the efficiency of sampling from this
distribution  
can be improved by noting that most possible $z_i$ values still have zero
probability.
Let ${\cal Z}$ represent the set of $z_i$ values having nonzero probability
for given values of $y_{i-1}$, $y_{i}$, and $v$, where $v$ denotes the
branch being threaded.  There are two cases to consider, a main case and a
special case.  We will denote the corresponding subsets of $z_i$ values
${\cal Z}_1$ and 
${\cal Z}_2$, with ${\cal Z} = {\cal Z}_1 \cup {\cal Z}_2$.  Recall that
$z_i = (w_i, u_i)$ and $y_{i} = (x_i, t_i)$, where $x_i$ and $w_i$ are
branches in $T^{n-1}_{i-1}$ and $T^{n-1}_{i}$, respectively, and $u_i$ and
$t_i$ are time points from the set ${\cal P} = \{s_0, \dots, s_K\}$.  
In the main case, the recombination occurs on the new branch $v$.  Here,
the recombination time $u_i$ must be at least as recent as both the old and
new re-coalescence times, $t_{i-1}$ and $t_i$.  Thus, ${\cal Z}_1 =
\{(v,u_i) \;|\; u_i \in {\cal P}, \, u_i \leq \min(t_{i-1},
t_i)\}$.  Notice that $|{\cal Z}_1| \leq K+1$.

The special case occurs when the recombination occurs not on the new
branch, $v$, but instead on $x_{i-1}$, the branch to which $v$ re-coalesces
at position $i-1$.  A recombination on branch $x_{i-1}$, below the point at
which $v$ joins it, followed by a re-coalescence of $x_{i-1}$ to $v$
(meaning that $x_i = x_{i-1}$) will
produce a signature exactly like the symmetric case of a recombination on
$v$ followed by a re-coalescence to $x_{i-1}$ (Supplementary Figure \ref{fig:recomb-cases}), so this
scenario must also be considered.  This case can only
occur when $x_{i-1} = x_i$ and in the interval of time between the start of
branch $x_i$ and $\min(t_{i-1}, t_i)$.  Recombinations on other branches
need not be considered, because the existence of such a recombination would
imply that $R_i \ne \emptyset$, contrary to our assumption.
Hence,
\begin{equation}
{\cal Z}_2 = \begin{cases}
\{(x_i,u_i) \;|\; u_i \in {\cal P}, \, u_i \geq s_k, \, u_i \leq \min(t_{i-1}, t_i)\}
& x_{i-1} = x_i \\ 
\emptyset & \text{otherwise},
\end{cases}
\end{equation}
where $s_k$ is the time point of the child node of branch $x_i$.  As
with ${\cal Z}_1$, $|{\cal Z}_2| \leq K+1$. 

By enumerating the elements of ${\cal Z}$, it is possible to sample
each $z_i$ in $O(K)$ time.  The same 
approach can be used to enable calculation of the $a^i_{l,m}$ values
(equation \ref{eqn:transition-prob}) in $O(K)$ time.  

\subsection{Data Preparation}

\subsubsection{Simulated Data}
Except where noted otherwise, simulations were performed under the
full coalescent-with-recombination model \cite{Hudson1983}.  After
generation of  
local trees, sequence alignments were generated using a finite-sites
Jukes-Cantor model \cite{JUKECANT69}.  All simulations were
performed using custom computer programs.
Our standard simulation scheme involved the generation of 
of twenty 1-Mb sequences, assuming an effective
population size of $N=$ 10,000,
a mutation rate of $1.8\times 10^{-8}$ mutations/site/generation,
% didn't we decrease this?
and mutation-to-recombination rate ratios of $\mu/\rho \in \{1, 2, 4, 6\}$
(i.e., recombination rates of $\rho \in \{1.8, 0.9, 0.45, 0.3\} \times
10^{-8}$ events/site/generation).  One hundred replicate data sets were
generated for each choice of $\mu/\rho$. 
Alternative parameter values were used in certain cases, as noted in the text
and figure captions.

% flesh this out.
% something about dividing into tree generation and data generation?  
% something about generation under SMC

\subsubsection{Real Data}

Information about human polymorphisms came from the ``69 Genomes'' data set
from Complete Genomics (CG)
(\url{http://www.completegenomics.com/public-data/69-Genomes}). 
For each individual considered, we recorded the diploid
genotype call reported for each position in the hg19 (Genome Reference
Consortium Human Build 37) reference genome using CG's `masterVar'
files.  We
considered both ``SNPs'' and ``length-preserving substitutions'' in the
masterVar file, and also noted positions where CG could not
confidently assign a genotype. All other positions were assumed to be 
homozygous for the allele reported in the reference genome.

Borrowing from our previous work on demography inference
\cite{Gronau2011}, we applied several filters to these data to reduce the
impact of technical errors from alignment, sequencing, genotype inference,
and genome assembly.  These filters include simple repeats, recent segmental
duplications, and transposable elements. We phased the data
using SHAPEIT v2 \cite{DELAETAL13}, guided by the pedigree information
describing the relationships among the 69 individuals. After phasing,
we removed the child in each trio, as well as all but the four grandparents
in the 17-member CEU pedigree, leaving 54 unrelated individuals in our
data set. From this set, we further filtered all CpG sites, 
sites with more than two observed alleles, and sites with a CG no-call in 
any of the 54 individuals.

%Filters were
%applied to eliminate repetitive sequences, recent duplications, 
%CpG sites,
%and regions not showing conserved synteny with outgroup genomes.  

%We also recorded the positions at which CG could not
%confidently assign a genotype for subsequent masking (see below).  
%We eliminated simple repeats, recent transposable
%elements, recombination hotspots \cite{1KGCONS10}, and also excluded all position
%pairs having a ``CG'' dinucleotide in any of the human samples or the
%outgroups, to avoid biases from hypermutable CpGs. As a further caution,
%we excluded position pairs with CX in an outgroup and YG in human, to avoid
%potential ancestral CpGs. We also excluded recent segmental duplications and
%regions in each outgroup genome not showing conserved synteny with the
%human reference sequence, also as described by \cite{GRONETAL11}. Finally,
%we excluded any additional regions found in the ``black list'' filter reported
%in \cite{ENCOCONS12}.
In order to
account for region-specific variation in recombination and mutation
rates, we used the HapMap phase II recombination map
\cite{Frazer2007}, and a mutation rate map estimated from alignments
of several primate genomes, including chimpanzee (panTro2), orangutan
(ponAbe2), and rhesus Macaque (rheMac2) \cite{GRONETAL12}.  Mutation
rates were scaled to have an average of $1.26\times10^{-8}$
mutations/generation/site and were averaged over 100kb non-overlapping
windows. This value was obtained by assuming a genome-wide average of
$1.8\times10^{-8}$ mutations/generation/site, and observing a $30\%$ 
reduction in nucleotide diversity when the CpG filter was applied.

\begin{figure}[h!]
\begin{center}
\includegraphics[width=4.35in]{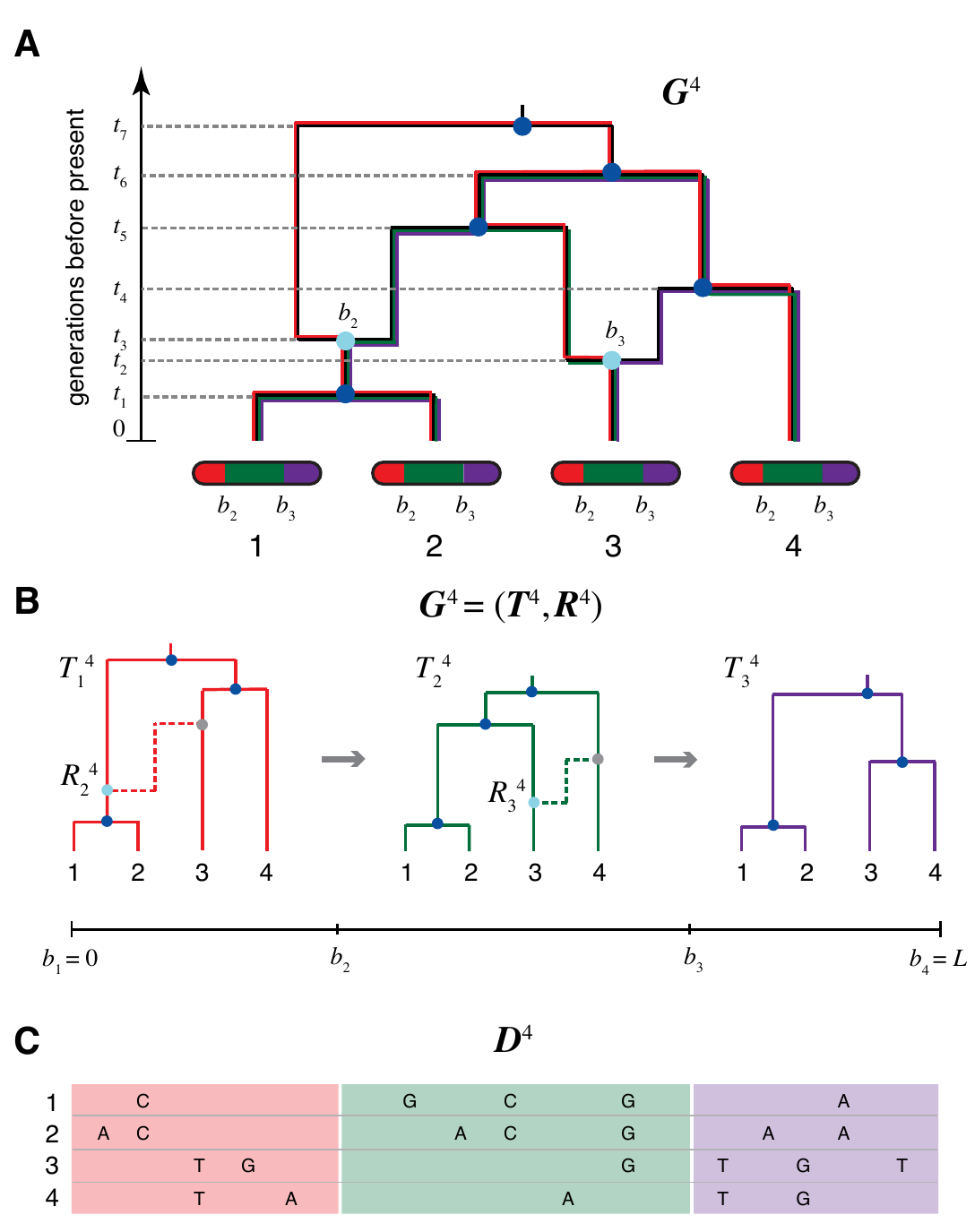}
\caption{{\bf An ancestral recombination graph (ARG) for four sequences.}
\label{fig:arg}
}
\end{center}
\end{figure}

Calls of hypervariable and invariant regions were obtained 
from the CG FTP site ({\url{ftp://ftp2.completegenomics.com}). Copy number
  variant calls for each individual were obtained from a 
file named cnvDetailsDiploidBeta, which was extracted from an
ASM-VAR-files tar archive.

\subsection{Data Analysis}

To sample ARGs genome-wide, we split each sequence alignment into
non-overlapping segments of 2 Mb, flanked on each side by 100 kb of overlapping
sequence.  We chose a core set of 12 individuals (24 haplotypes) 
randomly such that each major population group was represented.  We then
used \wv to sample ARGs for these genomes, assuming a population
size of $N=11,534$, $K=20$ time steps, and a maximum time of
$s_K=1,000,000$ generations.  Our prior estimate of $N$ was based on 
an empirical estimate of 
$4N\mu \approx \pi = 5.8 \times 10^{-4} $ from the CG sequence data,
and an assumption of $\mu =
1.26\times10^{-8}$ mutations per site per generation for
non-CpG sites (see previous section).
This initial step involved 500 sampling
iterations, consisting of 100 initial iterations under an infinite sites
assumption, and 400 iterations with the full finite sites model.  The final
sample from this initial step was used as a starting point for threading in
the remaining 
genomes. Once these were threaded, we applied \wv with infinite sites for 100
iterations, followed by 2400 iterations with the finite sites
model. Samples were recorded every 10 iterations for the final 2000
iterations, for a total of 200 samples.  For our genome-wide analyses, we
integrated the separate 2.2 Mb analyses by setting a switchpoint at the
middle of each overlapping 100 kb segment, in order to
minimize boundary effects at the analyzed sites.

To compute the neutral CDFs in Figure \ref{fig:przeworski}, we used a set
of putatively neutral regions obtained by removing all GENCODE (v15) genes plus 
1000 bp flank on either side of each exon, as well as all mammalian
phastCons elements plus  
100 bp of flanking sequence. From the remaining portion of the genome, we sampled 1000 sets 
of 69 regions with the same distribution of lengths as the non-CpG regions 
identified by \cite{LEFFETAL13}.

To estimate the allele age at each polymorphic site, we considered all
local genealogies sampled at that position, discarding any sampled
genealogies that required more than one mutation to explain the observed
data. In addition, we required that all 
of the retained genealogies implied the same derived allele, excluding
positions that violated this condition from our analysis. For the remaining
cases, we estimated the allele age for each sample as the average age of
the branch on which the mutation leading to the derived allele was assumed
to occur by parsimony, and averaged this value across samples.

\subsubsection{Reconstruction of Population Phylogeny}

%CONSENSUS
%
For the estimation of population phylogenies using PhyloNet, we
began by extracting genealogies from two loci per $\sim$2-Mb block, such that each
genealogy was approximately 1 Mb from the next one.  We then 
computed a consensus tree at each locus using a standard majority rule for edges
across the sampled local trees.  Here we considered every
100th sample from our \wv sampling run (21 trees per locus). In addition, we
collapsed identical adjacent local 
consensus trees into a single tree.  In the end, our analysis considered
2,304 trees from 1,376 $\sim$2-Mb blocks.
%SUBSAMPLING
%
Next, we reduced each tree to a single representative of each of the 11
represented populations by selecting one haploid sample
per population at random, and extracting the subtree 
spanning these 11 leaves.
%
% PHYLONET
%
We ran PhyloNet version 3.5.0 on these 11-leaf consensus
trees using the InferNetwork\_parsimony option and specifying a maximum
number of hybridization nodes in the range 0--5.
%
% FIG
%
%The population phylogeny shown in Fig.\ \ref{fig:poptrees}A describes the
%tree reconstructed when specifying no hybridization nodes, and the network
%shown in Fig.\ \ref{fig:poptrees}B describes the network inferred when
%requiring no more than 5 hybridization nodes, indicating for each of the
%hybridization nodes the stage in which it appears in the network.
%
% MORE
%
Identical phylogenies and networks were obtained for four different random
choices of haploid samples per population.

\enlargethispage{\baselineskip}

\section{Acknowledgments}

%This project was supported by a David and Lucile Packard Fellowship for
%Science and Engineering (to A.S.), NIH/NIGMS grant GM102192 (to A.S.), and
%postdoctoral fellowships from the Cornell Center for Comparative and
%Population Genomics (to M.D.R. and I.G.).  In addition, 
Our real data
analysis was enabled by use of the Extreme Science and Engineering
Discovery Environment (XSEDE), which is supported by National Science
Foundation grant number OCI-1053575 and hosted by the Texas Advanced
Computing Center (TACC) at The University of Texas at Austin.  We thank
Asger Hobolth, Thomas Mailund, Graham Coop, Richard Durbin,
Gerton Lunter, Gil McVean, Bob Griffiths, and many others for helpful
discussions.

%=============================================================================
% bib

\newpage
\singlespacing
\raggedright
\setstretch{1} 
%\bibliography{manuscript,compbio}

\clearpage

%=============================================================================

\setstretch{1.5}

\section*{Figure Legends}

\noindent {\bf Fig.\ \ref{fig:arg}}:
{\bf An ancestral recombination graph (ARG) for four sequences.}  (A) Going
backwards in time (from bottom to top), the graph shows how lineages that
lead to modern-day chromosomes (bottom) either ``coalesce'' into common
ancestral lineages (dark blue circles), or split into the distinct parental
chromosomes that were joined (in forward time) by recombination events
(light blue circles).  Each coalescence and recombination event is
associated with a specific time (dashed lines), and each recombination
event is also associated with a specific breakpoint along the chromosomes
(here, $b_2$ and $b_3$).  Each non-recombining interval of the sequences
(shown in red, green, and purple) corresponds to a ``local tree'' embedded
in the ARG (shown in matching colors).  Recombinations cause these trees to
change along the length of the sequences, making the correlation structure
of the data set highly complex.  The ARG for four sequences is denoted
$\V{G}^4$ in our notation.  (B) Representation of $\V{G}^4$ in terms of a
sequence of local trees $\V{T}^4$ and recombination events $\V{R}^4$.  A
local tree $T_i^4$ is shown for each nonrecombining segment in colors
matching those in (A).  Each tree, $T_i^4$, can be viewed as being
constructed from the previous tree, $T_{i-1}^4$, by placing a recombination
event along the branches of $T_{i-1}^4$ (light blue circles), breaking the
branch at this location, and then allowing the broken lineage to
re-coalesce to the rest of the tree (dashed lines in matching colors; new
coalescence points are shown in gray).  Together, the local trees and
recombinations provide a complete description of the ARG.  The Sequentially
Markov Coalescent (SMC) approximate the full coalescent-with-recombination
by assuming that $T_i^n$ is statistically independent of all previous
trees given $T_{i-1}^n$.  (C) An alignment of four sequences, $\V{D}^4$,
corresponding to the linearized ARG shown in (B).  For simplicity, only the
derived alleles at polymorphic sites are shown.  The sequences are assumed
to be generated by a process that samples an ancestral sequences from a
suitable background distribution, then allows each nonrecombining segment
of this sequence to mutate stochastically along the branches of the
corresponding local tree.  Notice that the correlation structure of the
sequences is fully determined by the local trees; that is, $\V{D}^n$ is
conditionally independent of the recombinations $\V{R}^n$ given the local
trees $\V{T}^n$.\vspace{2ex}

\noindent {\bf Fig.\ \ref{fig:arghmm}}: 
\textbf{The ``threading'' operation.}  The threading operation adds an
$n$th sequence to an ARG of $n-1$ sequences under a discretized version of
the SMC (the DSMC) that requires all coalescence and recombination events
to occur precisely at pre-defined time points, $s_0, \dots s_K$ (horizontal
dashed lines).  In this example, the fourth sequence has been removed from
ARG $\V{G}^4$ from Figure \ref{fig:arg}, leaving a tree with $n-1=3$ leaves
at each position $i$ ($T^{n-1}_i$; shown in black).  The fourth sequence
(shown in red) is re-threaded through the remaining portion of the ARG by a
two-step process that first samples a coalescence point $y_i$ for this
sequence at each $T^{n-1}_i$ (dark blue points), thereby defining a new
tree $T^n_i$, and second, samples a recombination point $z_i$ to reconcile
each adjacent pair of trees, $(T^{n}_{i-1}, T^{n}_i)$ (light blue points).
For simplicity, only the distinct local trees for the four nonrecombining
segments (after threading) are shown.  The gray box highlights the pair of
trees immediately flanking the breakpoint $b_3$.  Notice that the first
recombination from Figure \ref{fig:arg} is retained (dark gray nodes and
dashed line in left-most tree).  In general, new recombinations are
prohibited at the locations of ``given'' recombinations $R^{n-1}$ (see
text).  Note that it is possible for the attachment point of the $n$th
sequence in the local trees to move due to old recombinations as well as
new ones (not shown in this example).\vspace{2ex}

\noindent {\bf Fig.\ \ref{fig:graphmodels}}:
\textbf{Graphical models for Discretized Sequentially Markov Coalescent
  (DSMC) models}.(A) Full DSMC model for $n$ samples with local trees,
$\V{T}^n = (T_1^n, \dots T_L^n)$, recombinations, $\V{R}^n = (R_1^n, \dots
R_L^n)$, and alignment columns, $\V{D}^n = (D_1^n, \dots D_L^n)$.
Together, $\V{T}^n$ and $\V{R}^n$ define an ancestral recombination graph,
$\V{G}^n$.  Solid circles indicate observed variables and empty circles
indicate latent variables.  Arrows indicate direct dependencies between
variables and correspond to conditional probability distributions described
in the text.  Notice that the $R_i^n$ variables can be integrated out of
this model, leading to the conventional graph topology for a hidden Markov
model.  (B) The same model as in (A), but now partitioning the latent
variables into components that describe the history of the first $n-1$
sequences ($\V{T}^{n-1}$ and $\V{R}^{n-1}$) and components specific to the
$n$th sequence ($\V{Y} = (y_1, \dots, y_L)$ and $\V{Z} = (z_1, \dots,
z_L)$).  The $\V{T}^{n-1}$ and $\V{R}^{n-1}$ variables are represented by
solid circles because they are now ``clamped'' at specific values.  A
sample of $(\V{Y}, \V{Z})$ represents a threading of the $n$th sequence
through the ARG. (C) Reduced model after elimination of $\V{Z}$ by
integration, enabling efficient sampling of coalescent threadings $\V{Y}$.
This is the model used by the first step in our two-step sampling approach.
In the second step, the $\V{Z}$ variables are sampled conditional on
$\V{Y}$, separately for each $z_i$.  In this model, the grouped nodes have
complex joint dependencies, leading to a heterogeneous state space and
normalization structure, but the linear conditional independence structure
of an HMM is retained. \vspace{2ex}

\noindent {\bf Fig.\ \ref{fig:sim-eval-multipanel}}:
\textbf{Simulation results.} (A) Recovery of global features of simulated
ARGs from sequence data.  This plot is based on sets of 20 1-Mb sequences
generated under our standard simulation parameters (see Methods) with
$\mu/\rho=2$ (see Supplementary Figure \ref{fig:sim-stats-eval} for
additional results).  From left to right are shown true ($x$-axis) versus
inferred ($y$-axis) values of the log joint probability (the logarithm of
equation \ref{eqn:dsmc-lik}), the total number of recombinations, and the
total branch length of the ARG. Each data point in each plot represents one
of 100 simulated data sets.  In the vertical dimension, circles represent
averages across 100 sampled ARGs based on the corresponding data sets,
sampled at intervals of 10 after a burn-in of 200 iterations, and error
bars represent the interval between the 2.5 and 97.5 percentiles.  In the
second and third plots, circles are interpretable as posterior expected
values and error bars as 95\% Bayesian credible intervals. (B)~Posterior
mean TMRCA (dark red line, with 95\% credible intervals in light red)
versus true TMRCA (black line) along a simulated genomic segment of 1 Mb.
This plot is based on a single representative data set of 20 1-Mb sequences
generated under our standard simulation parameters with $\mu/\rho=6$ (see
Supplementary Figure \ref{fig:tmrca-full} for additional
results). \vspace{2ex}

\noindent {\bf Fig.\ \ref{fig:metagene-sweeps}}:
\textbf{Measures of genetic variation near protein-coding genes and partial
  selective sweeps}.} Shown (from top to bottom) are nucleotide diversity
($\pi$), time to most recent common ancestry (TMRCA), and relative TMRCA
halflife (RTH) for the 13 individuals (26 haploid genomes) of European
descent (CEU and TSI populations) in the Complete Genomics data set
(similar plots for African population are shown in Supplementary Figure
\ref{fig:metagene-sweeps-afr}).  Nucleotide diversity $\pi$ was computed as
the average rate of nucleotide differences per site across all pairs of
chromosomes, whereas sitewise values of the TMRCA and RTH were computed by
averaging over local trees sampled by \wv.  (A) Estimates for 17,845
protein-coding genes from the Consensus Coding Sequence (CCDS) track in the
UCSC Genome Browser (hg19).  Estimates for noncoding regions were computed
by averaging in a sliding window of size 300 bp then averaging across
genes.  Estimates for coding exons were computed by first averaging over
fourfold degenerate (4d) sites of each exonic type (first, middle, last),
then averaging across genes (see Methods).  Only 4d sites were considered
to focus on the influence of selection from linked sites rather than direct
selection but the plots are similar when all sites are included (data not
shown).  ``First exon'' is taken to begin at the annotated start codon and
``last exon'' to end at the stop codon, so that both exclude untranslated
regions.  The TMRCA is measured in thousands of generations.  RTH is ratio
of the time required for the first 50\% of lineages to find a most recent
common ancestor to the full TMRCA (see Supplementary Figure
\ref{fig:rth-conceptual}).  Error bars (dashed lines for noncoding regions)
indicate 95\% confidence intervals as estimated by bootstrapping over
regions.  (B) Similar plots for 255 100-kb regions predicted to have
undergone partial selective sweeps in the CEU population based on the iHS
statistic \cite{VOIGETAL06}.  In this case, all measures are computed in a
sliding window of 10,000 bases.  Notice that both protein-coding genes and
putative selective sweeps display substantial reductions in nucleotide
diversity, but the genes show a much more prominent reduction in TMRCA,
whereas the sweeps show a much more prominent reduction in RTH.  These
signatures are consistent with a dominant influence from background
selection rather than hitchhiking in protein-coding genes (see text).
\vspace{2ex}

\noindent {\bf Fig.\ \ref{fig:alleleAge}}:
\textbf{Mean allele age as a function of annotation class and derived
  allele frequency.} (A) Estimated age of derived allele in generations,
averaged across polymorphic sites of various annotation classes. Estimates
were derived from ARGs sampled by \wv based on the Complete Genomics data
set (see Methods). Error bars represent one standard deviation above and
below the mean. Neut = putatively neutral sites; 4d = fourfold degenerate
sites in coding regions; CNS = conserved noncoding sequences identified by
phastCons; PPh:\{Benign,PosDam,ProbDam\} = missense mutations identified by
PolyPhen-2 as ``benign'', ``possibly damaging'', or ``probably damaging'',
respectively; CV:\{NonPath,Path\} = mutations in ``nonpathogenic''
(categories 1--3) or ``pathogenic'' (categories 4 \& 5) classes in the
ClinVar database, respectively.  (B) Similar plot with categories further
divided by derived allele frequencies (DAF) in numbers of chromosomes out
of 108.  Error bars represent 95\% confidence intervals, as assessed by
bootstrapping.  In categories that combine multiple frequencies (e.g.,
4--5, 6--8), a subsampling strategy was used to ensure that the relative
contributions of the different frequencies matched those of the Neut class.
Estimates for DAF $>$20 were excluded due to sparse data.  Notice that ages
generally increase with DAF, as expected (see Supplementary Figure
\ref{fig:alleleAgeSim}), but at a considerably reduced rate in categories
under strong selection.  \vspace{2ex}

\noindent {\bf Fig.\ \ref{fig:poptrees}}:
\textbf{Human population phylogenies inferred from sampled ancestral
  recombination graphs.} Phylogenetic networks for the eleven populations
represented in the Complete Genomics data set were reconstructed using the
PhyloNet program \cite{THANNAKH09,YUETAL13}.  As input to PhyloNet, we used
2,304 local trees extracted from the ARG at approximately 1 Mb intervals,
with one randomly sampled chromosome per population (see Methods).  (A)
Population phylogeny inferred in the absence of hybridization/admixture,
showing the expected primary relationships among populations.  (B)
Population networks inferred when between one and five hybridization nodes
are allowed.  Populations inferred to be admixed are indicated by gray
lines and the inferred hybridization nodes are shown as gray circles.
Numbers indicate the order in which these nodes appear.  For example, when
one hybridization node is allowed, the MKK population is inferred to be
admixed, and when two are allowed, the MXL population is also inferred to
be admixed.  The inferred network is consistent with other recent studies
in many respects, but PhyloNet is unable to reconstruct the precise
topology of the complex subnetwork consisting of the GIH, MXL, PUR, CEU,
and TSI populations (see text).  Population names follow the convention
used by the HapMap 3 and 1000 Genomes projects: CHB = Han Chinese in
Beijing, China; JPT = Japanese in Tokyo, Japan; GIH = Gujarati Indians in
Houston, Texas; MXL = Mexican ancestry in Los Angeles, California; PUR =
Puerto Ricans in Puerto Rico; CEU = Utah residents with Northern and
Western European ancestry from the Centre d'Etude de Polymorphisme Humain
(CEPH) collection; TSI = Toscani in Italy; MKK = Maasai in Kinyawa, Kenya;
LWK = Luhya in Webuye, Kenya; ASW = African ancestry in Southwest USA; YRI
= Yoruba in Ibadan, Nigeria.\vspace{2ex}

\clearpage

\section*{Supplementary Figure Legends}

\noindent {\bf Supplementary Fig.\ \ref{fig:sim-dsmc}}:
\textbf{ARGs simulated under Discretized Sequentially Markov Coalescent
  model are similar to those simulated under continueous models.}ARGs were
simulated using the coalescent-with-recombination (red), Sequentially
Markov Coalescent (green), and Discretized Sequentially Markov Coalescent
(blue).  Three versions of the DSMC were considered: ones with with $K=40$
(dark blue), $K=20$ (medium blue), and $K=10$ (light blue) time intervals.
In all cases, we assumed $s_K=$ 200,000 generations.  Our standard
simulation parameters were used (see Methods) except that sequences were of
length 100 kb (rather than 1 Mb) to save in computation.  (A) Numbers of
recombinations at four different recombination rates corresponding to
$\mu/\rho = 1, 2, 4, 6$ (in reverse order).  To make the comparison fair,
recombinations between nonancestral sequences (which are disallowed by the
SMC/DSMC) are excluded in the case of the coalescent-with-recombination.
However, ``diamond'' or ``bubble'' recombinations (ones that are
immediately reversed by coalescence events, going backwards in time) were
included, so any distortion from excluding these events in the SMC/DSMC is
reflected in the figure.  (B) Numbers of segregating sites at three
different effective population sizes with $\mu/\rho=1$.\vspace{2ex}

\noindent {\bf Supplementary Fig.\ \ref{fig:leaf-trace}}:
\textbf{Illustration of ``leaf trace.''} An example leaf trace (highlighted
in gray) is shown for a hypothetical 10-kb genomic segment and six haploid
sequences.  The ARG for these sequences contains two local trees (shown to
left and right) separated by a single recombination event (red circle and
arrow).  In the leaf trace, each sequence is represented by a line, and
these lines are ordered and spaced according to the local tree at each
position.  Spacing between adjacent lines is proportional to time to most
recent common ancestry of associated sequences.  Nonrecombining genomic
intervals are reflected by blocks of parallel lines.  Recombinations lead
to changes in spacing and/or order and produce vertical lines in the
plot. Notice that aspects of the leaf ordering are arbitrary, because the
two children between each ancestral node can be exchanged without altering
the meaning of the diagram.  In addition, this visualization device applies
to a single ARG and does not easily generalize to distributions of possible
ARGs.  For our genome browser tracks, we use the single most likely ARG
sampled by \wv as the basis for the plots.  Finally, note that the lines in
the plot can be colored in various ways.  In our current tracks, they are
colored according to the population origin of each haploid
sequence.\vspace{2ex}

\noindent {\bf Supplementary Fig.\ \ref{fig:sim-converge}}:
\textbf{Convergence of \wv with simulated data.}  When the number of
sequences exceeds 6--8, the Metropolis-Hastings algorithm and subtree
threading operation are needed for \wv to have acceptable convergence
properties.  This plot shows results for 20 1-Mb sequences, generated under
our standard simulation parameters with $\mu/\rho=2$ (Methods).  Here the
measure of convergence is the difference between the number of inferred
recombination events and the number of true recombination events.  Other
measures show similar patterns.\vspace{2ex}

\begin{figure}[h!]
\begin{center}
\includegraphics{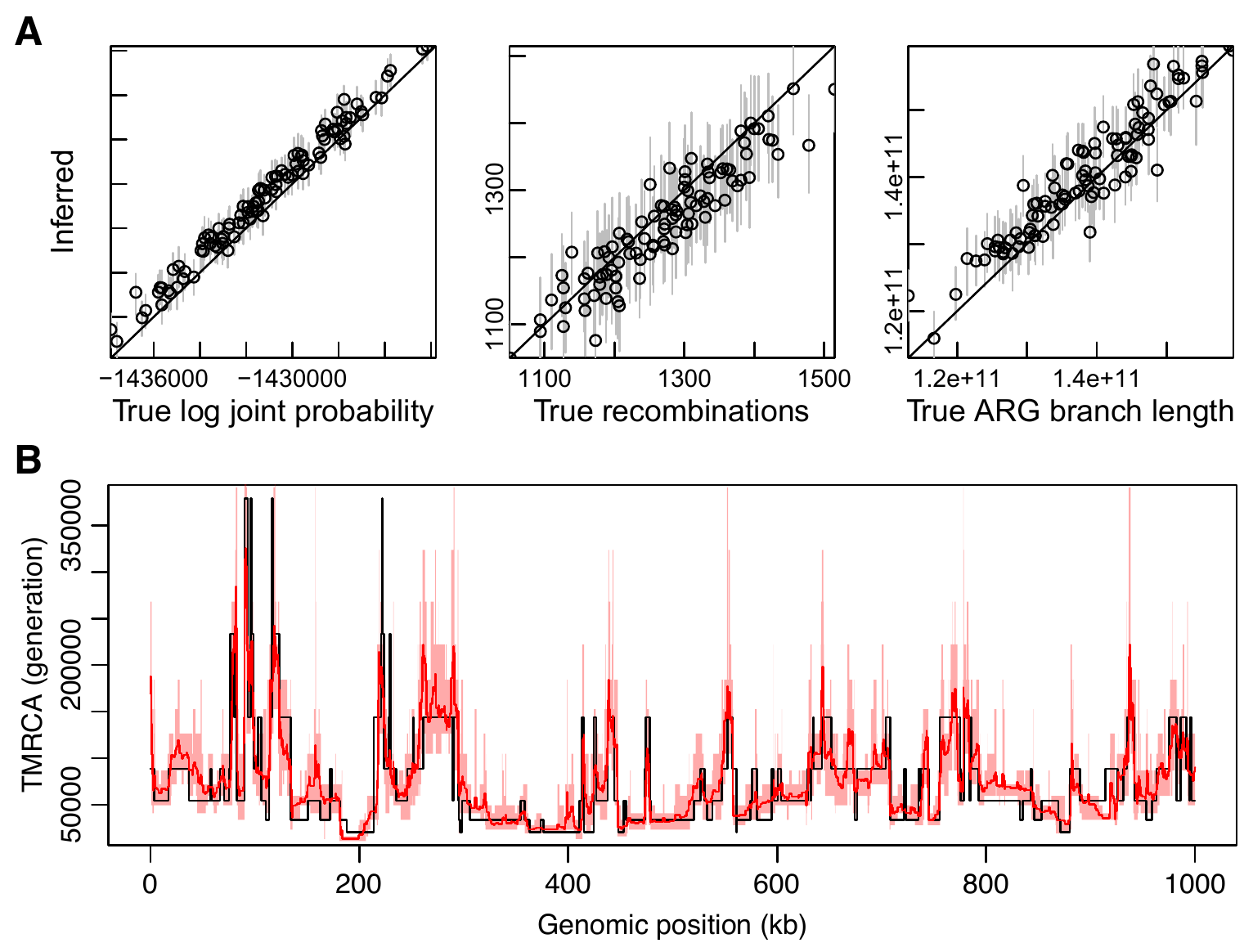}
\caption{\textbf{Simulation results.}
\label{fig:sim-eval-multipanel}   
}
\end{center}
\end{figure}

\noindent {\bf Supplementary Fig.\ \ref{fig:sim-stats-eval}}:
\textbf{Recovery of global features of simulated data for various values of
  $\mu/\rho$.}  This figure is the same as Figure
\ref{fig:sim-eval-multipanel}A, except that it shows results for four
different values of the mutation-to-recombination rate ratio, ranging from
$\mu/\rho=1$ (bottom row) to $\mu/\rho=6$ (top row).  The second row from
the bottom (with $\mu/\rho=2$) is identical to Figure
\ref{fig:sim-eval-multipanel}A.  Notice that high values of $\mu/\rho$ lead
to reduced variance in all estimates, owing to larger numbers of mutations
per local genealogy, but that the estimates remain reasonably accurate in
all cases.  However, there does appear to be a slight tendency to
under-estimate the number of recombinations, particularly at low values of
$\mu/\rho$, probably due to approximations inherent in the DSMC (see text).
Note that these are generated by the full coalescent with recombination,
not the DSMC.  \vspace{2ex}

\noindent {\bf Supplementary Fig.\ \ref{fig:tmrca-full}}:
\textbf{Recovery of TMRCA along simulated sequences for various values of
  $\mu/\rho$.}  This figure is the same as Figure
\ref{fig:sim-eval-multipanel}B except that it shows results for four
diffrent values of the mutation-to-recombination rate ratio, ranging from
$\mu/\rho=1$ (bottom panel) to $\mu/\rho=6$ (top panel).  Each panel
represents one randomly selected simulated data set.  Pearson's correlation
coefficients ($r$) for true vs.\ estimated TMRCAs across all local trees
are shown in the top right corner of each panel.  As expected, the quality
of the estimates generally improves with $\mu/\rho$, but this example
suggests there is limited improvement above $\mu/\rho=4$.\vspace{2ex}

\noindent {\bf Supplementary Fig.\ \ref{fig:sim-recomb-map}}:
\textbf{Recovery of recombination rates from simulated data.} We simulated
an alignment of 100 sequences with $N=10,000$ and $\mu=2.5\times10^{-8}$,
allowing for variable recombination rates based on estimates along the
human genome.  Despite the assumption in the prior of a constant
recombination rate of $\rho=1.16\times10^{-8}$, the posterior mean estimate
of the average number of recombinations in a 1 kb sliding window (red line)
correlates well with the true recombination rates used during simulation
(black line).  Notice that recombination hotspots are clearly identifiable
by peaks in the inferred rates but the magnitudes of these peaks are
dampened by the use of a uniform prior.  Only recombinations that produced
changes in tree topology (the class that is detectable by our methods) were
considered for the plot of the true recombination rate.  \vspace{2ex}

\noindent {\bf Supplementary Fig.\ \ref{fig:alleleAgeSim}}:
\textbf{Estimating ages of derived alleles in simulated data.} (A,C,E,G)
Inferred allele age correlates well with true allele age according to both
Pearson's ($r$) and Spearman's rank ($r_s$) correlation coefficients.
Correlation is strongest for high mutation/recombination rate ratios. Ages
were estimated by calculating the midpoint of the branch on which the
mutation was inferred to occur, under an infinite sites model, and
averaging across sample from the posterior distribution.  Points are
colored on a spectrum from blue to green in proportion to derived allele
frequencies.  (B,D,F,G) Allele frequency has significantly lower
correlation with true allele age, implying that the ARG will enable much
better estimates of allele age than allele frequencies alone.  Ages are
measured in generations before the present.  Our standard simulated data
sets were used (Methods).\vspace{2ex}

\noindent {\bf Supplementary Fig.\ \ref{fig:sim-top-eval}}:
\textbf{Recovery of local tree topologies.} Sequences were simulated under
the coalescent-with-recombination using our standard parameters (Methods),
ARGs were inferred using \wv, then 100 equally spaced local trees were
extracted from the sampled ARGs.  The topologies of these trees were
compared with the true trees generated during simulation at corresponding
positions in the alignment.  We compared \wv with the heuristic {\em
  Margarita} program \cite{Minichiello2006} by two measures: (A) branch
correctness (one minus the normalized Robinson-Foulds (RF) distance
\cite{Robinson1981}) and (B) Maximum Agreement Subtree (MAST) percentages
(the size of the largest leaf-set such that induced subtrees are
topologically equivalent, expressed as a percentage of the total number of
leaves), across a range of mutation to recombination rate ratios
($\mu/\rho$).  In all cases, \wv produced significantly more accurate trees
than {\em Margarita}. \vspace{2ex}

\noindent {\bf Supplementary Fig.\ \ref{fig:sim-uncertainty}}:
\textbf{Local tree branch posterior probabilities inferred by \wv
  accurately reflect their probability of correctness.}  The branch
posterior probabilities found by \wv (red) more accurately reflect the
probability of the branch being correct than the frequency at which {\em
  Margarita} (blue) infers a branch.  For each method, branches were binned
by their posterior probability (windows of 5\%) and compared against their
frequency of branch correctness.  Shaded regions represent the 95\%
binomial confidence interval.  This plot is based on our standard simulated
data set with $\mu/\rho=6$.  Posterior probabilities for \wv are based on
1000 samples from the Markov chain, and the probabilities for {\em
  Margarita} reflect 100 independent samples.  \vspace{2ex}

\noindent {\bf Supplementary Fig.\ \ref{fig:rth-conceptual}}:
\textbf{Illustration of relative TMRCA halflife (RTH).} Expected
genealogies under (A) neutral drift, (B) background selection, and (C) a
partial selective sweep.  In each panel, the arrows to the left indicate
the complete TMRCA ($T$) and the ``half TMRCA'' ($H$), that is, the minimum
time required for half of all lineages to find a single most recent common
ancestor.  The relative TMRCA halflife (RTH) is defined by the ratio $H/T$.
Because background selection (B) should primarily reduce the overall rate
of coalescence, in a manner more or less homogeneous with respect to time,
it is expected to have little effect on the RTH.  Partial sweeps (C),
however, will tend to produce a ``burst'' of coalescent events following a
causal mutation (red circle), leading to reduced values of $H$.
Nevertheless, because some lineages escape the sweep, the full TMRCA $T$ is
likely to remain similar to its value under neutrality.  As a result, the
RTH will be reduced.  \vspace{2ex}

\begin{figure}[h!]
\begin{center}
\includegraphics[width=5.5in]{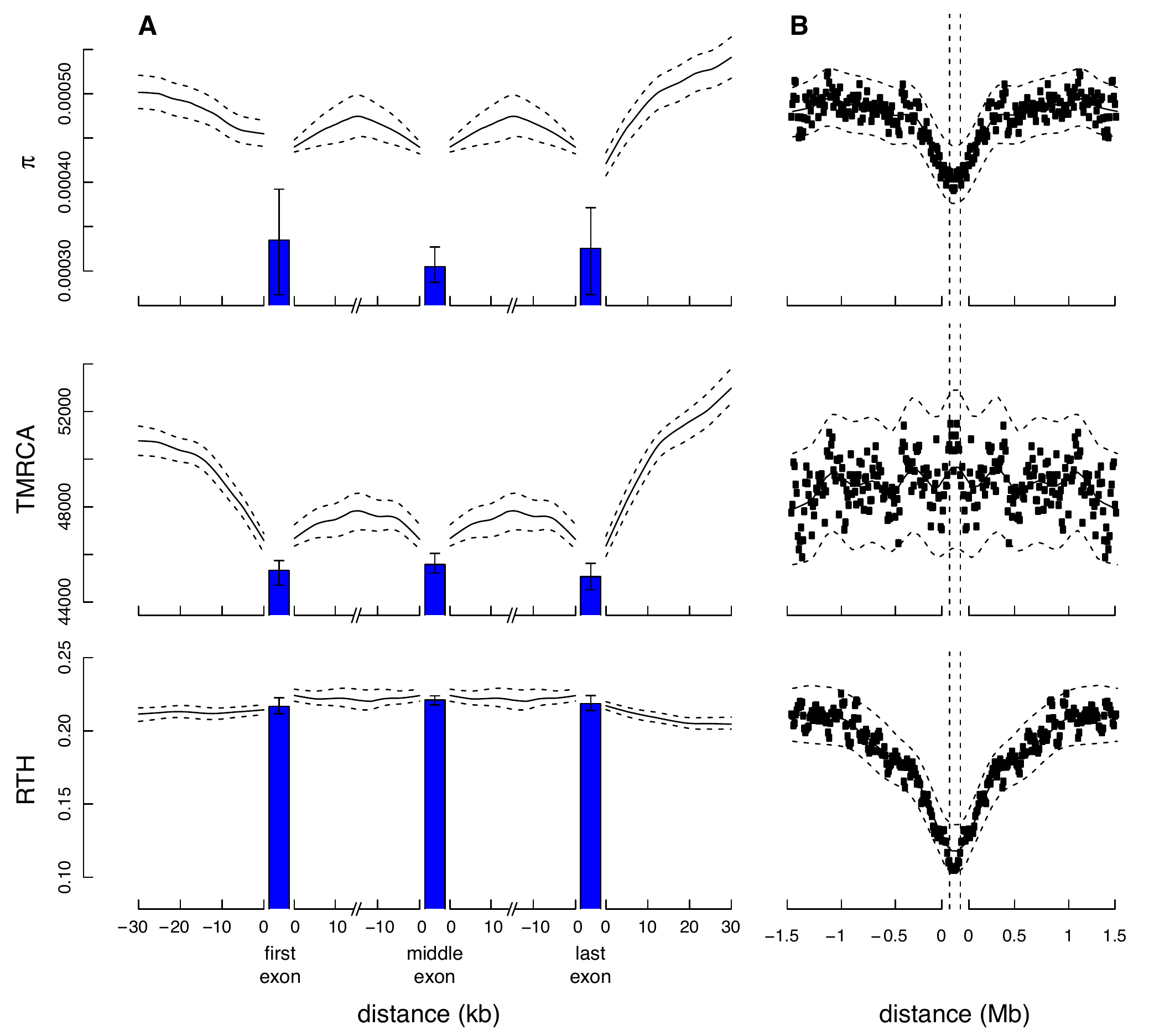}
\caption{\textbf{Measures of genetic variation near protein-coding genes
    and partial selective sweeps}.  }
\label{fig:metagene-sweeps}
\end{center}
\end{figure}

\noindent {\bf Supplementary Fig.\ \ref{fig:metagene-sweeps-afr}}:
\textbf{Measures of genetic variation near protein-coding genes and partial
  selective sweeps for African populations}. This figures is identical to
Figure \ref{fig:metagene-sweeps} except that it shows results for 17
African individuals or 34 haploid genomies (from the YRI, MKK, and LWK
populations).  Panel (B) is based on 271 100-kb regions predicted to have
undergone partial selective sweeps in the YRI population based on the iHS
statistic \cite{VOIGETAL06}.  \vspace{2ex}

\noindent {\bf Supplementary Fig.\ \ref{fig:hla}}:
\textbf{Time to most recent common ancestry (TMRCA) in the human leukocyte
  antigen (HLA) region. } Genome browser track displaying the sitewise time
to most recent common ancestry (TMRCA) estimated by \wv based on the
Complete Genomics individual human genome sequence data (track is available
at \url{http://genome-mirror.bscb.cornell.edu}, assembly hg19). The human
leukocyte antigen (HLA) region on human chromosome 6 contains many genomic
intervals with extremely elevated expected TMRCAs, including four of the
top 20 10-kb regions in the genome (highlighted here in gold; see
descriptions in Table \ref{tab:highTmrca}).  The red line indicates the
posterior mean of the TMRCA (estimated by averaging over the sampled local
trees) and the blue lines above and below indicate a Bayesian 95\% credible
interval.  \vspace{2ex}

\noindent {\bf Supplementary Fig.\ \ref{fig:kcne4}}:
\textbf{\wv tracks near {\em KCNE4}.} Shown is a $\sim$10-kb peak in the
estimated TMRCA about 20 kb downstream of the {\em KCNE4} gene (shown in
blue), which encodes a potassium voltage-gated channel strongly expressed
in the embryo and adult uterus.  The peak overlaps two ChIP-seq-supported
transcription factor binding sites analyzed by Arbiza et al.\
\cite{ARBIETAL13} (``INSIGHT Regulatory Selection'' track).  The four
tracks below the TMRCA plot show that the region in question displays
elevated rates of both low-frequency ($<$10\% derived allele frequency;
shown in blue) and high-frequency ($\geq$10\%; shown in red) polymorphisms
in the Complete Genomics data set, despite that divergence-based estimates
of the mutation rate are at or below the genome-wide average (average
values are indicated by horizontal black lines).  \wv explains these
observations by inferring one of the deepest average TMRCAs in the human
genome (\#5 in Table \ref{tab:highTmrca}).  Additional tracks show no
indication of copy number variation or recent duplications in this region.
The leaf trace indicates that the signal for a deep TMRCA is driven by
individuals from African populations (shown in green; the European and East
Asian populations are shown in blue and red, respectively), suggesting that
this region may contain ancient haplotypes specific to Africa.
\vspace{2ex}

\noindent {\bf Supplementary Fig.\ \ref{fig:bcar3}}:
\textbf{\wv tracks near {\em BCAR3}.} Shown is a large region of elevated
TMRCA in an intron of the {\em BCAR3} gene, which is involved in the
development of anti-estrogen resistance in breast cancer.  One 10-kb
segment of this region has an average expected TMRCA of 377,017
generations, or approximately 9.4 My (\#9 in Table \ref{tab:highTmrca}).
As in the previous example, this region shows elevated polymorphism rates
but average or below-average mutation rates and overlaps ChIP-seq-supported
transcription factor binding sites (INSIGHT track) \cite{ARBIETAL13}.
Again, the regions of extreme TMRCA do not seem to be explained by copy
number variation or recent duplications.  In this case, however, the leaf
trace demonstrates that the ancient haplotypes are distributed across all
three major population groups (African=green, European=blue, East
Asian=red). \vspace{2ex}

\noindent {\bf Supplementary Fig.\ \ref{fig:tulp4}}:
\textbf{\wv tracks near {\em TULP4}.} Another large region of elevated
TMRCA upstream of the {\em TULP4} gene, which is thought to be involved in
ubiquitination and proteosomal degradation and has a possible association
with cleft lip.  One 10-kb segment has an average expected TMRCA of 345,382
generations (8.6 My; \#16~in Table \ref{tab:highTmrca}).  As in the
previous two examples, this region has elevated polymorphism rates but not
mutation rates, overlaps ChIP-seq-supported transcription factor binding
sites (INSIGHT track), and does not seem to be an artifact of copy number
variation or recent duplications.  \vspace{2ex}

\noindent {\bf Supplementary Fig.\ \ref{fig:przeworski}}:
\textbf{Distribution of TMRCAs in regions predicted to be under balancing
  selection.} Cumulative distribution functions (CDFs) are shown for the
125 regions identified by Leffler et al.\ \cite{LEFFETAL13} based on
segregating haplotypes shared between humans and chimpanzees (black
circles), the subset of 69 loci containing no shared polymorphisms in CpG
dinucleotides (red circles) and a collection of 69 putatively neutral
regions having the same length distribution.  Neutral regions consisted of
noncoding regions from which known genes, binding sites, and conserved
elements had been removed (see \cite{ARBIETAL13}).  Notice the pronounced
shift toward larger TMRCAs in the regions predicted to be under balancing
selection, and a slightly more pronounced shift for the subset not
containing CpGs (which are less likely to have undergone parallel mutations
on both lineages).  TMRCAs are measured in generations, as in all other
figures and tables. \vspace{2ex}

\noindent {\bf Supplementary Fig.\ \ref{fig:frem3}}:
\textbf{\wv tracks near locus containing segregating haplotypes shared in
  humans and chimpanzees.} Elevated TMRCA corresponding to a region
identified by Leffler et al.\ \cite{LEFFETAL13} between the {\em FREM3} and
{\em GYPE} genes (\#11 in Table \ref{tab:lefflerTmrca}; see black square in
track at bottom).  The shared polymorphisms in this region are in strong
linkage disequilibrium with eQTLs for {\em GYPE}, a paralog of {\em GYPA},
which may be under balancing selection.  The leaf trace indicates that the
ancient haplotypes are shared across major human population groups
(African=green, European=blue, East Asian=red).  \vspace{2ex}

\begin{figure}[h!]
\begin{center}
\includegraphics[width=6in]{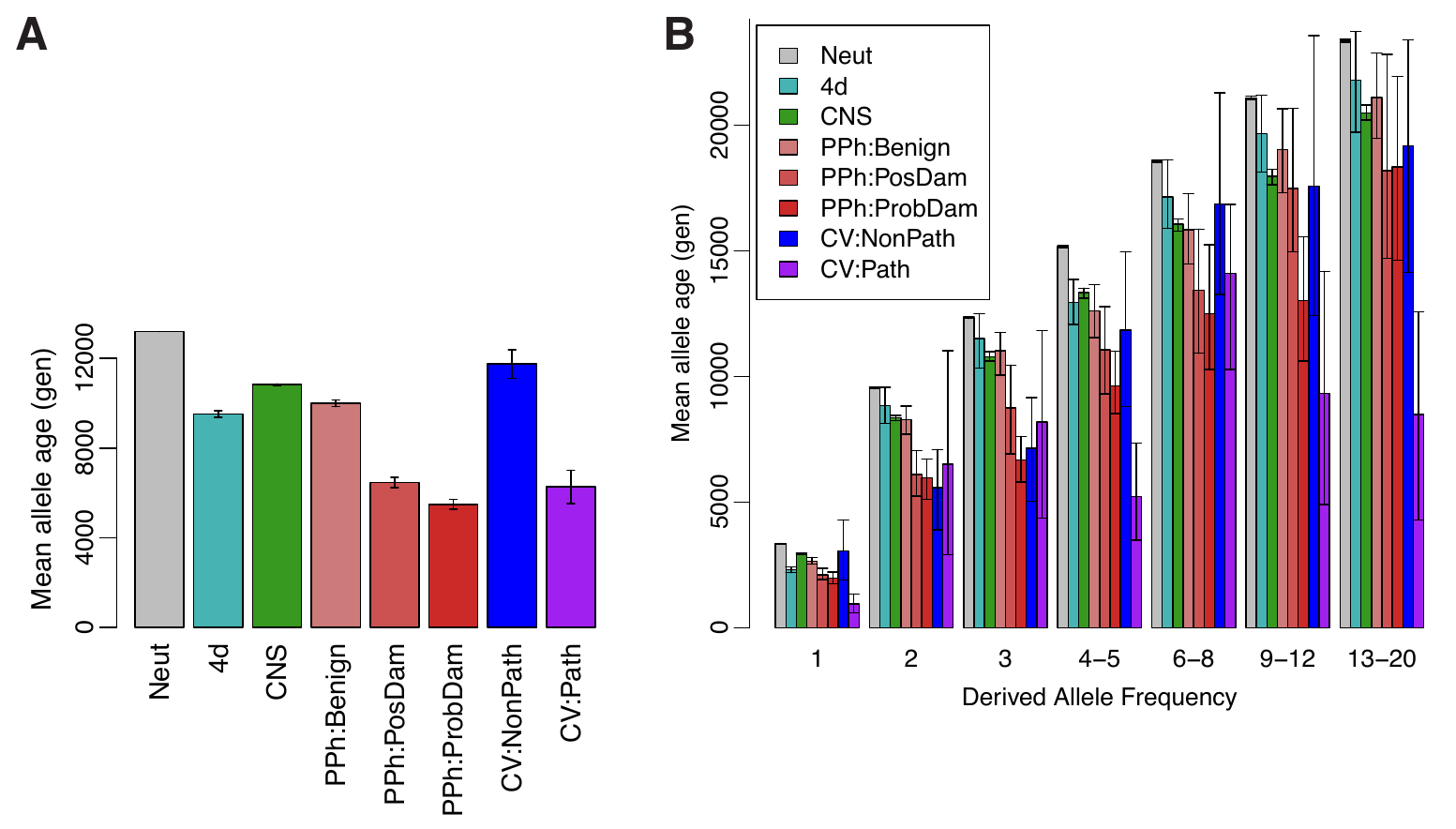}
\caption{\textbf{Mean allele age as a function of annotation class and
    derived allele frequency.}
}
\label{fig:alleleAge}
\end{center}
\end{figure}

\noindent {\bf Supplementary Fig.\ \ref{fig:alleleAgeDiff}}:
\textbf{Reduction in mean allele age as a function of annotation class and
  derived allele frequency.}  This figure shows the same information as
Figure \ref{fig:alleleAge}B, but instead of plotting absolute values of the
estimated allele ages, it plots the estimated {\em reduction} in allele age
relative to neutrality, which is defined as the differences between the
estimated age for each annotation type and the estimate for the
corresponding neutral class (in generations).  This representation shows
clearly that the reduction in allele age increases with allele frequency
much more rapidly for annotation classes under strong selection than for
those under weak selection.  The contrast between the nearly neutral
classes (4d, PPh:Benign, CV:NonPath) and the strongly selected classes
(PPh:ProbDam, CV:Path) is particularly striking.  This difference can be
understood as follows.  Reductions in allele age at nearly neutral sites
will primarily be a consequence of selection at linked sites, which, to a
first approximation, will decrease the local effective population size.
This will have the effect of approximately re-scaling allele ages by a
constant factor across all ages, making the reduction in age roughly
proportional to the absolute age.  Mutations under strong direct selection,
by contrast, will spend disproportionally less time at higher frequencies,
making their reductions in age at high frequencies disproportionally larger
than those for nearly neutral mutations (see \cite{KIEZETAL13}).  This
effect will occur even in the absence of dominance ($h = \frac12$), but it
could be exascerbated by dominance, which will tend to make low-frequency
alleles invisible to direct selection.  In any case, this plot shows that
selection from linked sites can produce comparable, or even larger,
reductions in age than direct selection at low allele frequencies, but at
high frequencies, direct selection tends to dominate in age
reduction. \vspace{2ex}

\noindent {\bf Supplementary Fig.\ \ref{fig:recomb-cases}}:
\textbf{Cases for new recombination $z_i$ given re-coalescence point
  $y_i$.} (A) In the main case, the recombination $z_i$ (blue point) occurs
on the branch that is being threaded into the ARG ($v$; shown in red).
After a recombination on this branch, a re-coalescence can occur at any
point $y_i$ (green points) in the local tree $T^{n-1}_i$ such that $y_i$ is
at least as old as $z_i$.  Therefore, when enumerating the possible $z_i$
consistent with a given $y_i$, one must consider all points on branch $v$
at least as recent as $y_i$.  This set is denoted ${\cal Z}_1$ in the text.
(B) There is an additional special case to consider when branch $v$
coalesces to the same branches of $T^{n-1}_i$ at positions $i-1$ and $i$,
that is, when $x_{i-1} = x_i$.  In this case, it is possible that the
recombination $z_i$ (blue point) occurs not on the new branch $v$ but on
$x_i$ (black branch) at a time point no older than the re-coalescence time
$y_i$ (green points).  A recombination of this kind will leave an identical
signature to the symmetric case of a recombination on $v$ in the same time
interval followed by a re-coalescence of $v$ to $x_i$.  Therefore, when
enumerating the possible $z_i$ consistent with a given $y_i$ such that
$x_{i-1} = x_i$, one must also consider the set ${\cal Z}_2$ consisting of
all $z_i$ on $x_i$ such that $z_i$ is at least as recent as $y_i$.  Notice
that, in both (A) and (B), the tree excluding $v$ is unchanged by all
recombination and coalescence scenarios $(z_i, y_i)$ under consideration,
i.e., $T^{n-1}_{i-1} = T^{n-1}_{i}$ (black branches).  \vspace{2ex}

\clearpage

\begin{table}[h!]
\begin{center}
\caption{{\bf Key to notation}}
\label{tab:notation}
\vspace{3ex}

\begin{minipage}{6.5in}
\begin{center}
\begin{tabular}{lp{5.5in}}
\hline\\[-1.5ex]
\multicolumn{2}{c}{{\bf Population Genetic Parameters}}\\[1ex]
\hline\\[-.5ex]
$\mu$ & Mutation rate, in events per site per generation \\
$\rho$ & Recombination rate, in events per site per generation \\
$N$ & Effective population size, in number of individuals\footnote{Model allows for
  a separate $N_l$ for each time interval $l$ but all analyses in this
  paper assume a constant $N$ across time intervals.}\\
$\Theta$ & Full parameter set, $\Theta = (\mu, \rho, N)$\\[2ex]
\hline\\[-1.5ex]
\multicolumn{2}{c}{{\bf Time Discretization}}\\[1ex]
\hline\\[-.5ex]
$K$ & Total number of time intervals (user-defined)\\
$s_j$ & Time point $j$ ($0 \leq j \leq K$), defining a boundary between time
intervals (generations before present)\\
$\Delta s_j$ & Length of $j$th time interval, $\Delta s_j = s_{j+1}-s_j$\\
$s_{j+\frac12}$ & Midpoint of $j$th time interval\\
$B(T, j)$ & Set of branches in a tree $T$ associated with time interval
$j$\\
$B_j$ & Number of branches associated with time interval $j$, $B_j =
|B(T,j)|$ (with $T$ determined by context) \\
$A(T, j)$ & Set of ``active'' branches at time point $j$ \\
$A_j$ & Number of ``active'' branches at time point $j$, $A_j = |A(T,j)|$ (with $T$ determined by context)\\[2ex]
\hline\\[-1.5ex]
\multicolumn{2}{c}{{\bf Ancestral Recombination Graph}}\\[1ex]
\hline\\[-.5ex]
$L$ & Length of analyzed sequence alignment in nucleotides \\
$n$ & Number of sequences in alignment \\
$D_i^n$ & Alignment column at $i$th position; cumulatively, $\V{D}^n = (D^n_1, \dots, D^n_L)$ \\
$T_i^n$  & Local tree for $i$th position; cumulatively, $\V{T}^n = (T_1^n,
\dots, T_L^n)$ \\ 
$R_i^n$ & Recombination point between $i-1$st and $i$th position;
cumulatively, $\V{R}^n = (R_2^n, \dots, R_L^n)$\\
$\V{G}^n$ & Full ARG for $n$ sequences, $\V{G}^n = (\V{T}^n, \V{R}^n)$ \\
$y_i = (u_i, t_i)$ & Coalescence point for threaded sequence at $i$th
position, defined by a 
branch $u_i$ and a time point $t_i$; cumulatively, $\V{Y}=(y_1,\dots,y_L)$ \\
$z_i = (w_i, u_i)$ & Recombination point for threaded sequence between
positions $i-1$ and $i$, defined by a
branch $w_i$ and a time point $u_i$; cumulatively,
$\V{Z}=(z_2,\dots,z_L)$\\[2ex] 
\hline\\[-1.5ex]
\multicolumn{2}{c}{{\bf Hidden Markov Model}}\\[1ex]
\hline\\[-.5ex]
$a^i_{l,m}$ & Transition probability from state $l$ to state $m$ between
position $i$ and $i+1$\\
$\pi_l$ & Initial state probability for state $l$\\
$b^i_l(D_i^n)$ & Emission probability for alignment column $D_i^n$ in state $l$
at position $i$\\[2ex]
\hline
\end{tabular}
\end{center}
\end{minipage}
\end{center}
\end{table}
\clearpage

\begin{table}[h!]
\begin{center}
\caption{\textbf{Top twenty 10 kb regions in the human genome by estimated TMRCA.}}
\label{tab:highTmrca}
\vspace{1ex}

\begin{minipage}{6.5in}
\begin{center}
\begin{small}
\begin{tabular}{rlrrrrrrp{2in}}
\hline
{\bf \#} & {\bf Chr}\footnote{Genomic coordinates in hg19 assembly.  The genome was
  simply partitioned into nonoverlapping 10 kb intervals in hg19
  coordinates.} & {\bf Start} & {\bf End} & {\bf TMRCA}\footnote{Posterior 
  expected TMRCA in generations, averaged across unfiltered genomic positions in region.} & {\bf
  Poly/kb}\footnote{Number of polymorphisms in Complete Genomics dataset
  in region per kilobase of
  unfiltered sequence.} & {\bf Npoly}\footnote{Normalized polymorphism
  rate: number of
  polymorphisms per unfiltered kilobase divided first by the local
  mutation rate (as estimated from divergence to nonhuman primate outgroup
  genomes) then by the average of the same polymorphism/divergence ratio in
  designated neutral regions.  The resulting value can be
  interpreted as a fold increase in the mutation-normalized polymorphism
  rate compared with the expectation under neutrality.} & {\bf
  CNV}\footnote{Possible copy number variant (CNV), based on Complete
  Genomics ``hypervariable'' or ``invariant'' labels (see
  Methods).  Polymorphism rates in these
  regions may be over-estimated.} & {\bf Comments}\\
\hline 
1 & chr4 & 190590001 & 190600000 & 615775 & 16.6 & 32.8 & $\checkmark$ & Part
of large intergenic region near telomere of
long arm of chr 4 (see \cite{HODGEYRE10}) \\
2 & chr5 & 21560001 & 21570000 & 503311 & 16.2 & 5.1 & $\checkmark$ & Intron of {\em GUSBP1}\\
3 & chr3 & 97930001 & 97940000 & 479803 & 16.4 & 5.3 & & Intergenic region in
cluster of olfactory receptor genes\\
4 & chr6 & 57270001 & 57280000 & 479504 & 13.7 & 28.0 & $\checkmark$ & Intron of {\em PRIM2} \\
5 & chr2 & 223940001 & 223950000 & 449728 & 19.8 & 4.3 & & Intergenic region
downstream of {\em KCNE4}\\
6 & chr5 & 21550001 & 21560000 & 412679 & 14.2 & 4.4 & $\checkmark$ & Intron of {\em
GUSBP1} \\
7 & chr6 & 57220001 & 57230000 & 399887 & 16.2 & 12.8 & $\checkmark$ & Intron of
{\em PRIM2} \\
8 & chr6 & 29680001 & 29690000 & 380228 & 15.3 & 10.0 & & Intergenic region
upstream of {\em HLA-F}\\
9 & chr1 & 94220001 & 94230000 & 377017 & 8.0 & 4.2 & & Intron of {\em BCAR3}\\
10 & chr8 & 123070001 & 123080000 & 375128 & 15.3 & 4.2 & & Intron of {\em BC052578} \\
11 & chr11 & 55670001 & 55680000 & 374537 & 12.0 & 4.3 & & Intergenic region
between {\em TRIM51} and {\em OR5W2}\\
12 & chr6 & 29950001 & 29960000 & 371110 & 17.6 & 7.6 & & Intergenic region
between {\em HLA-A} and {\em HLA-J}\\
13 & chr17 & 64010001 & 64020000 & 367842 & 8.6 & 5.5 & & Intron of {\em CEP112}\\
14 & chr6 & 29670001 & 29680000 & 365313 & 15.8 & 10.1 & & Intergenic region
upstream of {\em HLA-F} \\
15 & chr11 & 55690001 & 55700000 & 361088 & 11.5 & 4.1 & & Intergenic region
between {\em OR5W2} and {\em OR5I1} \\
16 & chr6 & 158680001 & 158690000 & 345382 & 10.4 & 4.8 & & Intergenic region
upstream of {\em TULP4}\\
17 & chr6 & 29720001 & 29730000 & 341797 & 12.4 & 8.0 & & Intergenic region
between {\em HLA-F} and {\em HLA-G}\\
18 & chr17 & 43790001 & 43800000 & 335647 & 11.2 & 5.0 & & Intron of {\em CRHR1}\\
19 & chr6 & 8470001 & 8480000 & 325656 & 10.1 & 4.5 & & Intron of noncoding RNA {\em
LOC100506207}\\
20 & chr4 & 141920001 & 141930000 & 325570 & 12.1 & 3.2 & & Intron of {\em RNF150}\\

\hline
\end{tabular}
\end{small}
\end{center}
\end{minipage}
\end{center}
\end{table}
\clearpage

\begin{table}[h!]
\begin{center}
\caption{\textbf{Top twenty regions of shared human/chimpanzee haplotypes
    by estimated TMRCA.}
}
\label{tab:lefflerTmrca}
\vspace{1ex}

\begin{minipage}{6.5in}
\begin{center}
\begin{small}
\begin{tabular}{rlrrrrrrp{1.6in}}
\hline
{\bf \#} & {\bf Chr}\footnote{Genomic coordinates in hg19 assembly.} & {\bf Start} & {\bf End} & {\bf TMRCA}\footnote{Posterior
  expected TMRCA in generations, averaged across unfiltered genomic positions in region.} & {\bf Poly/kb}\footnote{Number of polymorphisms in Complete Genomics dataset
  in region per kilobase of
  unfiltered sequence.}  & {\bf
  Npoly}\footnote{Normalized polymorphism rate: number of
  polymorphisms per unfiltered kilobase divided first by the local
  mutation rate (as estimated from divergence to nonhuman primate outgroup
  genomes) then by the average of the same polymorphism/divergence ratio in
  designated neutral regions (see Methods).  The resulting value can be
  interpreted as a fold increase in the mutation-normalized polymorphism
  rate compared with the expectation under neutrality.} & {\bf
  CNV}\footnote{Possible copy number variant (CNV), based on Complete
  Genomics ``hypervariable'' or ``invariant'' labels (see
  Methods).  Polymorphism rates in these
  regions may be inflated.  Few of
these regions were identified in the Leffler et al.\ data set, probably
because the authors were careful to filter out duplicated regions from
their analysis \cite{LEFFETAL13}.} & {\bf Comments} \\
\hline 
1 & chr7 & 47799979 & 47803415 & 307590 & 10.5 & 2.9 & & First exon/intron of
{\em LINC00525}\\
2 & chr4 & 56144164 & 56148467 & 256051 & 14.4 & 4.0 & & Upstream of {\em SRD5A3}\\
3 & chr5 & 8022829 & 8024476 & 249553 & 9.3 & 2.0 & & Downstream of {\em MTRR}\\
4 & chr3 & 143684547 & 143688535 & 235598 & 9.8 & 2.9 & & Upstream of {\em C3orf58}\\
5 & chr9 & 99546087 & 99550934 & 233492 & 8.6 & 2.6 & & Upstream of {\em ZNF510}\\
6 & chr18 & 58437379 & 58439410 & 228782 & 8.5 & 1.8 & & Distally upstream of {\em MC4R}\\
7 & chr8 & 134404327 & 134405512 & 227555 & 16.4 & 3.7 & & Downstream of {\em ST3GAL1}\\
8 & chr21 & 22045484 & 22048252 & 215718 & 12.2 & 2.5 & & Downstream of {\em LINC00320}\\
9 & chr7 & 45252745 & 45257527 & 201522 & 13.5 & 4.3 & & Downstream of {\em RAMP3} \\
10 & chr2 & 241121578 & 241124345 & 200321 & 16.1 & 3.0 & $\checkmark$ & Upstream of {\em OTOS}\\
11 & chr4 & 144654907 & 144662554 & 182348 & 11.9 & 2.5 & & Upstream of {\em FREM3}\\
12 & chr3 & 36203964 & 36205036 & 173655 & 15.5 & 2.9 & & Upstream of {\em STAC}\\
13 & chr2 & 101276944 & 101278537 & 173448 & 14.0 & 2.9 & & Downstream of {\em PDCL3} \\
14 & chr1 & 157716093 & 157718074 & 170583 & 10.1 & 2.4 & & Exon and introns of
{\em FCRL2}\\
15 & chr14 & 22320920 & 22323473 & 159251 & 13.8 & 2.4 & & Intron of {\em TCRA}\\
16 & chr14 & 88803535 & 88805909 & 155431 & 8.9 & 2.2 & & Upstream of {\em KCNK10}\\
17 & chr20 & 5337103 & 5340864 & 149816 & 11.1 & 2.8 & & Upstream of {\em PROKR2}\\
18 & chr4 & 57919549 & 57920587 & 146684 & 17.5 & 4.9 & & Intron of {\em IGFBP7}\\
19 & chr14 & 86147042 & 86149069 & 143608 & 10.1 & 2.1 & & Downstream of {\em FLRT2}\\
20 & chr11 & 81489342 & 81492793 & 143222 & 10.2 & 1.8 & & Downstream of {\em BC041900}\\

\hline
\end{tabular}
\end{small}
\end{center}
\end{minipage}
\end{center}
\end{table}
\clearpage

%===========================================================================

\clearpage

\clearpage

\clearpage

\clearpage

\clearpage

\clearpage

\clearpage

%%%%%%%%%%%%%%%%%%%%%%%%%%%%%%%%%%%%%%%%%%%%%%%%%%%%%%%%%%%%%%%%%%%%%%%%%%%

% List supplemental figures with an S
\setcounter{figure}{0}
\renewcommand{\thefigure}{S\arabic{figure}}
\renewcommand{\figurename}{Supplementary Figure}

\begin{figure}[h!]
\begin{center}
\includegraphics[width=6in]{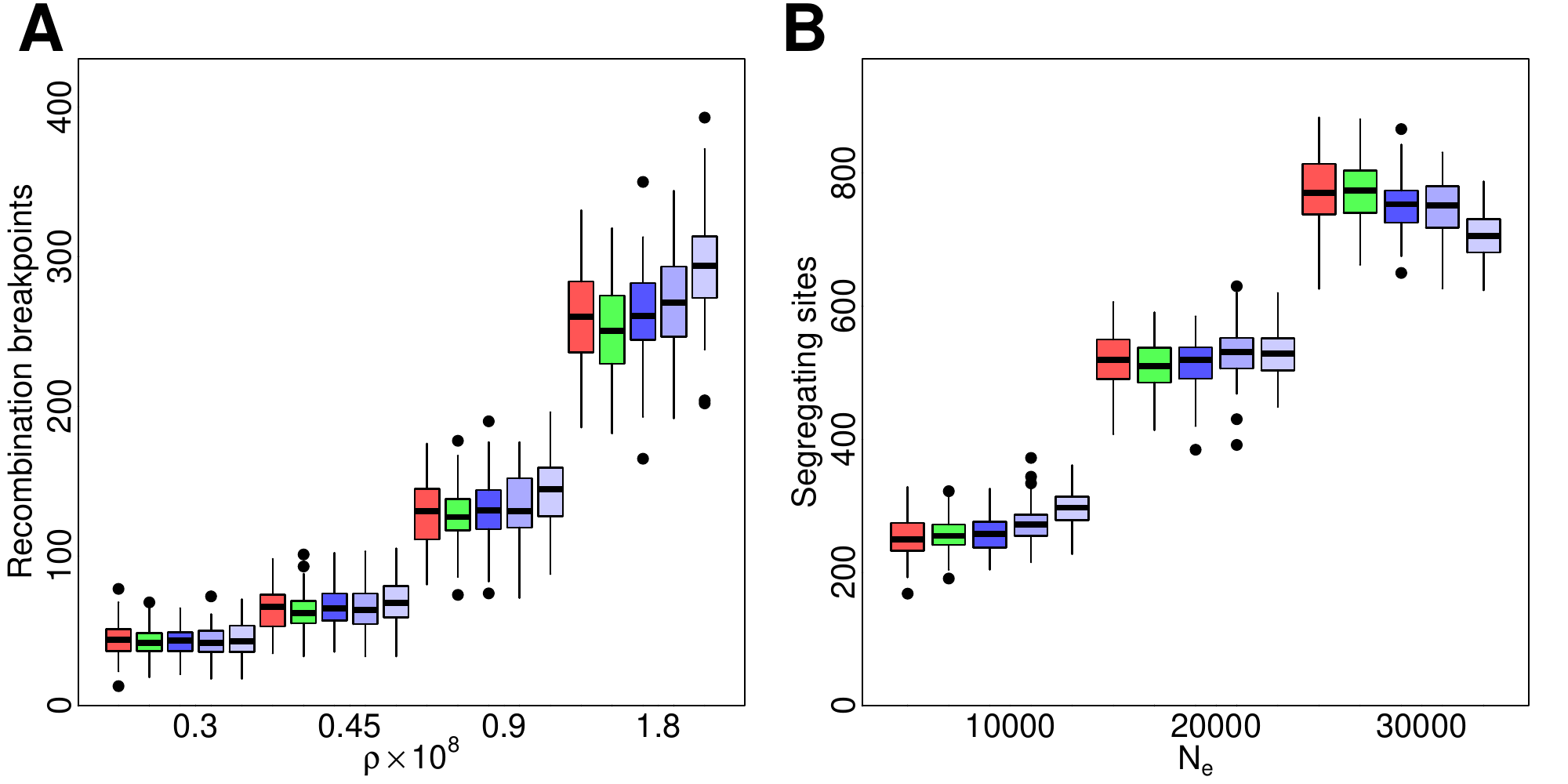}
\caption{\textbf{ARGs simulated under Discretized Sequentially Markov
    Coalescent model are similar to those simulated 
    under continueous models.}
\label{fig:sim-dsmc}
}
\end{center}
\end{figure}
\clearpage

\begin{figure}[h!]
\begin{center}
\includegraphics[width=6in]{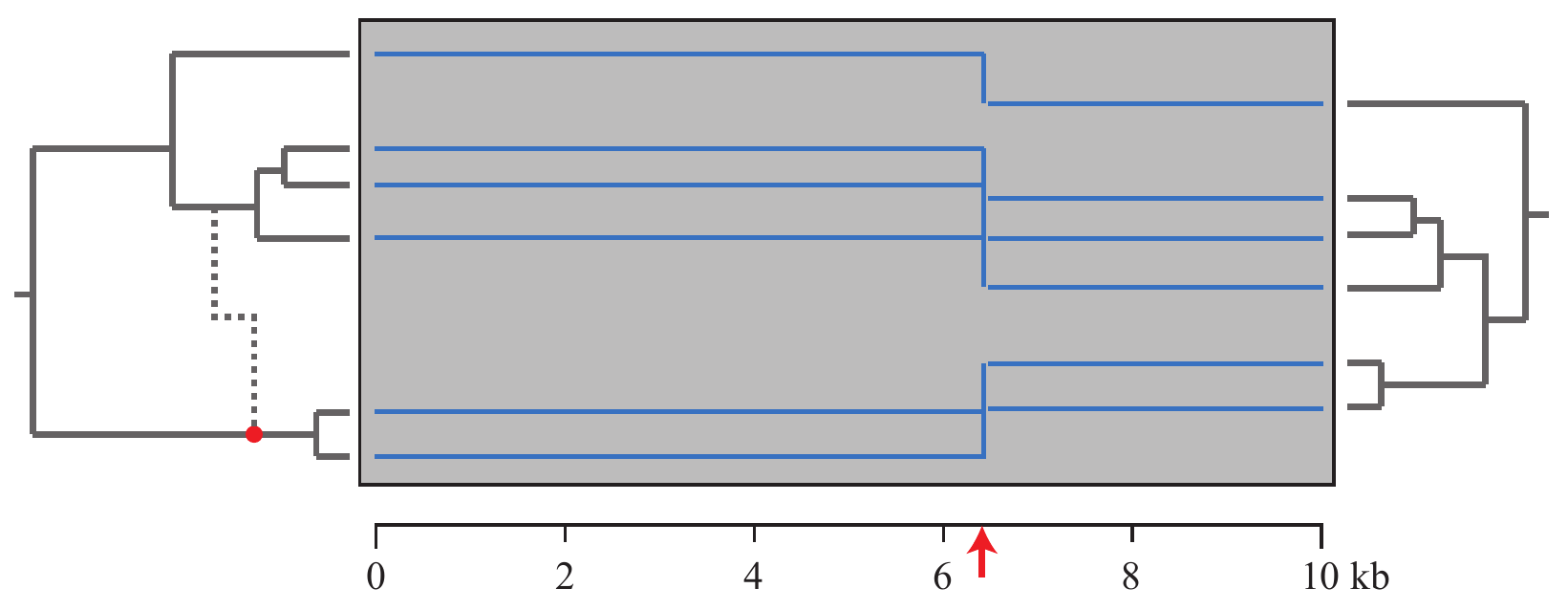}
\caption{\textbf{Illustration of ``leaf trace.''}}
\label{fig:leaf-trace}
\end{center}
\end{figure}
\clearpage

\begin{figure}[h!]
\begin{center}
\includegraphics[width=6in]{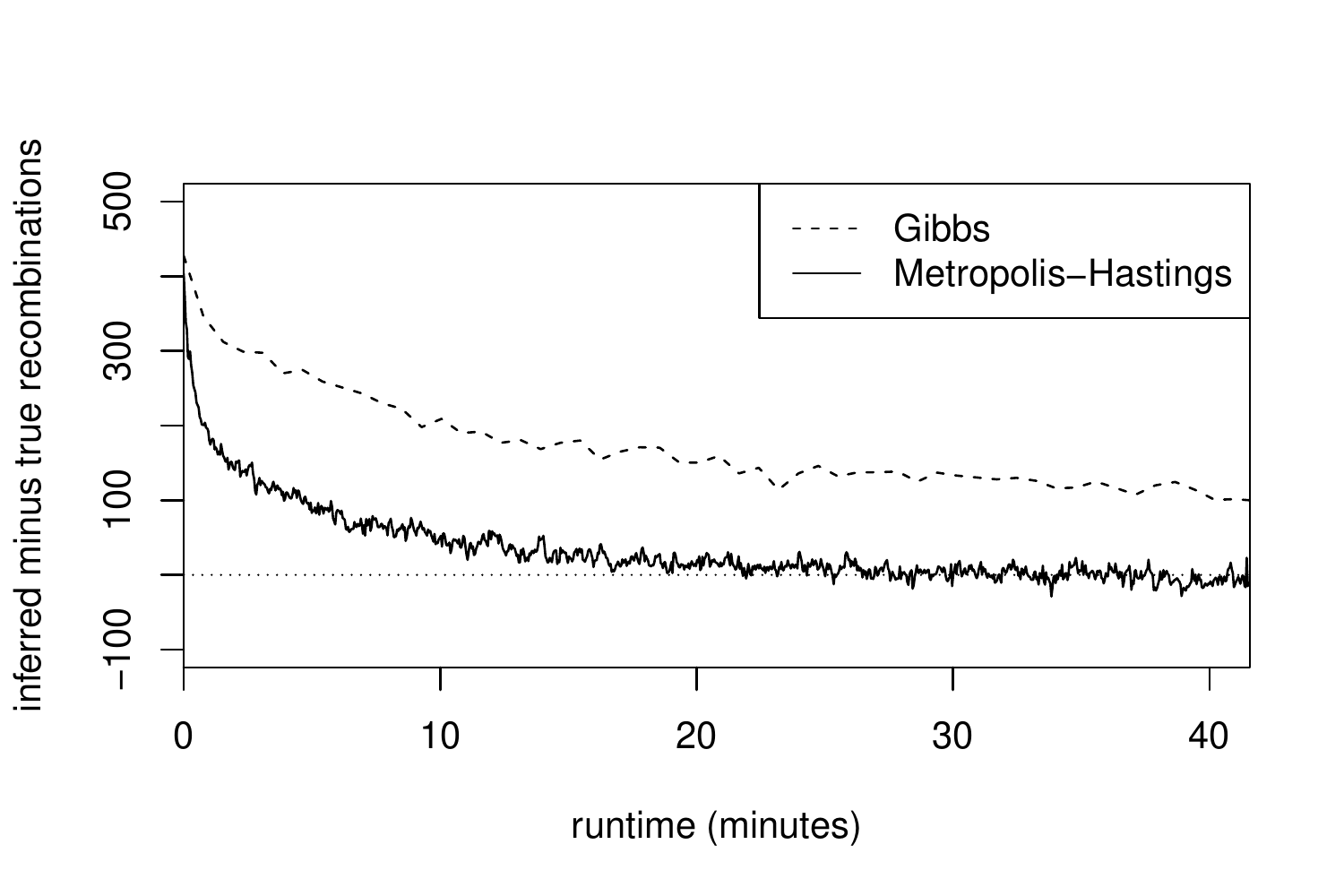}
\caption{\textbf{Convergence of \wv with simulated data.}  
\label{fig:sim-converge}
}
\end{center}
\end{figure}
\clearpage

\begin{figure}[h!]
\begin{center}
\includegraphics[width=6in]{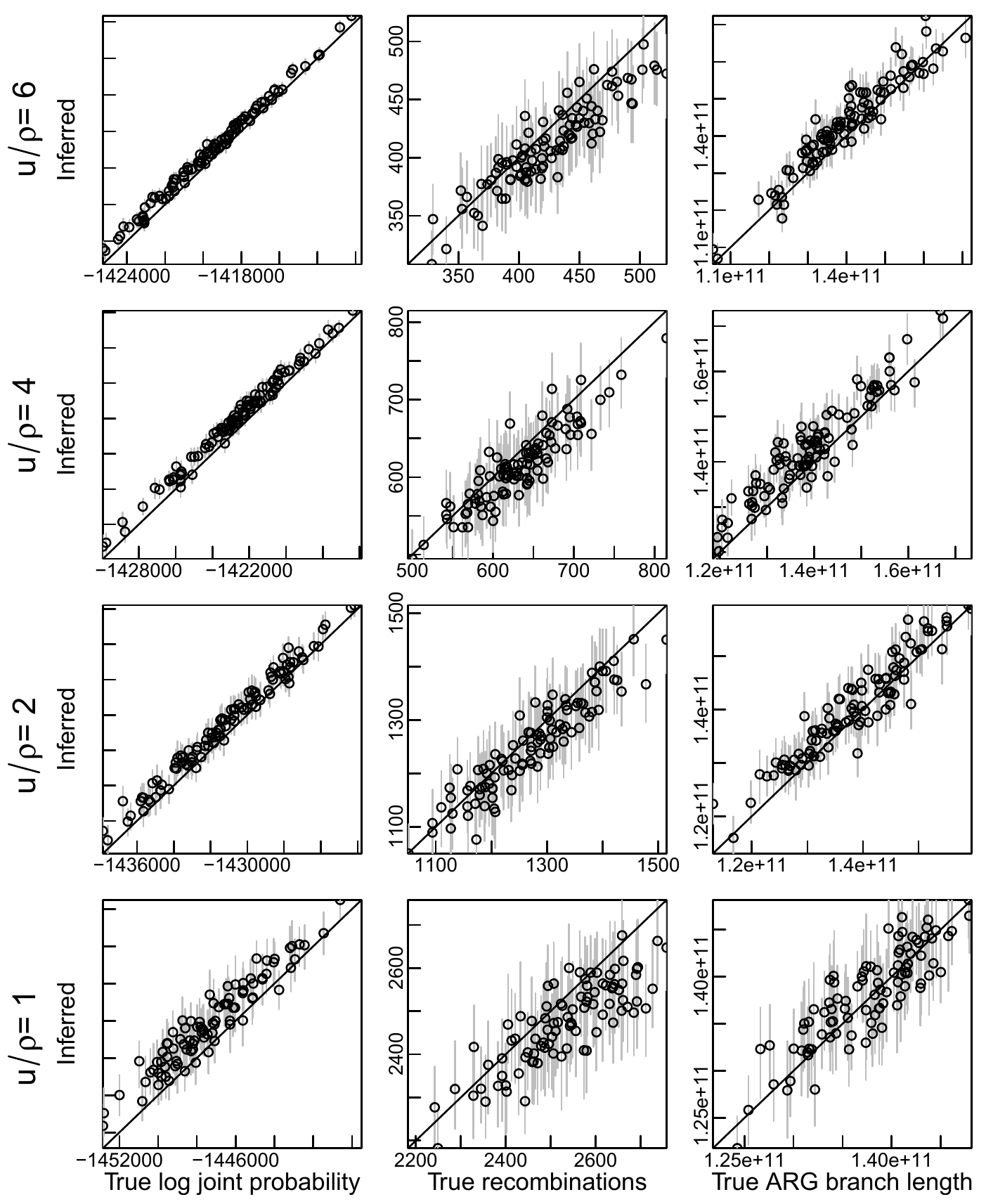}
\caption{\textbf{Recovery of global features of simulated data for various
    values of $\mu/\rho$.}  
\label{fig:sim-stats-eval}
}
\end{center}
\end{figure}
\clearpage

\begin{figure}[h!]
\begin{center}
\includegraphics[width=6in]{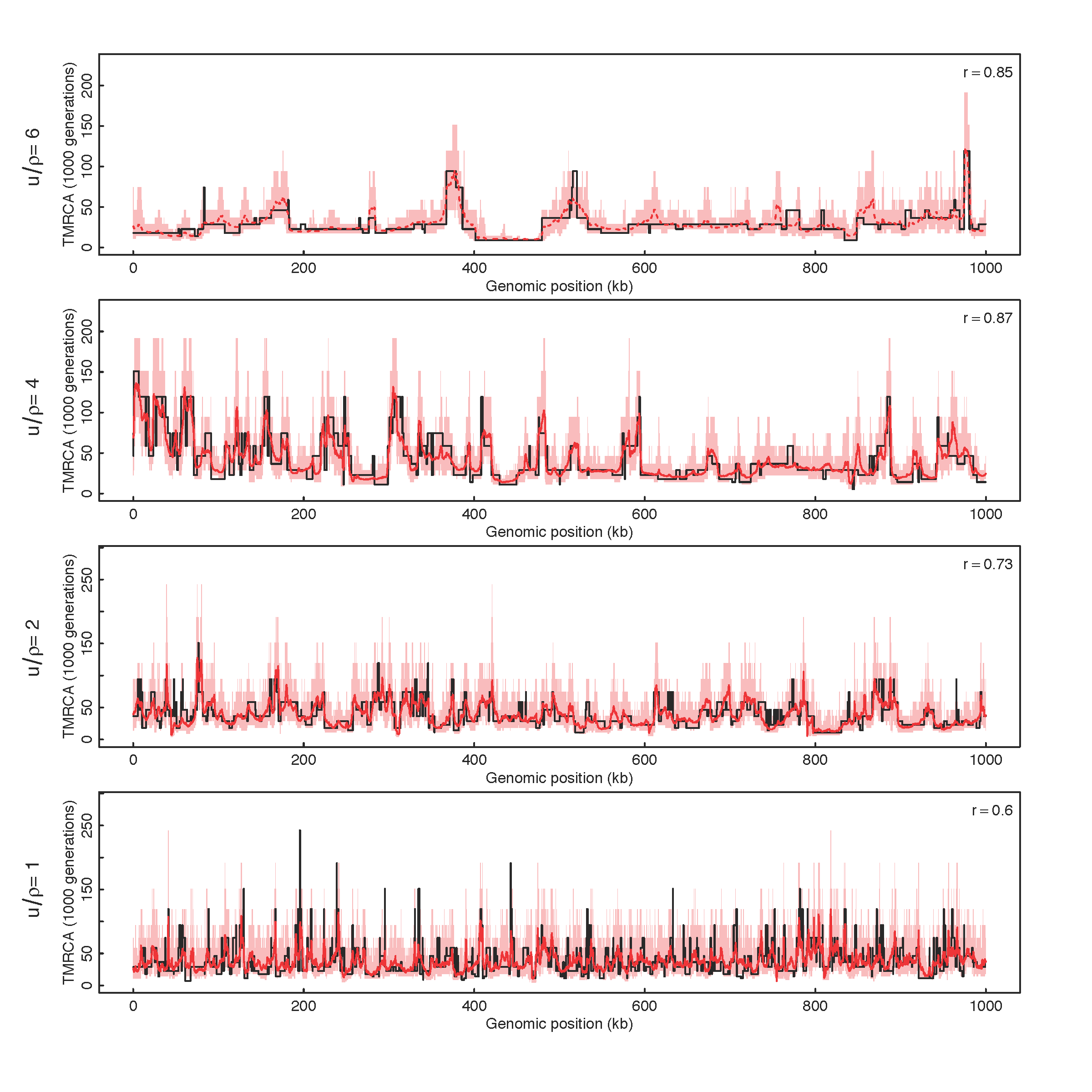}
\caption{\textbf{Recovery of TMRCA along simulated sequences for various
    values of $\mu/\rho$.}   
\label{fig:tmrca-full}
}
\end{center}
\end{figure}
\clearpage

\begin{figure}[h!]
\begin{center}
\includegraphics[width=6in]{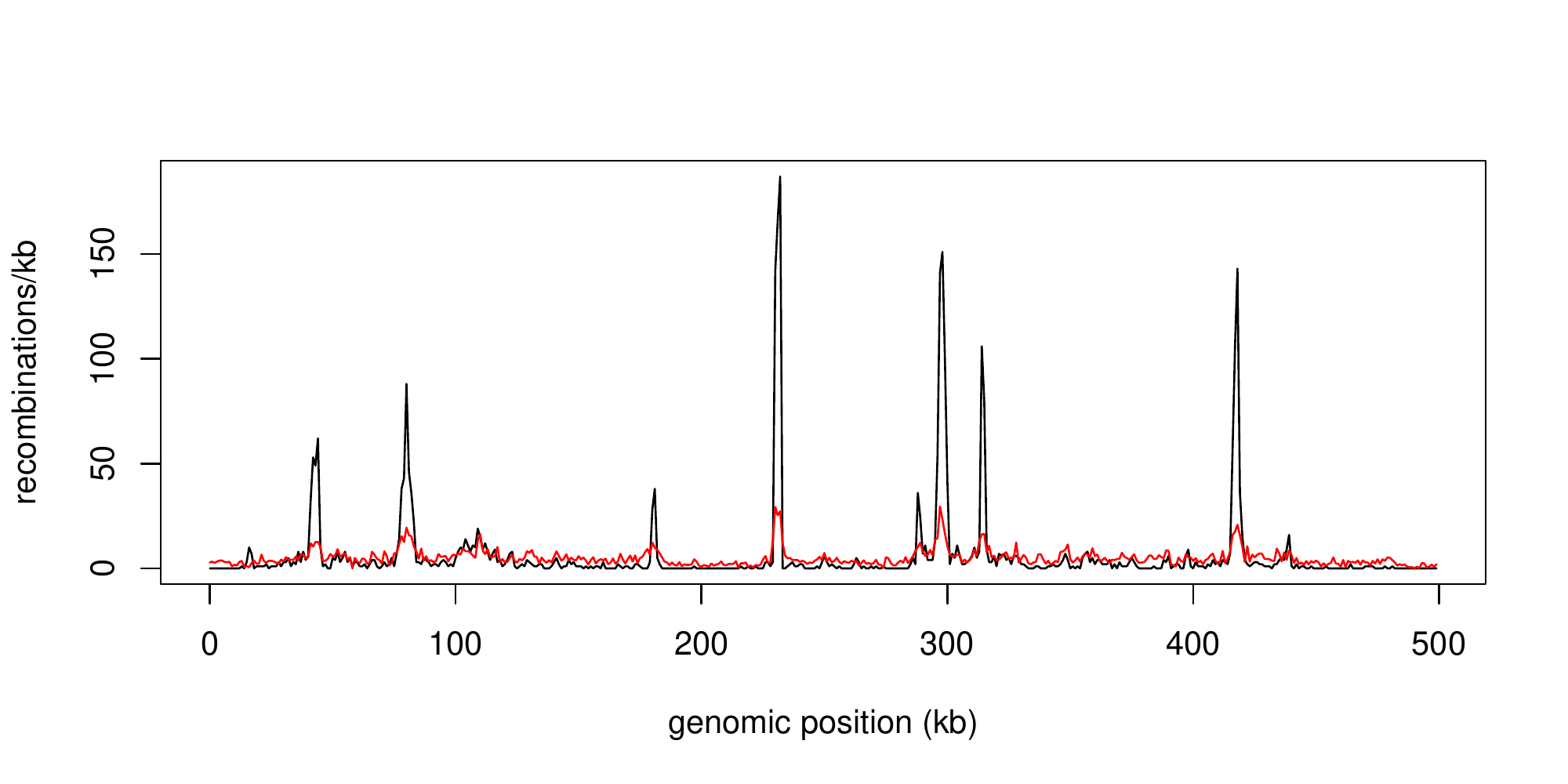}
\caption{\textbf{Recovery of recombination rates from simulated
    data.}  
\label{fig:sim-recomb-map}
}
\end{center}
\end{figure}
\clearpage

\begin{figure}[h!]
\begin{center}
\includegraphics[width=4in]{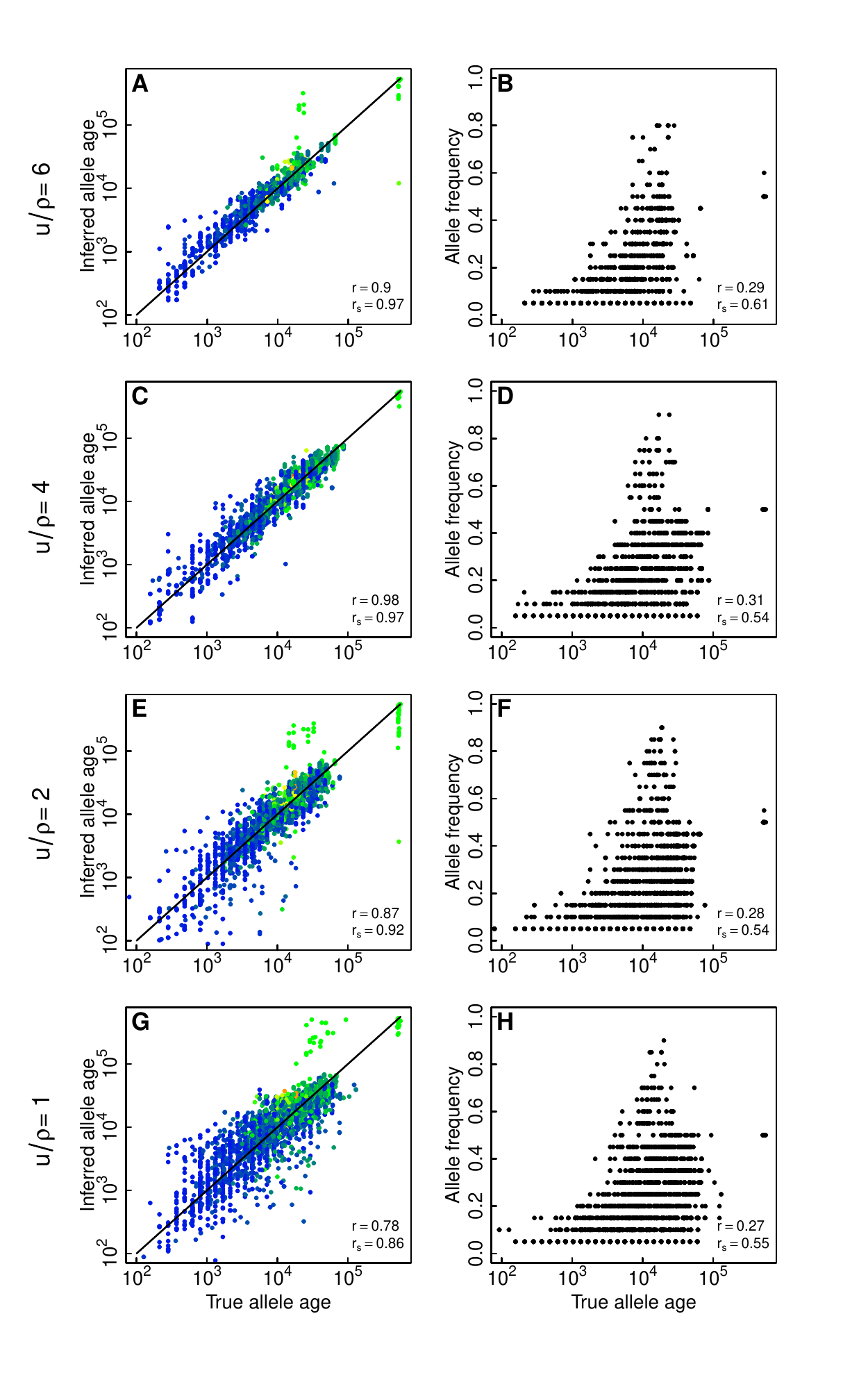}
\caption{\textbf{Estimating ages of derived alleles in simulated
    data.} 
\label{fig:alleleAgeSim}
}
\end{center}
\end{figure}
\clearpage

\begin{figure}[h!]
\begin{center}
\includegraphics[width=6in]{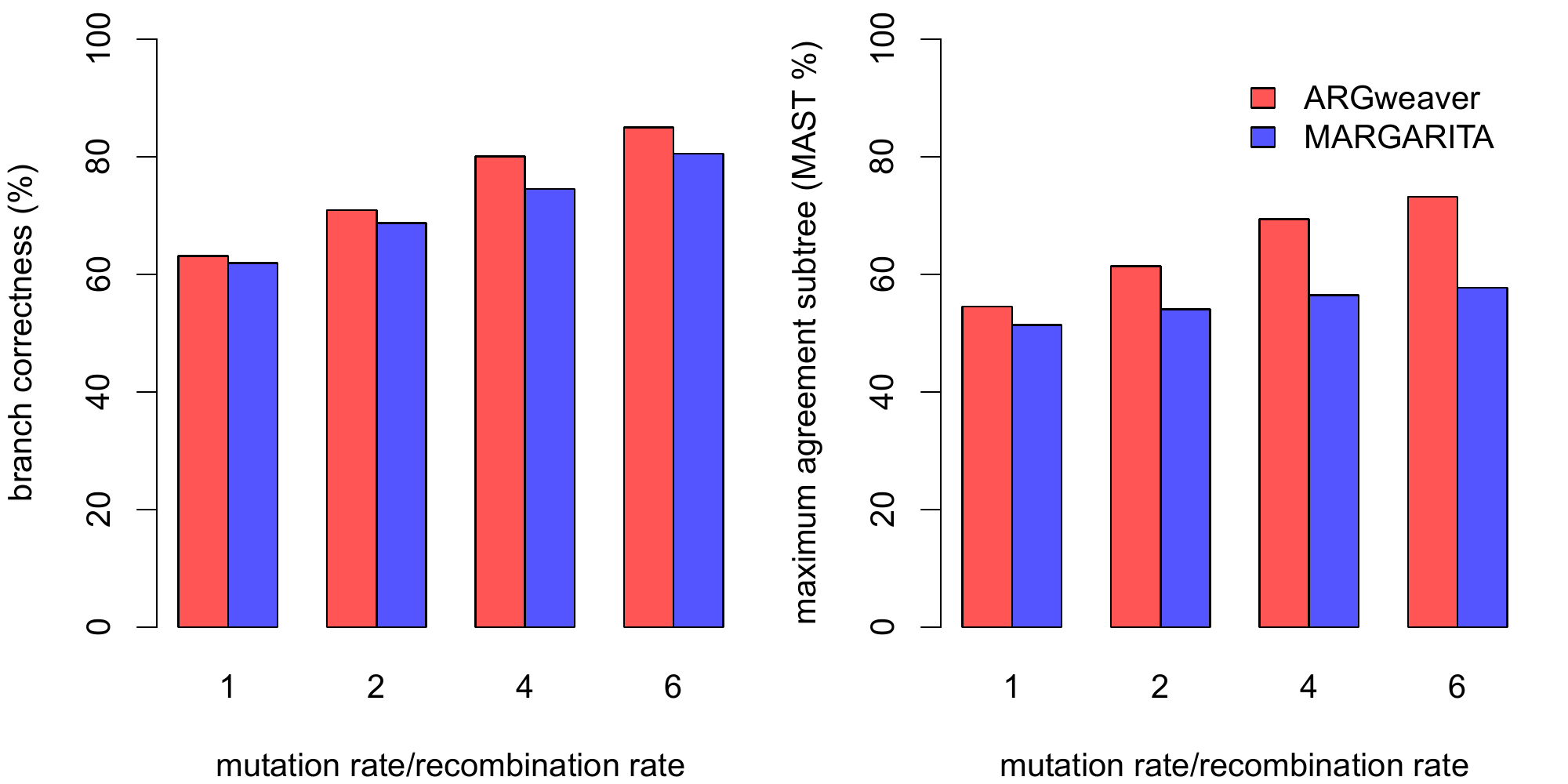}
\caption{\textbf{Recovery of local tree topologies.}  
\label{fig:sim-top-eval}
}
\end{center}
\end{figure}
\clearpage

\begin{figure}[h!]
\begin{center}
\includegraphics[width=4in]{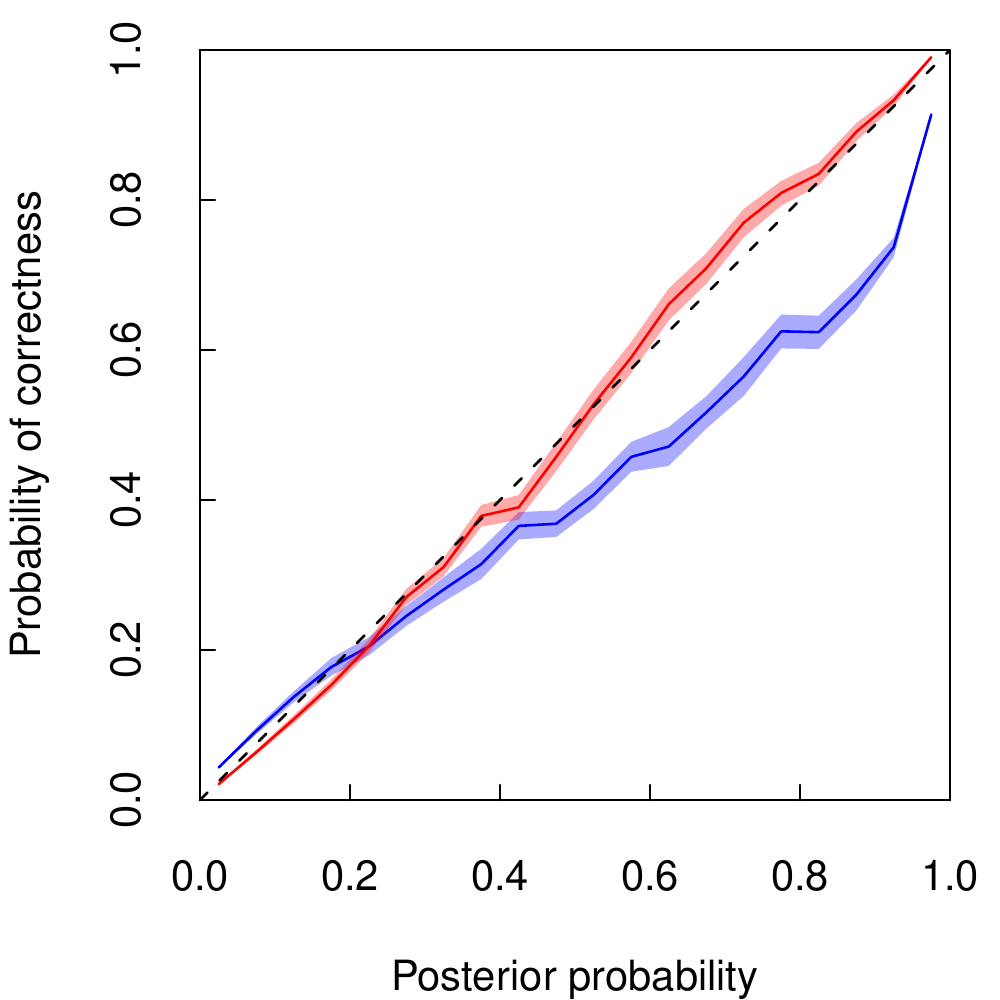}
\caption{\textbf{Local tree branch posterior probabilities inferred by
\wv accurately reflect their probability of correctness.}  
\label{fig:sim-uncertainty}
}
\end{center}
\end{figure}
\clearpage

\begin{figure}[h!]
\begin{center}
\includegraphics[width=6.5in]{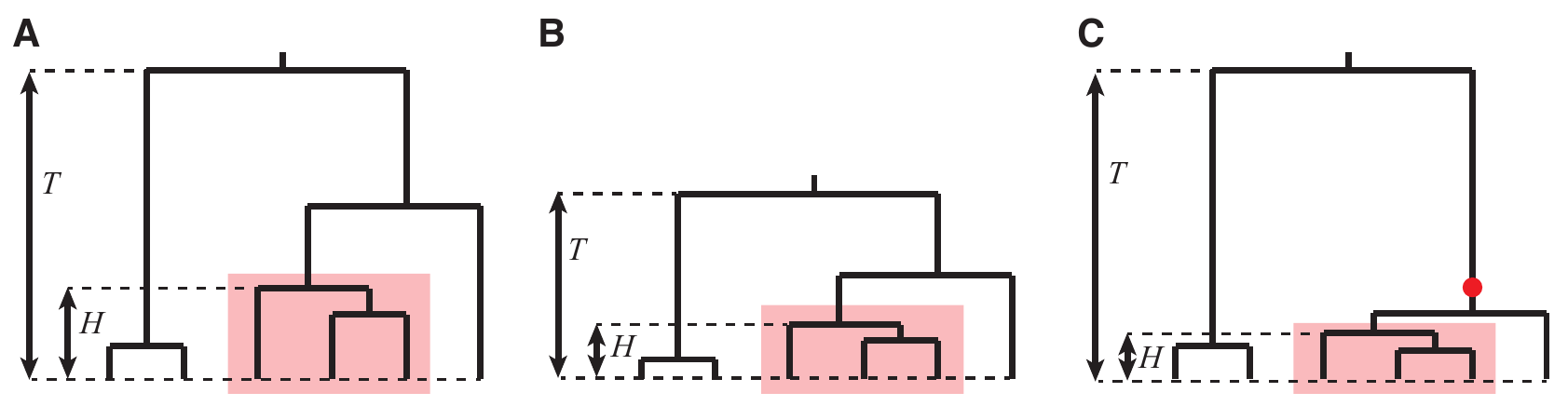}
\caption{\textbf{Illustration of relative TMRCA halflife (RTH).}  \label{fig:rth-conceptual}
}
\end{center}
\end{figure}
\clearpage

\begin{figure}[h!]
\begin{center}
\includegraphics[width=5.5in]{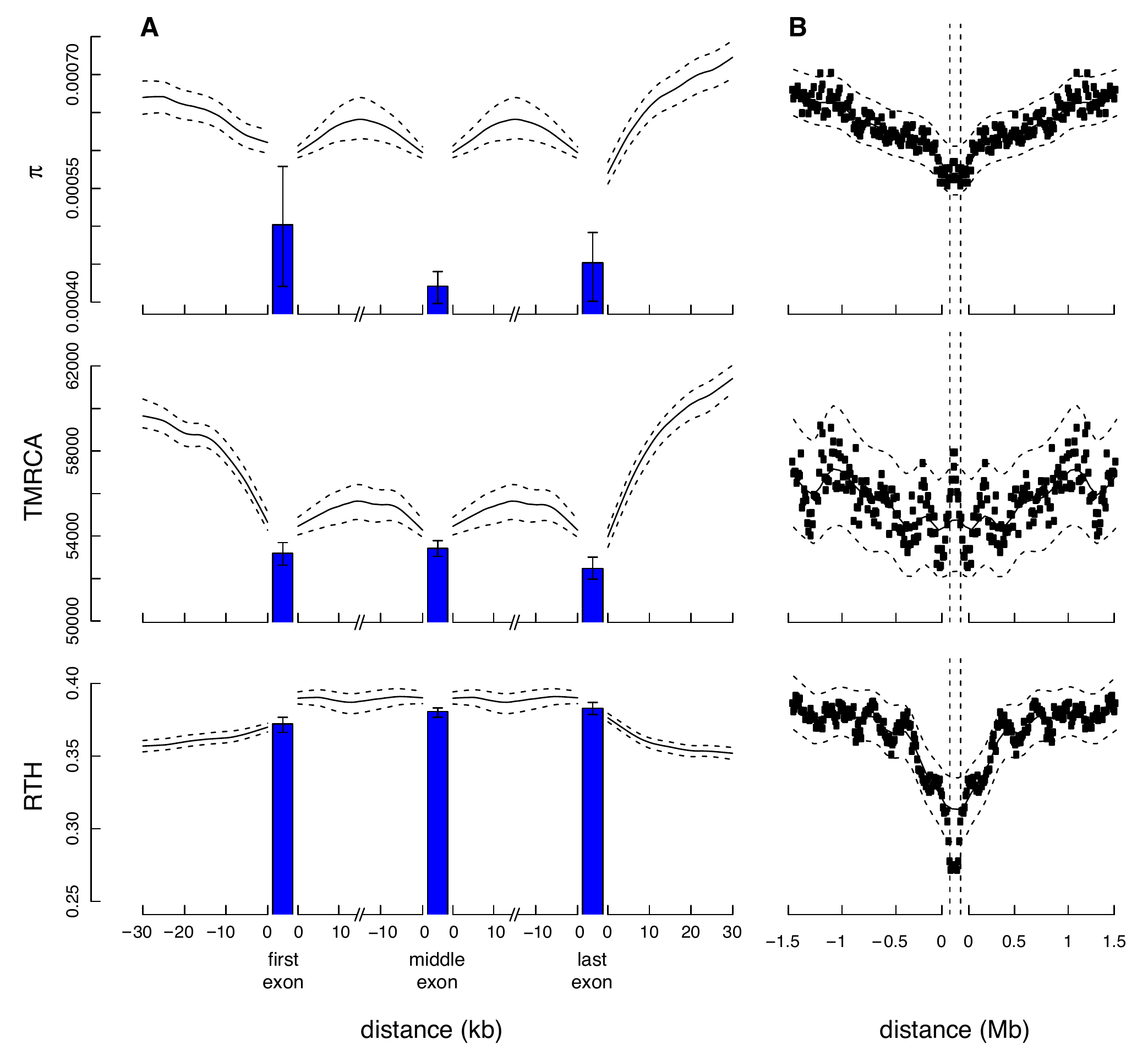}
\caption{\textbf{Measures of genetic variation near protein-coding genes and
  partial selective sweeps for African populations}.  }
\label{fig:metagene-sweeps-afr}
\end{center}
\end{figure}
\clearpage

\begin{figure}[h!]
\begin{center}
\includegraphics[width=6.5in]{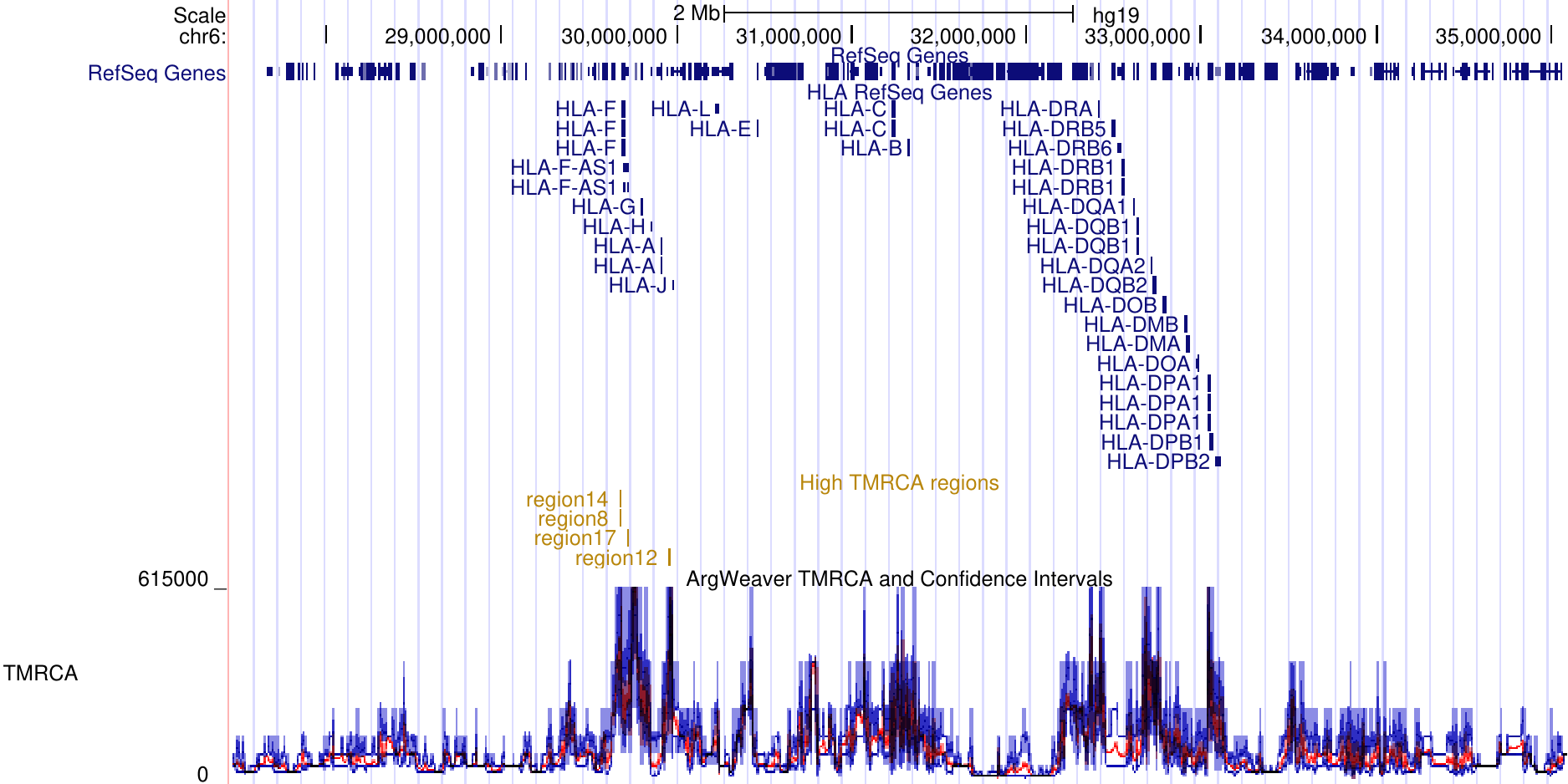}
\caption{\textbf{Time to most recent common ancestry (TMRCA) in the human
    leukocyte antigen (HLA) region. }}
\label{fig:hla}
\end{center}
\end{figure}
\clearpage

\begin{figure}[h!]
\begin{center}
\includegraphics[width=6.5in]{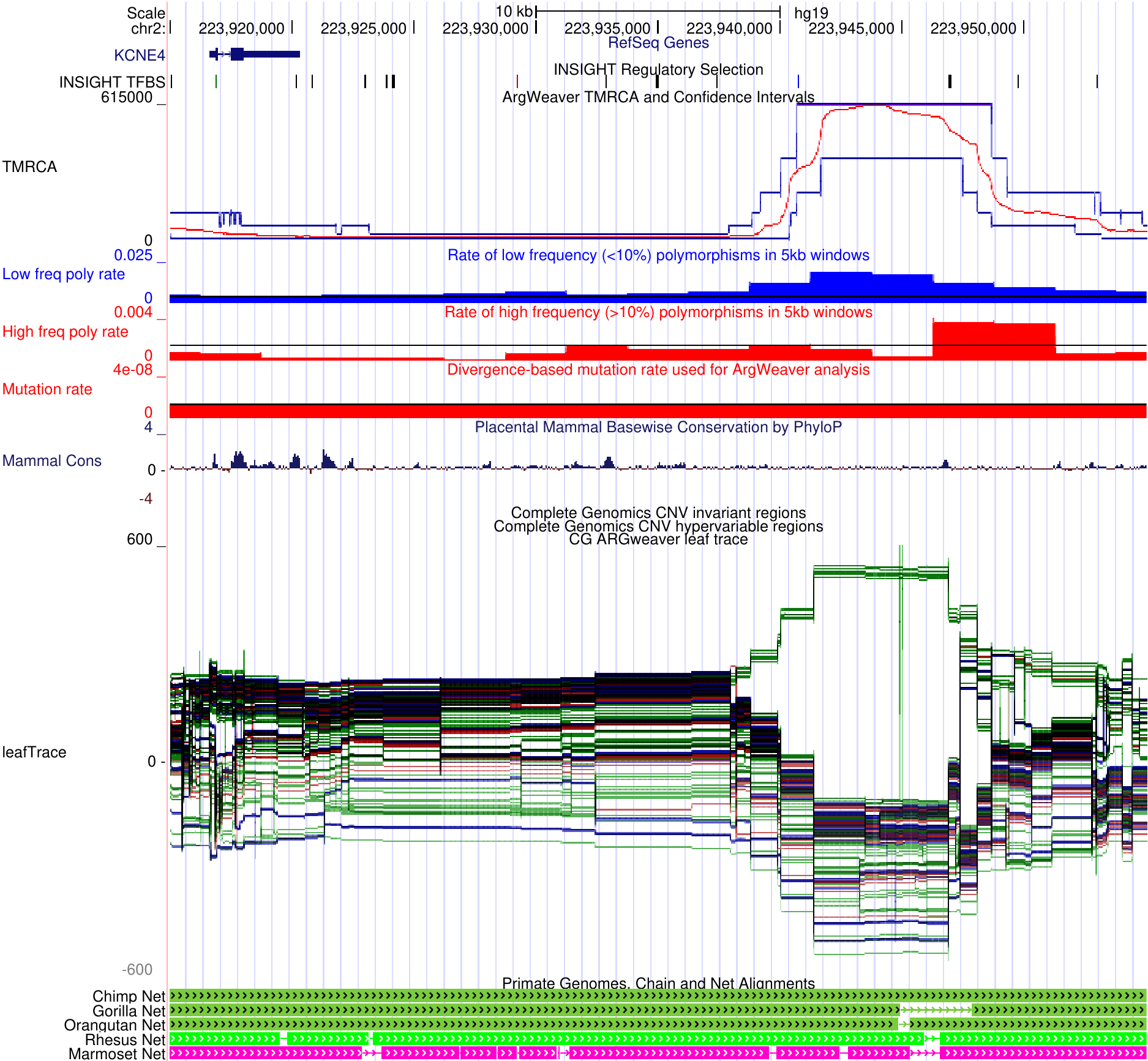}
\caption{\textbf{\wv tracks near {\em KCNE4}.} }
\label{fig:kcne4}
\end{center}
\end{figure}
\clearpage

\begin{figure}[h!]
\begin{center}
\includegraphics[width=5.5in]{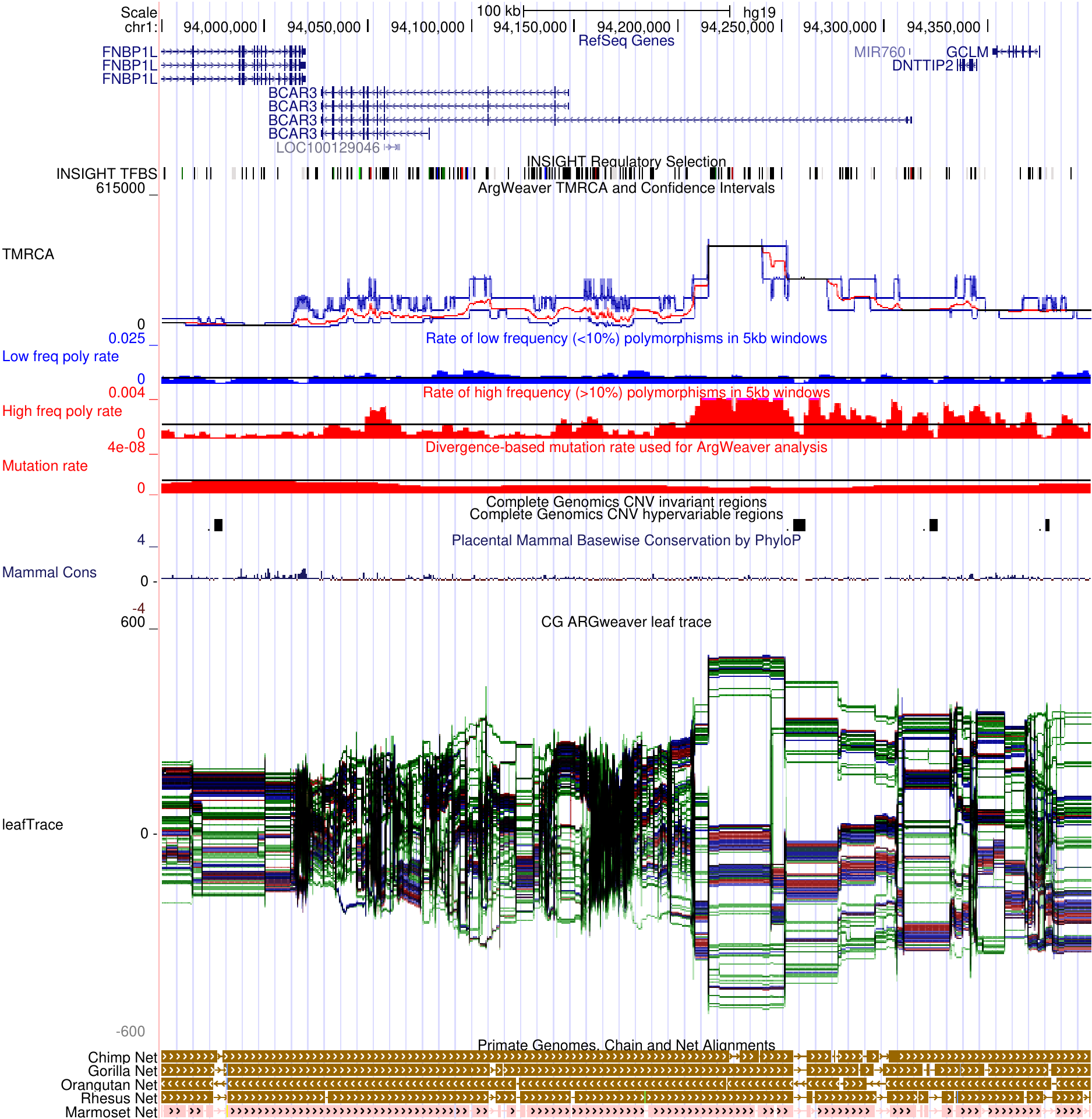}
\caption{\textbf{\wv tracks near {\em BCAR3}.}}
\label{fig:bcar3}
\end{center}
\end{figure}
\clearpage

\begin{figure}[h!]
\begin{center}
\includegraphics[width=5.5in]{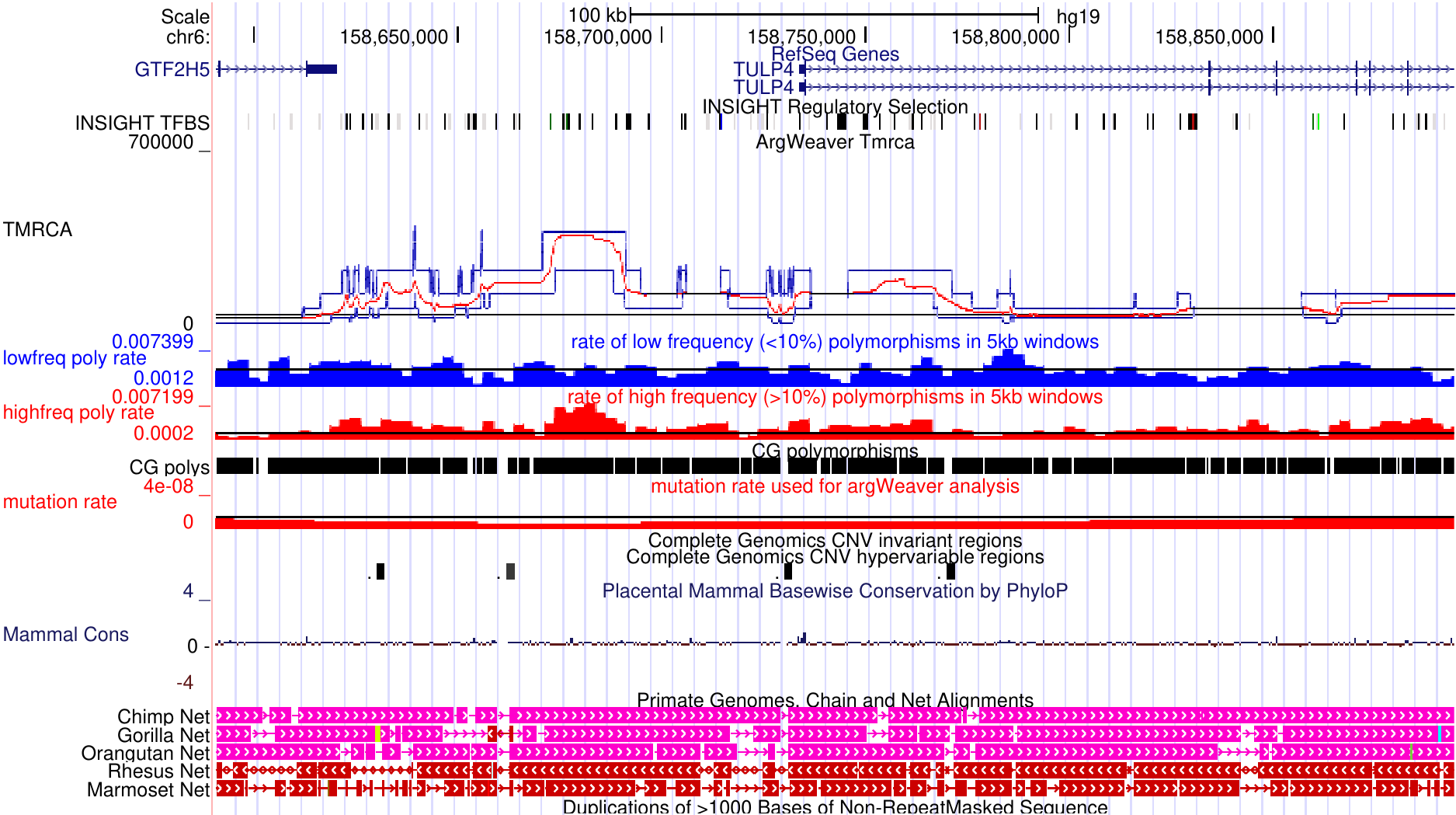}
\caption{\textbf{\wv tracks near {\em TULP4}.}  }
\label{fig:tulp4}
\end{center}
\end{figure}
\clearpage

\begin{figure}[h!]
\begin{center}
\includegraphics[width=3in]{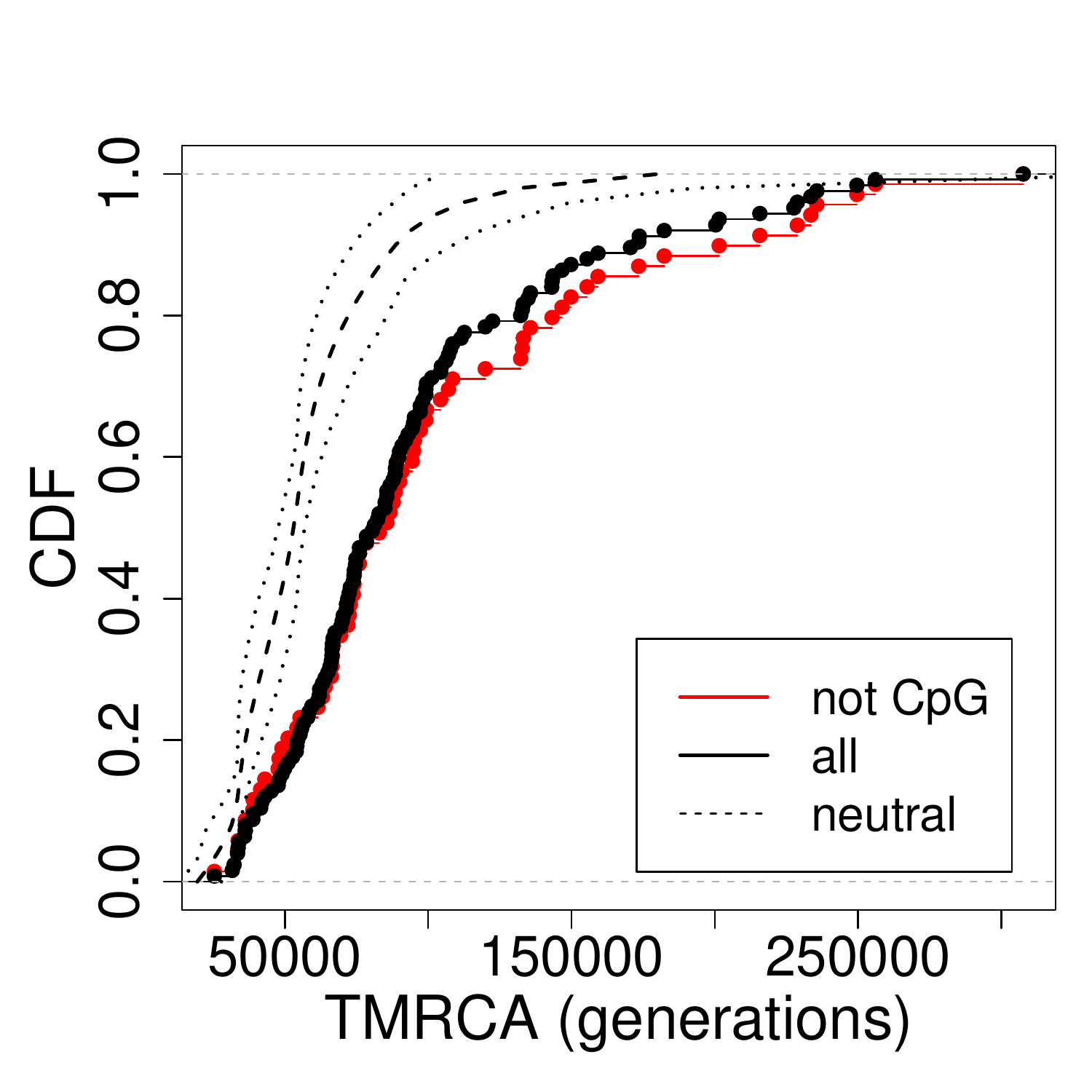}
\caption{\textbf{Distribution of TMRCAs in regions predicted to be under
    balancing selection.}  }
\label{fig:przeworski}
\end{center}
\end{figure}
\clearpage

\begin{figure}[h!]
\begin{center}
\includegraphics[width=6.5in]{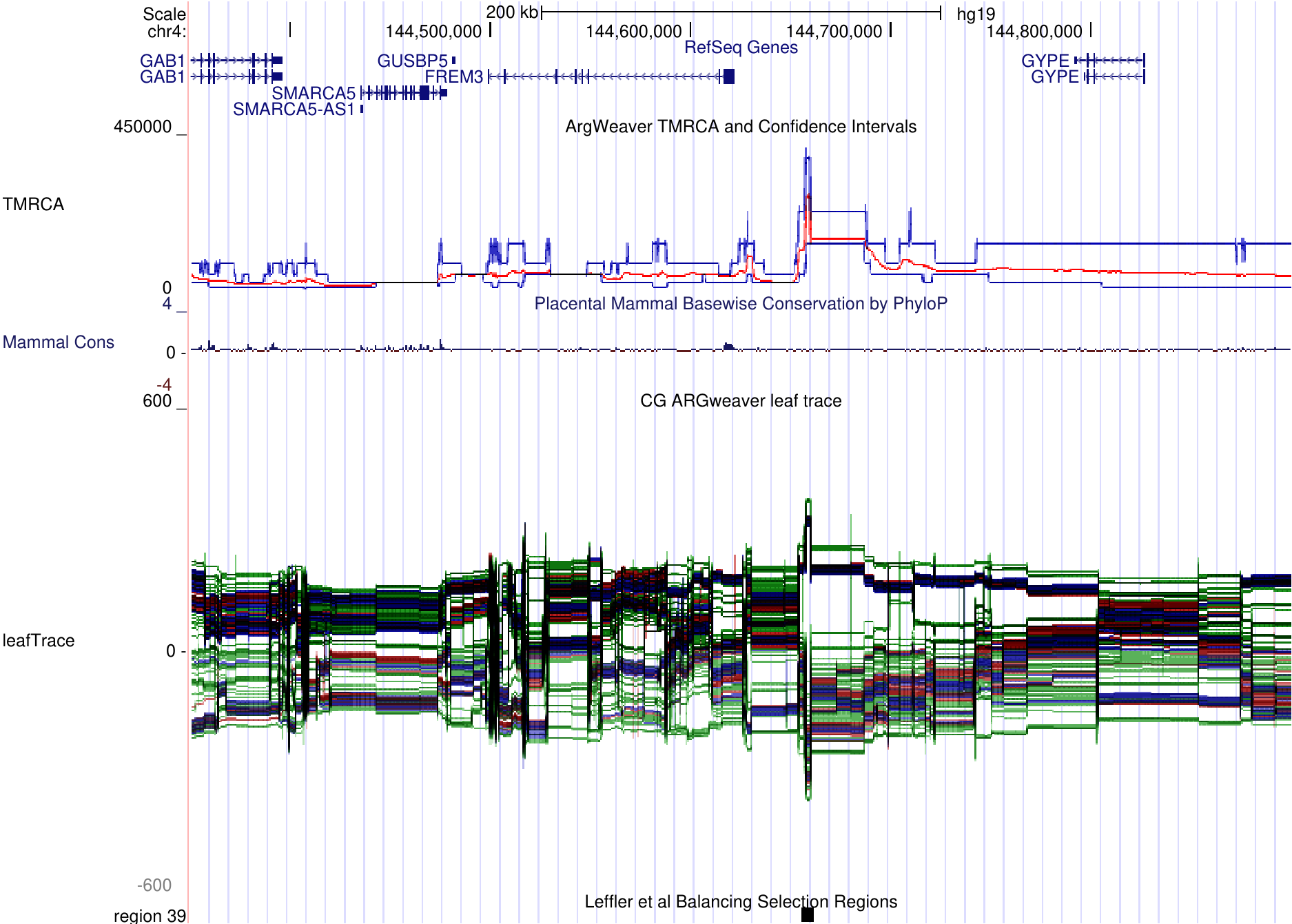}
\caption{\textbf{\wv tracks near locus
    containing segregating
    haplotypes shared in humans and chimpanzees.} }\label{fig:frem3}
\end{center}
\end{figure}
\clearpage

\begin{figure}[h!]
\begin{center}
\includegraphics[width=4in]{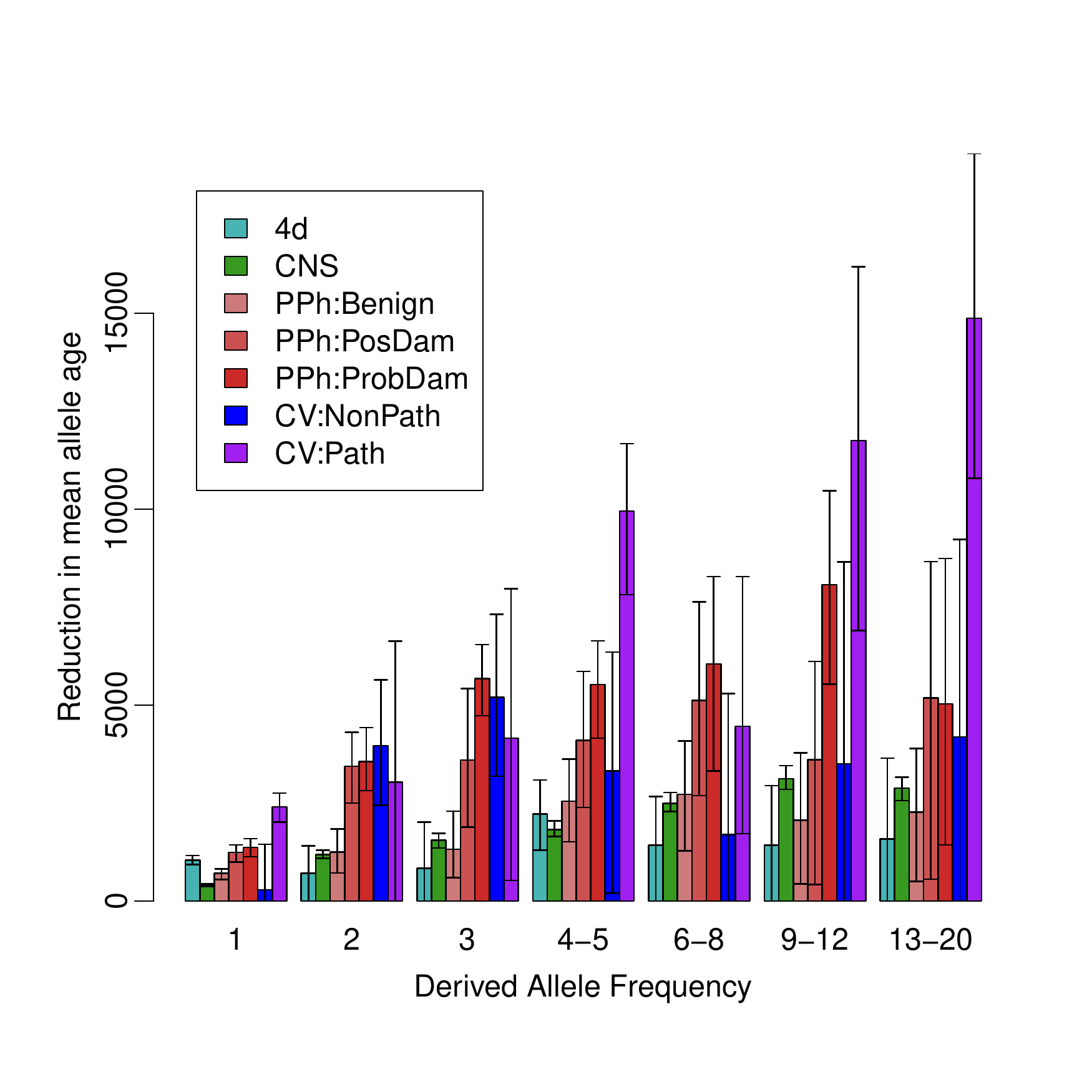}
\caption{\textbf{Reduction in mean allele age as a function of annotation
    class and derived allele frequency.}  
}
\label{fig:alleleAgeDiff}
\end{center}
\end{figure}
\clearpage

\begin{figure}[h!]
\begin{center}
\includegraphics[width=4in]{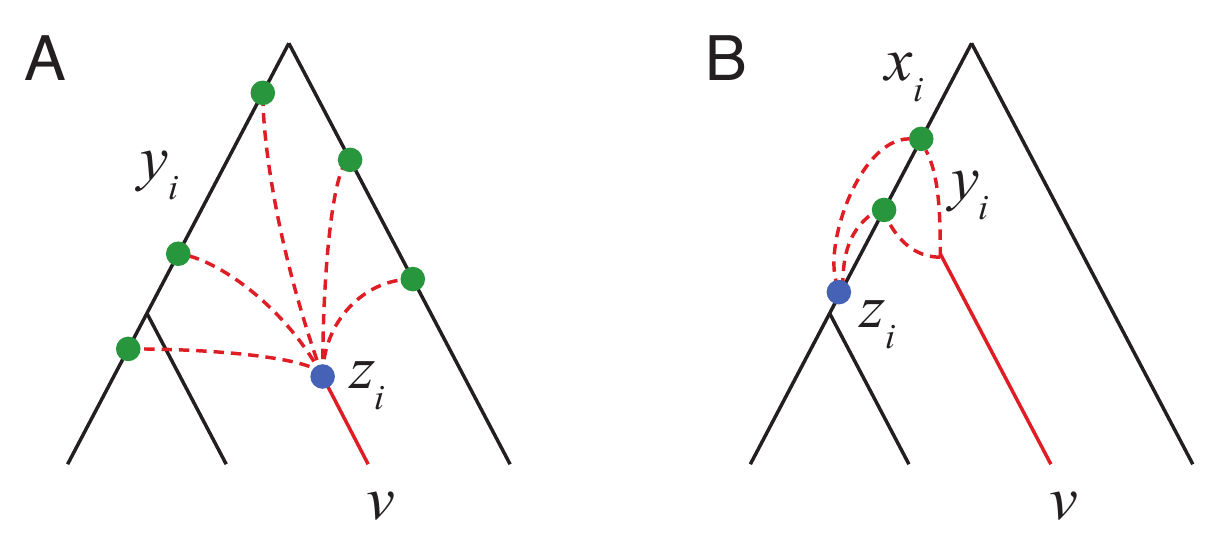}
\caption{\textbf{Cases for new recombination $z_i$ given
    re-coalescence point $y_i$.}  }
\label{fig:recomb-cases}
\end{center}
\end{figure}

\clearpage
%\doublespacing
\setstretch{1.5}

%=============================================================================
\begin{flushleft}
{\LARGE
\textbf{Genome-wide inference of ancestral recombination graphs}}\\[1ex]
{\Large \textbf{Supplementary Information: Text S1}}
\\[1ex]
Matthew D.\ Rasmussen, 
Melissa J.\ Hubisz,
Ilan Gronau,
Adam Siepel
\\[1ex]
Department of Biological Statistics and Computational Biology, 
Cornell University, Ithaca, New York 14853, USA
\end{flushleft}

%=============================================================================

\section*{Supplementary Methods}

\subsection*{Calculation of transition probabilities}

\begin{figure}
\begin{center}
\includegraphics[width=6.5in]{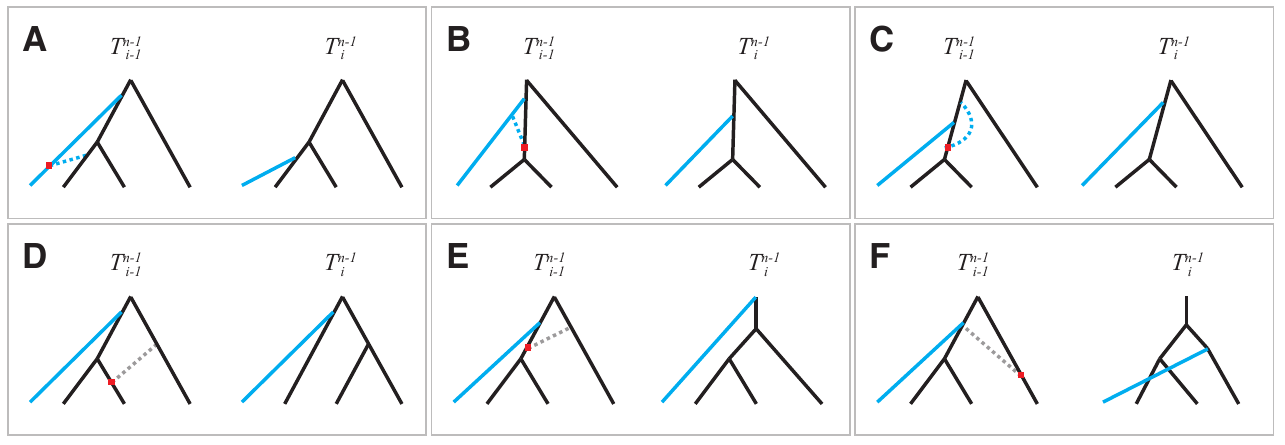}
\caption{{\bf Examples of thread transitions.} (A) When no old
  recombinations are present ($R^{n-1}_i = \emptyset$) the thread state
  $y_i$ can change if a new recombination $z_i$ added.  If this
  recombination is added to the new branch $v$ (blue), that branch may
  re-coalesce anywhere else in the local tree.  Alternatively, the new
  recombination can be placed on the branch associated with the previous
  thread state, $x_{i-1}$, in which case the recoalescence must occur on
  the new branch $v$ (B) or on branch $x_{i-1}$ (C).  When an old
  recombination is present ($R^{n-1}_i \not= \emptyset$), the associated
  SPR operation often does not effect the thread location.  However, it can
  cause the thread state to change its branch (D) or its time (E) in a
  deterministic manner.  (F) If the recoalescence point is the same as the
  thread point, the thread state can change to recombination-bearing branch
  $w_i$ and any time within the interval between the recombination and
  recoalescence times, $t_i \in [u_i,t]$.  The thread state can also change
  to the branch above the recoalescence point and keep the same time (not
  shown).}
\label{fig:cases-illust}
\end{center}
\end{figure}

The general formula for the transition probabilities of the HMM (equation
\ref{eqn:transition-prob}) can be simplified and its evaluation can be made
more efficient by recognizing several distinct scenarios for the joint
configuration of the previous local tree $T_{i-1}^n$, the current local
tree $T_i^n$, and the recombination $R_{i}^n$.  We will consider two main
cases, corresponding to the 
presence ($R_i^{n-1} \ne \emptyset$) and absence ($R_i^{n-1} = \emptyset$)
of ``old'' (previously sampled) recombinations, respectively (see Figure~\ref{fig:cases-illust}).  In addition,
we will consider three subcases of each
of these main cases.  Throughout this section, we will use the notation
$y_{i-1} = (x_{i-1}, t_{i-1})$ and $y_i = (x_i, t_i)$ to indicate the
previous and current coalescence points for the resampled branch,
respectively, with each $x_i$ indicating a branch and each $t_i$ a time
point.  We will assume $y_{i-1}$ is indexed by $l$ and $y_i$ by $m$, and we
will assume their time points are indexed by $a$ and $b$, respectively
(i.e., $t_{i-1}=s_a$, $t_i=s_b$).  In
addition, $z_i=(w_i, u_i)$ will denote a new recombination between
positions $i-1$ and $i$, with $w_i$ indicating a branch and $u_i$ a time
point in $T_{i-1}^n$.  We will use $v$ to indicate the new branch that is
being threaded into the ARG.  We will assume the single sequence threading
operation, so $v$ must be an external branch, but the subtree threading
setting is very similar (as discussed in later sections).  We will also use
the notation $S(u)$ to indicate the index of the time point associated with
a node $u$ in a local tree.

\subsubsection*{Major Case \#1: No Old Recombinations}

Let us first consider the case in which there are no old recombinations,
$R_i^{n-1} = \emptyset$.  Because of the restriction of at most one
recombination per genomic position, this is the only case in which a new
recombination is possible ($z_i \ne \emptyset$).  The three subcases for
the transition probabilities are as follows:

% I THINK WE DO WANT TO SWITCH TO AN i-1 FORMULATION OF EQUATION 20

\begin{enumerate}
\item {\bf Recoalescence to different branches:} $x_{i-1} \ne x_i$.  In
  this subcase, the recombination must have occurred on the new branch,
  $w_i = v$, as discussed in the section
  entitled ``Sampling a Recombination 
  Threading'' 
  in the main text.  In addition,
  the time of the recombination, $u_i$, must range between 0 and the
  minimum of $t_{i-1}$ and $t_i$.  Thus,
\begin{align}
a^{i-1}_{l,m} &= \sum_{z_{i}}
P(\bar{R}_{i}^{n-1}, z_{i} \;|\; \bar{T}_{i-1}^{n-1}, y_{i-1}=l) \;
P(\bar{T}_{i}^{n-1}, y_{i}=m \;|\; \bar{R}_{i}^{n-1}, z_{i},
\bar{T}_{i-1}^{n-1}, y_{i-1}=l) \notag \\
&= \sum_{k=0}^{\min(a,b)}
P(\bar{R}_{i}^{n-1}=\emptyset, z_{i}=(v, s_k) \;|\; \bar{T}_{i-1}^{n-1}, y_{i-1}=l) \;
P(\bar{T}_{i}^{n-1}, y_{i}=m \;|\; \bar{R}_{i}^{n-1}=\emptyset, z_{i}=(v, s_k),
\bar{T}_{i-1}^{n-1}, y_{i-1}=l) 
\label{eqn:diff-branch}
\end{align}
where, for simplicity, we drop the explicit conditioning on the model
parameters $\rho$ and $N$.  Notice that this sum has no more than $K+1$
terms. 

As in the general case, the first term in equation \ref{eqn:diff-branch} is given
by equation \ref{eqn:recomb} and the second term by equation
\ref{eqn:recoal-master} 
from the main text.  However, these equations simplify 
in this case.  Because 
we have assumed that $R_i^{n-1} = \emptyset$ and we can assume that
$z_i=(v,s_k)$ represents a valid recombination, we can write the following
in place of equation~\ref{eqn:recomb},
\begin{equation}
P(\bar{R}_{i}^{n-1}=\emptyset, z_{i}=(v, s_k) \;|\; \bar{T}_{i-1}^{n-1},
y_{i-1}=l) = 
\frac1{A'_k} \cdot \frac{B_k \, \Delta s_k}{C} \cdot \left[1 - \exp(- \rho
  |T^n_{i-1}|)\right],
\label{eqn:recomb-simple}
\end{equation}
where, 
\begin{equation}
A'_k = \begin{cases}
2 & \text{if }s_k=s_r \\
A_k & \text{otherwise}
\end{cases}
\label{eqn:A-prime}
\end{equation}
and all other terms are as defined for equation \ref{eqn:recomb}.
Similarly, equation \ref{eqn:recoal-master} simplifies in this case to,
\begin{equation}
P(\bar{T}_{i}^{n-1}, y_{i}=m \;|\; \bar{R}_{i}^{n-1}=\emptyset, z_{i}=(v, s_k),
\bar{T}_{i-1}^{n-1}, y_{i-1}=l) 
 =   \frac1{A_b^{(-v)}} P(s_b \;|\; v,\, s_k,\, T^n_{i-1})
\label{eqn:recoal-master-simple}
\end{equation}
where $P(s_b \;|\; v,\, s_k,\, T^n_{i-1})$ is given by equations \ref{eqn:old-recomb}--\ref{eqn:recomb-norm}.

\item {\bf Recoalescence to same branch at different times: } $x_{i-1} =
  x_i$, $t_{i-1} \ne t_i$.  In this case, the recombination may have
  occurred either on the new branch $v$ or on the recoalescence branch
  $x_i$.  If the recombination occurred on branch $v$, then, as above, its
  time index can range 
  between 0 and the minimum of $a$ and $b$ (the indices of $t_{i-1}$ and $t_i$,
  respectively).   If it occurred on
  branch $x_i$, then its time index can range between the time index at which
  $x_i$ came into existence, which is given by $S(x_i)$, and the  minimum
  of $t_{i-1}$ and $t_i$.  Thus,
\begin{small}
\begin{align}
a^{i-1}_{l,m} &= \sum_{z_{i}}
P(\bar{R}_{i}^{n-1}, z_{i} \;|\; \bar{T}_{i-1}^{n-1}, y_{i-1}=l) \;
P(\bar{T}_{i}^{n-1}, y_{i}=m \;|\; \bar{R}_{i}^{n-1}, z_{i},
\bar{T}_{i-1}^{n-1}, y_{i-1}=l) \notag \\
&= \sum_{k=0}^{\min(a,b)}
P(\bar{R}_{i}^{n-1}=\emptyset, z_{i}=(v, s_k) \;|\; \bar{T}_{i-1}^{n-1}, y_{i-1}=l) \;
P(\bar{T}_{i}^{n-1}, y_{i}=m \;|\; \bar{R}_{i}^{n-1}=\emptyset, z_{i}=(v, s_k),
\bar{T}_{i-1}^{n-1}, y_{i-1}=l) \notag \\
&\qquad + \sum_{k=S(x_i)}^{\min(a,b)}
P(\bar{R}_{i}^{n-1}=\emptyset, z_{i}=(x_i, s_k) \;|\; \bar{T}_{i-1}^{n-1}, y_{i-1}=l) \;
P(\bar{T}_{i}^{n-1}, y_{i}=m \;|\; \bar{R}_{i}^{n-1}=\emptyset, z_{i}=(x_i, s_k),
\bar{T}_{i-1}^{n-1}, y_{i-1}=l) 
\end{align}
\end{small}

As in case (1), the first term in each of these sums is given by equation
\ref{eqn:recomb-simple} and the second term is given by equation
\ref{eqn:recoal-master-simple}.  Each of these sums also has no more than
$K+1$ terms. 

\item {\bf Recoalescence to same branch at same time:} $x_{i-1} =
  x_i$, $t_{i-1} = t_i$.  This case is similar to the previous one, except that
  it must also allow for the possibility of no recombination between
  positions $i-1$ and $i$ ($z_i = \emptyset$).  Thus,
\begin{small}
\begin{align}
a^{i-1}_{l,l} &= \sum_{z_{i}}
P(\bar{R}_{i}^{n-1}, z_{i} \;|\; \bar{T}_{i-1}^{n-1}, y_{i-1}=l) \;
P(\bar{T}_{i}^{n-1}, y_{i}=l \;|\; \bar{R}_{i}^{n-1}, z_{i},
\bar{T}_{i-1}^{n-1}, y_{i-1}=l) \notag \\
&= \exp\left(-\rho|T_{i-1}^n|\right) \notag \\
&\qquad + \sum_{k=0}^{a}
P(\bar{R}_{i}^{n-1}=\emptyset, z_{i}=(v, s_k) \;|\; \bar{T}_{i-1}^{n-1}, y_{i-1}=l) \;
P(\bar{T}_{i}^{n-1}, y_{i}=m \;|\; \bar{R}_{i}^{n-1}=\emptyset, z_{i}=(v, s_k),
\bar{T}_{i-1}^{n-1}, y_{i-1}=l) \notag \\
&\qquad + \sum_{k=S(x_i)}^{a}
P(\bar{R}_{i}^{n-1}=\emptyset, z_{i}=(x_i, s_k) \;|\; \bar{T}_{i-1}^{n-1}, y_{i-1}=l) \;
P(\bar{T}_{i}^{n-1}, y_{i}=m \;|\; \bar{R}_{i}^{n-1}=\emptyset, z_{i}=(x_i, s_k),
\bar{T}_{i-1}^{n-1}, y_{i-1}=l) 
\end{align}
\end{small}
\end{enumerate}

\subsubsection*{Major Case \#2: Old Recombinations}

The other major case to consider is
when a recombination is already given, $R^{n-1}_i = (w_i,
s_k)\not= 
\emptyset$.  Our modeling assumptions prohibit a new recombination in this
case, so it must be true that $z_i=\emptyset$.  Thus,
\begin{align}
a^{i-1}_{l,m} &= \sum_{z_{i}}
P(\bar{R}_{i}^{n-1} = (w_i, s_k), z_{i} \;|\; \bar{T}_{i-1}^{n-1}, y_{i-1}=l) \;
P(\bar{T}_{i}^{n-1}, y_{i}=m \;|\; \bar{R}_{i}^{n-1}= (w_i,
s_k), z_{i},
\bar{T}_{i-1}^{n-1}, y_{i-1}=l) \notag \\
&= 
P(\bar{R}_{i}^{n-1}= (w_i,
s_k), z_{i}=\emptyset \;|\; \bar{T}_{i-1}^{n-1}, y_{i-1}=l) \;
P(\bar{T}_{i}^{n-1}, y_{i}=m \;|\; \bar{R}_{i}^{n-1}= (w_i,
s_k), z_{i}=\emptyset,
\bar{T}_{i-1}^{n-1}, y_{i-1}=l)
\end{align}
Because we can assume in this setting that $\bar{R}_{i}^{n-1}$ represents a valid
recombination, the first term has a form similar to that of equation
\ref{eqn:recomb-simple}, that is,
\begin{equation}
P(\bar{T}_{i}^{n-1}, y_{i}=m \;|\; \bar{R}_{i}^{n-1}= (w_i,
s_k), z_{i}=\emptyset,
\bar{T}_{i-1}^{n-1}, y_{i-1}=l) = 
\frac1{A'_k} \cdot \frac{B_k \, \Delta s_k}{C} \cdot \left[1 - \exp(- \rho
  |T^n_{i-1}|)\right],
\label{eqn:recomb-simple2}
\end{equation}
where $A'_k$ is given by equation \ref{eqn:A-prime} and all other terms are
as defined for equation \ref{eqn:recomb}.  Similarly, the second 
term has a form similar to that of equation \ref{eqn:recoal-master-simple},
\begin{equation}
P(\bar{T}_{i}^{n-1}, y_{i}=m \;|\; \bar{R}_{i}^{n-1}= (w_i,
s_k), z_{i}=\emptyset,
\bar{T}_{i-1}^{n-1}, y_{i-1}=l) 
 =   \frac1{A_b^{(-w_i)}} P(s_b \;|\; w_i,\, s_k,\, T^n_{i-1}).
\label{eqn:recoal-master-simple2}
\end{equation}

The calculation of these transition probabilities can
be further simplified by considering three subcases.  In defining these
cases, we use the notation $(x_i', t_i')$ to indicate the recoalescence
point associated with the old recombination $R_i^{n-1} = (w_i, u_i)$.
Notice that, in the case of an old recombination, 
this is not the same as the state $y_i = (x_i, t_i)$, which represents the
new coalescence point for branch $v$ (not branch $w_i$).  Here, the time
index $b$ corresponds to the recoalescence time $t'_i$, that is, $s_b=t'_i$.

\begin{enumerate}

\item {\bf Deterministic case.}  If the previous state $y_{i-1} = (x_{i-1},
  t_{i-1})$ does not equal either the recoalescence point $(x'_i,
  t'_i)$ or the recombination point $z_i = (w_i, s_k)$, then the transition
  process is completely deterministic, meaning that there is only one
  transition with non-zero probability.  A series of well-defined rules
  identifies the state $y_i$ that must follow $y_{i-1}$ (Figure
  \ref{fig:deterministic-rules}).  Note that, because the HMM is
  unnormalized, the probability of the permitted state transition will
  generally not be equal to one.

% ACS: I think we can skip the stuff about the times of the new nodes 
%If the new branch coalesces to the same branch as the branch with the
%recombination, $x_{i-1} = w_i$, then the new branch's time $t_{i-1}=s_a$
%is considered the time of the parent of recombination node
%$s(p(y))=a$.  Otherwise, $s(p(y))$ can be computed from
%$T^{n-1}_{i-1}$.

\item {\bf Recombination-point case.}
If the previous state $y_{i-1}$ equals the recombination point $z_i= (w_i,
s_k)$, then the recombination
point can be either above the new branch $v$ or below it.  If the recombination
is above branch $v$, then the
new state must be the same as the old one, $y_i = y_{i-1}$.
%and $s(p(w_i))$ 
%is computed in the usual way from $T^{n-1}_{i-1}$.  
If, on the other hand, the recombination is below branch $v$, then branch
$x_{i-1}$ ``escapes'' and the new branch $v$ must coalesce up higher in the
tree (see Figure \ref{fig:deterministic-rules} for a similar calculation).
%In this case, we also have $s(p(w_i)) = a$.

\item {\bf Recoalescence-point case.}
If the previous state
equals 
the recoalescence point, $(x_{i-1}, t_{i-1}) = (x'_i, t'_i)$, we must
consider the possibility that the recombining branch $w_i$ 
recoalesces at $(x'_i, t'_i)$ as well as the possibility that $w_i$
recoalesces at any location along 
the new branch $v$.  The reason is that all such scenarios allow
$T^{n-1}_i$ to have 
the same configuration after removal of $v$.  Note that this case is
always distinct 
from case (2) because the recoalescence cannot
be on the same branch as the recombination (this would imply a {\em
  bubble}, which are not allowed in the SMC process).
The destination states $y_i$ that are relevant for this scenario are
$y_i=(x_{i-1}, t_i)$, $y_i=(p(w_i), t'_i)$, and $y_i=(w_i, t_l)$, where
$p(w_i)$ is the new parent of $w_i$ (and $x_{i-1}$), and $t_l \leq t'_i$ is
any valid recoalescing point along $v$.
Since the
recombination point $R^{n-1}_i = (w_i, s_k)$ is not the same as the state
$y_{i-1} = (x_{i-1}, t_{i-1})$ in this case,
there is no ambiguity about the location of the recombination.
%If the new state $y_i = (x_i, t_i=s_b)$, then the recoalescence time
%$s_j=s_b$, because $w_i$ either coalesces to the new branch $v$ or
%it coalesces at $s_a$ and the state time is not changed, $s_a=s_b$.
%Also, $s(p(y))$ is computed in the usual way from $T^{n-1}_{i-1}$.

\end{enumerate}

%=============================================================================

%Let $(y, s_k)$ denote the recombination point in the local tree
%$T^{n-1}_{i-1}$.  Let $(z, t_z)$ be the coalescing point for $y$.

\begin{figure}
{
\footnotesize
\singlespacing
\lstset{language=c++}
\lstset{escapechar=!}
\lstset{mathescape= true}
\begin{lstlisting}[frame=tblr]
function get_deterministic_transition($(x_{i-1}, t_{i-1})$, $(w_i, s_k)$, $(x'_i, t'_i)$, mapping) {
  if ($(x_{i-1},t_{i-1})$ == $(x'_i,t'_i)$ || $(x_{i-1},t_{i-1})$ == $(w_i, s_k)$) {
    // not a deterministic case
    return NULL
  } else if ($x_{i-1} \not= w_i$) {
    // SPR only removes a subset of descendents, if any
    // trace up from remaining leaf to find correct new state
    disrupt = false
    if ($x_{i-1}$.is_leaf()) {
      // SPR cannot disrupt leaf branch
      $x_i = x_{i-1}$
    } else {
      if ($w_i$ == $x_{i-1}$.children[0]) {
        // left child is not disrupted
        $x_i$ = mapping[$x_{i-1}$.children[0]]
        disrupt = true
      } else if ($w_i$ == $x_{i-1}$.children[1]) {
        // right child is not disrupted
        $x_i$ = mapping[$x_{i-1}$.children[1]]
        disrupt = true
      } else {
        // $x_i$ is not disrupted
        $x_i$ = mapping[$x_{i-1}$]
      }
    }
    // optionally walk up, if coalescence occurs under thread
    if ((x == $x_{i-1}$ && $t'_i$ < $t_{i-1}$) || ($x'_i$ == $x_i$ && $t'_i$ < $t_{i-1}$) || (disrupt && x == $x_i$ && t <= $t_{i-1}$))
      $x_i$ = $x_i$.parent
    return $(x_i, t_{i-1})$
  } else {
    // SPR is on same branch as thread
    if ($s_k$ > $t_{i-1}$) {
      // thread moves with SPR subtree
      return (mapping[$w_i$], $t_{i-1}$)
    } else {
      // SPR subtree moves out from underneath thread, therefore the new 
      // branch coalesces with the branch above the subtree
      parent = $w_i$.parent
      $t_i$ = parent.age
      other = $w_i$.sibling()
      $x_i$ = mapping[other]
      if (other == $x'_i$) $x_i$ = $x_i$.parent
      return ($x_i$, $t_i$)
    }
  }
}
\end{lstlisting}
}
\caption{{\bf Deterministic rules.} When a previous recombination is given
  ($R^{n-1}_i = (w_i, s_k)\not= \emptyset$), most transitions are
  deterministic and can be determined by the set of rules shown here.  The
  basic idea of this procedure is that the recombination $R^{n-1}_i$ and
  recoalescence $y_i=(x_i,t_i)$ together define a subtree pruning and
  regrafting (SPR) operation on the local tree $T^n_{i-1}$ such that the
  coalescence point $y_i$ of the new branch $v$ is unambiguous given
  $y_{i-1}$ and the other available information.  The variable {\tt
    mapping} maps nodes in $T^{n-1}_{i-1}$ to equivalent nodes in
  $T^{n-1}_i$.}
\label{fig:deterministic-rules}
\end{figure}

%=============================================================================
\subsubsection*{Dynamic programming}
A limiting step in the calculation of the transition probabilities
described in the previous sections is the evaluation of equation
\ref{eqn:old-recomb}.  A naive evaluation of this equation requires $O(K)$
time, resulting in a running time of $O(K^2)$ for the calculation of
individual transition probabilities.  

Let us re-express equation \ref{eqn:old-recomb} as,
\begin{equation}
P(s_j \;|\; w,\, s_k,\, T^n_{i-1}, \, \Theta) 
= \exp\left[ -C_{k,j-2} -
  \frac{B_{j-1}^{(-w)}\, \Delta s_{j-1,j-\frac12}}{2N_{j-1}} \right] \times \left[ 1- \exp
  \left( -\frac{B^{(-w)}_{j-1}\, \Delta s_{j-\frac12,j}}{2N_{j-1}} -
    \frac{B_{j}^{(-w)}\, \Delta s_{j,j+\frac12}}{2N_{j}}\right) \right],
\end{equation}
where,
\begin{equation}
C_{k,m} = \sum_{l=k}^{m} \frac{B_l^{(-w)}\, \Delta s_l}{2N_l}.
\end{equation}

Notice that,
\begin{align}
C_{k,m} &= \sum_{l=k}^{m} \frac{B_l^{(-w)}\, \Delta s_l}{2N_l}
\notag \\
&= \left(\sum_{l=0}^{m} \frac{B_l^{(-w)}\, \Delta s_l}{2N_l}\right) -
\left(\sum_{l=0}^{k-1} \frac{B_l^{(-w)}\, \Delta s_l}{2N_l}\right) \notag
\\
&= C_{0,m} - C_{0,k-1}.
\end{align}

The values of the form $C_{0,m}$ can be computed recursively in a
preprocessing step for $m = 0,\dots, K$, as follows:
\begin{equation}
C_{0,m} = \begin{cases}
0 & m=0 \\
C_{0,m-1} + \frac{B_m^{(-w)}\, \Delta s_m}{2N_m} & 1 \leq m \leq K.
\end{cases}
\end{equation}

Thus, after preprocessing, 
the evaluation of equation \ref{eqn:old-recomb} can be accomplished
in constant time and the calculation of individual transition probabilities
can be 
accomplished in $O(K)$ time.

%=============================================================================
\subsection*{Further optimization of forward algorithm}

Implemented in a direct manner, the forward algorithm would require
$O(L n^2 K^2)$ time.  However, by taking advantage of redundancies in 
the transition probabilities 
we can reduce this running time to $O(L n K^2)$.  The approach used here is
similar to that used by Paul et al.\ \cite{Paul2011}.

Recall that the forward algorithm computes a table of values of the form,
\begin{align}
f_{i,m} =& P(\V{T}_{1:i}^{n-1}, \V{R}_{1:i}^{n-1}, \V{T}_{1:i}^{n}, y_i=m
\;|\; \Theta) \notag \\
        =& b^i_m(D_i^n) \sum_{l} f_{i-1,l} \; a^{i-1}_{l,m},
\end{align}
where $b^i_m(D_i^n)$ is the emission probability for state $m$ and
alignment column $i$, and $a^{i-1}_{l,m}$ is the transition probability
from state $l$ to state $m$ at position $i-1$
(see section entitled ``Stochastic Traceback'' in main text for complete details). 

Again, let the state variable $y_i$ be defined by a branch $x_i$ and a
time $t_i$.  In addition, let
${\cal C}(x_i, j)$ be the index for state $y_i = (x_i, t_i=s_j)$, where $s_j$ is the
$j$th time point.  These indices define the orders of the rows and columns
of the 
transition matrix for position $i$, denoted $\V{A}_i$.  Now, observe that,
for many choices of consecutive states $l={\cal C}(x_{i-1}, j)$ and
$m={\cal C}(x_{i}, 
k)$, the transition probability $a^i_{l,m}$ does not depend on $x_{i-1}$
and $x_i$ but only depends on the time indices $j$ and $k$.  In particular, if
$x_{i-1} \ne 
x_i$, the transition probability is independent of the identity
of the branches, because of the symmetry among all branches at
each time point in the coalescent model.

These symmetries mean that the true dimensionality of $\V{A}_i$ is
considerably reduced. 
To exploit this reduced dimensionality, let us define a reduced transition
matrix $\V{A}'_i = \{a'^i_{j,k}\}$ indexed by the time points (i.e., $0
\leq j \leq K$ and $0 \leq k \leq K$), such that $a'^i_{j,k}$ gives the
transition probability from time point $j$ to time point $k$ assuming that
$x_i \ne 
x_{i+1}$.  We can now rewrite the recurrence in the forward algorithm as
follows, assuming that $k$ is the time point associated with index $m$:
\begin{align}
f_{i,m} =& b^i_m(D_i^n) \sum_{l} f_{i-1,l} a^{i-1}_{l,m} \notag \\
   =& b^i_m(D_i^n) \left[ 
   \left(\sum_{l : x_i \not= x_{i-1}} f_{i-1,l} a^{i-1}_{l,m}\right) + 
   \left(\sum_{l : x_i = x_{i-1}} f_{i-1,l} a^{i-1}_{l,m} \right)\right] \notag \\
=& b^i_m(D_i^n) \left[ 
\left(\sum_{j=0}^K \; 
     \sum_{l:l = {\cal C}(x_{i-1},j), x_i \not= x_{i-1}} f_{i-1,l} a'^{i-1}_{j,k}\right) +
   \left(\sum_{l : x_i = x_{i-1}} f_{i-1,l} a^{i-1}_{l,m} \right)\right] \notag \\
  =& b^i_m(D_i^n) \left[ 
\left(\sum_{j=0}^K a'^{i-1}_{j,k}
     \sum_{l:l = {\cal C}(x_{i-1},j), x_i \not= x_{i-1}} f_{i-1,l}\right) +
\left(\sum_{l : x_i = x_{i-1}} f_{i-1,l} a^{i-1}_{l,m} \right)\right] \notag \\
  =& b^i_m(D_i^n) \left[ 
\left(\sum_{j=0}^K a'^{i-1}_{j,k}
     f'_{i-1,j}\right) +
\left(\sum_{l : x_i = x_{i-1}} f_{i-1,l} a^{i-1}_{l,m} \right)\right]
%
%  =& b^i_m(D_i^n) \left[ 
%\left(\sum_{j=0}^K a'^{i-1}_{j,k} \left[
%     \sum_{l = {\cal C}(x_{i-1},j)} f_{i-1,l} - f_{i-1,l(x_i,j)} \right]\right) +
%\left(\sum_{l : x_i = x_{i-1}} f_{i-1,l} a^{i-1}_{l,m} \right)\right] \notag \\
%  =& b^i_m(D_i^n) \left[ 
%\left(\sum_{j=0}^K a'^{i-1}_{j,k} \left[ f'_{i-1,j} - f_{i-1,{\cal C}(x_i,j)} \right]\right) +
%\left(\sum_{l : x_i = x_{i-1}} f_{i-1,l} a^{i-1}_{l,m} \right)\right],
\end{align}
where
\begin{align}
f'_{i-1,j} =& \sum_{l:l = {\cal C}(x_{i-1},j), x_i \ne x_{i-1}} f_{i-1,l}
\notag \\
=& \left(\sum_{l:l = {\cal C}(x_{i-1},j)} f_{i-1,l}\right) - f_{i-1,{\cal C}(x_i,j)},
\end{align}
and $f_{i-1,{\cal C}(x_i,j)}$ is zero if the state $(x_i, s_j)$ does not exist.

Notice that the $f'_{i-1,j}$ terms can be reused in calculating $f_{i,m}$ for
all values of $m$.  As a result, 
computing each column of the forward table takes $O(n K^2)$ time instead of
$O(n^2K^2)$ time, and the total running time of the algorithm is reduced to
$O(L n K^2)$.

%Altogether we have
%\begin{align}
%F_{i,j} =& P(X_i | Y_i=j) \left[ 
%    \sum_{a=0}^K T'_{a,b} \left[ F'_{i-1,a} - F_{i-1,l(x_i,a)} \right]  + 
%    \sum_{l : x_i = x_{i-1}} T_{l,j} F_{i-1,l} \right] \notag \\
% =& P(X_i | Y_i=j) \left[ 
%    \sum_{a=0}^K T'_{a,b} F'_{i-1,a} - 
%    \sum_{a \in s(x_i)} T'_{a,b} F_{i-1,l(x_i,a)} + 
%    \sum_{l : x_i = x_{i-1}} T_{l,j} F_{i-1,l} \right] \notag \\
% =& P(X_i | Y_i=j) \left[ 
%    \sum_{a=0}^K T'_{a,b} F'_{i-1,a} +
%    \sum_{l = l(x_i, a)}
%    \left[ T_{l,j} F_{i-1,l} - T'_{a,b} F_{i-1,l} \right] \right] \notag \\
% =& P(X_i | Y_i=j) \left[ 
%    \sum_{a=0}^K T'_{a,b} F'_{i-1,a} +
%    \sum_{l = l(x_i, a)}
%    \left[ T_{l,j} - T'_{a,b} \right] F_{i-1,l} \right].
%\end{align}

%=============================================================================
\subsection*{Subtree sampling}
\label{sec:sample-internal-branches}

In this section, we outline our strategy for {\em subtree sampling} (i.e.,
resampling of internal branches in the local trees) in greater detail.
As described in the main text, subtree sampling is needed to enable efficient
mixing of the MCMC sampler with more than few sequences.  Unlike the
single-sequence threading operation, subtree sampling allows the ``deep
structure'' of the ARG to be perturbed in a reasonably efficient manner.

% subtree and main tree definitions
For each local tree $T^n_i$, imagine that one of the internal branches $v$
is removed, thus producing two trees: a {\em main tree} $T^{M,n}_i$ and a
{\em subtree} $T^{S,n}_i$.  The main tree has the same root node as the
original full tree $T^n_i$ and has a basal branch that extends to the
maximum time $s_K$.  The subtree $T^{S,n}_i$ has $v$ as its root node and
does not have a basal branch.  In this setting, the effect of the
recoalescence operation is to allow the partial local tree $(T^{M,n}_i,
T^{S,n}_i)$ to be reconnected into a full local tree.  This is accomplished
by allowing introducing a lineage leading to $v$ and allowing it to
recoalesce with the main tree.  Notice that this is a direct generalization
of the single sequence recoalescence operation.  In that case, $v$ is
required to be a leaf node, but in the general case, it is allowed to be
any node (other than the root) in the local tree.  As in the single
sequence case, we can denote the recoalescence point at site $i$ by $y_i =
(x_i, t_i)$.

% internal case is very similar to external case
Let the age of the subtree root $v$ be $s_q$.  Notice that the structures
of the main tree and the subtree below age $s_q$ do not affect the
coalescence rate of the new branch $v$.  Thus, resampling the
coalescence point for the internal branch is essentially the same as resampling
the coalescence point for an external branch.  
The only restriction is that the recoalesce point must be at least as old
as $s_q$.  This operation can be used within any local block having a
single local tree, i.e., for which $\V{T}^{n}_i = \V{T}^{n}_j$ for all $i$
and $j$.  
However, a new problem arises in the case in which the local trees differ
across sites, as discussed in the next section.

\subsubsection*{Resampling internal branches across multiple local blocks}

Internal branches can also be resampled across multiple local blocks.
This involves removing one branch from each tree in $\V{T^n}$ to
create a list of main trees $\V{T^{M,n}}$ and subtrees $\V{T^{S,n}}$.
A coalescence threading $\V{Y}$ can then be sampled to define how each
subtree recoalesces to the corresponding main tree, thereby defining a new
collection of complete local
trees $\V{T^n}$.  

The problem is that a poor choice of a series of internal branches to
remove and resample can result in a highly constrained threading
distribution.  To see why this is true, imagine that the selected series of
internal branches is such that the branch for each local block is
completely unrelated (e.g., in a different subtree of the full phylogeny)
to the previous one.  In this case, if a new recombination is sampled
within a local block during the subtree threading operation, that
recombination will have to be ``undone'' by the end of the local block to
allow the new local tree for that block to be reconciled with the main tree
and subtree for the next block.  Thus, any ``move'' in ARG space must
involve tightly coordinated sequences of recombinations that cancel one
another out, in a sense.  Because such sequences will be difficult to find,
there will be a strong tendency to 
simply resample the previous threading, and the sampler will not mix well.

The solution to this problem is to select sequences of internal branches that
are in some way mutually ``compatible,'' so that these constraints on the
reconciliation of local trees across blocks are relaxed.  It turns out that
it is sufficient to select sequences of branches
such that adjacent branches in the sequence
{\em share ancestry}, as defined below.

\begin{figure}
\begin{center}
\includegraphics[width=4.5in]{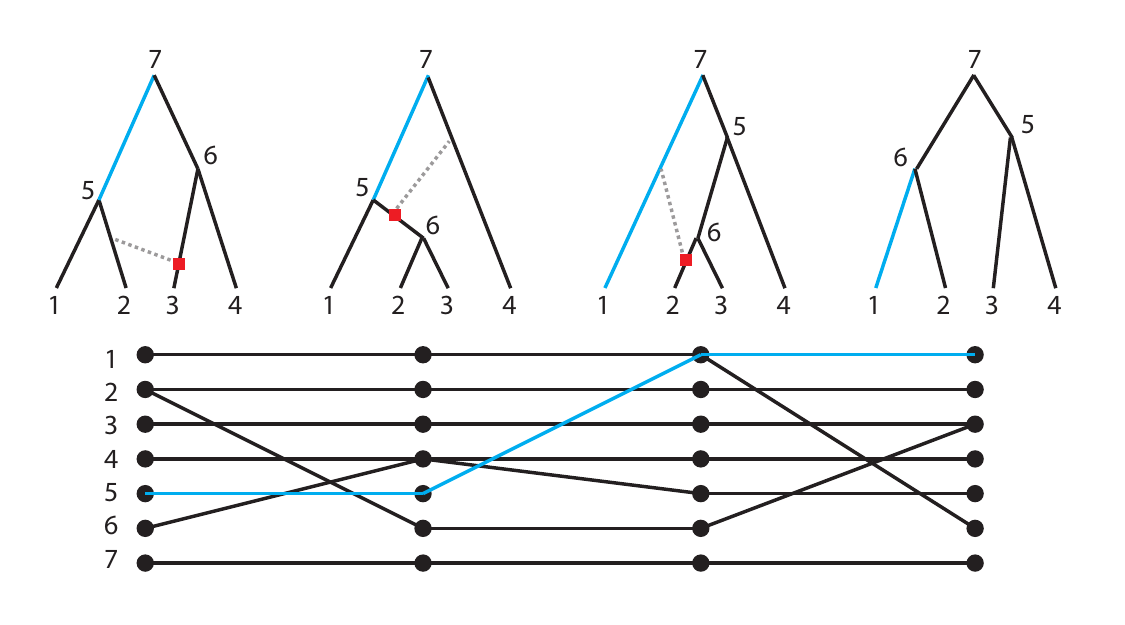}
\caption{{\bf Use of branch graph to select a series of internal branches for
    removal.}  The branch graph ${\cal B}$
  (bottom) describes which branches in the local trees (top) share
  ancestry.  Each node in the local trees has a corresponding node in
  the branch graph as indicated by the numbering scheme.  Directed
  edges connect nodes in neighboring blocks if, and only if, the
  corresponding nodes 
  in the local trees share ancestry.  A path along the branch graph 
  (blue at bottom) represents a valid series of branches (blue at top) in the
  local trees for removal and resampling.
 }
\label{fig:branch-graph}
\end{center}
\end{figure}

In order to identify
such sequences we use an auxiliary data structure called a {\em branch
  graph} (Figure \ref{fig:branch-graph}).
The branch graph
 ${\cal B}$ is derived
from the local trees $\V{T^n}$ and recombinations $\V{R^n}$.
To construct ${\cal B}$ we only need one local tree from each non-recombining
block.  Let this subset of trees and recombinations be represented by
the vectors $\V{T}$ and $\V{R}$, respectively (we will drop the superscript
for simplicity).  For each local tree
$T_i$ and node $v_{i,j} \in V(T_i)$, we create a node $u_{i,j} \in
V({\cal B})$.  Then we add a directed edge $(u_{i,j}, u_{i+1,k}) \in E({\cal
  B})$ if, and only if, the branches above $v_{i,j}$ and $v_{i+1,k}$ share
ancestry. 

We define ``shared ancestry'' as follows.
Let $z_{i+1} = (w_{i+1}, u_{i+1})$ represent the recombination point and $y_{i+1}
= (x_{i+1}, t_{i+1})$  
represent the recoalescing point leading from local tree $T_i$ to local
tree $T_{i+1}$.  In
addition, let $M$ be 
a mapping such that $M(v_{i,j}) = v_{i+1,k}$ if $v_{i,j}$ and $v_{i+1,k}$ 
represent precisely the same coalescent event in trees $T_i$ and $T_{i+1}$,
respectively.  Notice 
that the node above the 
recombination branch $w_{i+1}$ does not map to any node in $T_{i+1}$, that
is, $M(p(w_{i+1})) = \emptyset$.  Also, the new node created in $T_{i+1}$ by 
recoalescence, which we denote $v_{i+1}^+$, does not have any node mapping
to it.  However, all other 
nodes in 
$T_i$ and $T_{i+1}$ have a one-to-one mapping.  

Shared ancestry can occur in three ways.  Consider two arbitrary nodes in
adjacent local trees, $v_{i,j}$ and $v_{i+1,k}$.  First, if $M(v_{i,j}) \ne
\emptyset$ and $v_{i+1,k} \ne v_{i+1}^+$, then $v_{i,j}$ and $v_{i+1,k}$
share ancestry if, and only if, $M(v_{i,j}) = v_{i+1,k}$.  Second, if
$M(v_{i,j})=\emptyset$---meaning that $v_{i,j} = p(w_{i+1})$ is the node
that is eliminated by the recombination between $i$ and $i+1$---then
$v_{i,j}$ and $v_{i+1,k}$ share ancestry if, and only if, $v_{i+1,k}$ is
the {\em remaining child} of the eliminated node.  By ``remaining child''
we mean that $v_{i+1,k} = M(\text{sibling}(w_{i+1}))$ if
$\text{sibling}(w_i) \not= x_i$ or $v_{i+1,k} =
p(M(\text{sibling}(w_{i+1})))$ otherwise.  Finally, if $v_{i+1,k} =
v_{i+1}^+$, then $v_{i,j}$ and $v_{i+1,k}$ share ancestry if, and only if,
the recoalescence occurs on the branch above $v_{i,j}$, that is, $v_{i,j} =
x_i$.

These rules produce a graph ${\cal B}$ such that, for each local block $i$,
there is
one node (the one above which the recoalescence occurs) with an
out-degree of two, while all 
other nodes have an out-degree of one.  Similarly, for each local block,
there is
one node with in-degree of two (the remaining child of the node eliminated
by the recombination) and all other nodes have an in-degree of
one.  A directed path in the branch graph 
indicates a series of branches valid for removal.  

The number of directed paths indicates the number of possible ways to remove
internal branches according to this scheme.  This number can be computed in a
straightforward way using dynamic programming.  Let $P_{i,j}$ represent the
number of paths ends in block $i$ on branch $j$.  This value can be
computed recursively 
as follows:
\begin{align}
P_{i,j} = \begin{cases}
1, & \text{if } i = 1 \\
\sum_k P_{i-1,k} I[ (v_{i-1,k}, v_{i,j}) \in E({\cal B}) ], & \text{otherwise}.
\end{cases}
\end{align}
Thus, the total number of directed paths can be computed as $\sum_j
P_{m,j}$ where $m$ is the index of the last local block.  Paths can be sampled
uniformly using a standard traceback procedure.  Starting with the last
block $m$, the last node $j$ 
can be chosen with probability 
\begin{align}
\frac{P_{m,j}}{\sum_j P_{m,j}}.
\end{align}
Given a chosen node $v_{i,j}$, the next node $v_{i-1,k}$ in the traceback 
can be chosen with probability,
\begin{align}
\frac{P_{i-1,k}}{P_{i,j}}.
\end{align}

%=============================================================================
\subsection*{Gibbs and Metropolis-Hastings Sampling of ARGs}
\label{sec:gibbs}

Our goal is to sample ARGs $G^n$ from the posterior distribution
given the model parameters $\Theta = (\V{N}, \mu, \rho)$ and 
the data $D^n$, namely
\begin{align}
P(G^n \;|\; \Theta, D^n).
\end{align}

Using our threading method, we can define both Gibbs and generalized
Metropolis-Hastings 
Markov chain Monte Carlo (MCMC) methods for sampling ARGs.  Let
$g$ and $g'$ be 
two possible values for the random variable $G^n$.  Let $q(g \to g')$
give the probability of proposing $g'$ given $g$ under some
proposal procedure.  The Metropolis-Hastings algorithm requires that the
acceptance probability for the proposed move must be,
\begin{align}
A( g \to g' ) =& \min\left(1, \;
  \frac{P(G^n = g' \;|\; \Theta, D^n)}{P(G^n = g \;|\; \Theta, D^n)} \;
  \frac{q(g' \to g)}{q(g \to g')} \right).
\end{align}

Let us now consider a particular type of probabilistic
proposal procedure.  Let $S$ be a random
variable representing a random subgraph of an ARG $g$ and let $S(g)$ give a
restricted set of subgraphs of $g$.  Given a current ARG $g$, randomly
choose a subgraph $S=s$ and then sample from the posterior a new ARG
$g'$ in which the subgraph $s$ is held fixed (i.e., all changes occur
outside of $s$).  We can now write the proposal probability as,
\begin{align}
q(g \to g') =& \sum_{s \in S(g)} 
  P(S=s \;|\; G^n = g) \; P(G^n = g' \;|\; S=s, \Theta, D^n) \notag \\
=& \sum_{s \in S(g, g')} 
  P(S=s \;|\; G^n = g) \; P(G^n = g' \;|\; S=s, \Theta, D^n),
\end{align}
where we use the notation $S(g, g')$ to indicate $S(g) \cap S(g')$, thereby
enforcing the constraint that the sampled subgraph $s$ must belong to the
restricted sets for both the original ARG $g$ and the proposed ARG $g'$.
This proposal probability can be further simplified as follows:
\begin{align}
q(g \to g') =& \sum_{s \in S(g, g')} 
  P(S=s \;|\; G^n = g) \; \frac{P(G^n = g', S=s \;|\; \Theta, D^n)}
{\sum_{h : s \in S(h)} P(G^n=h, S=s \;|\; \Theta, D^n)} \notag \\
=& \sum_{s \in S(g, g')} 
   P(S=s \;|\; G^n = g) \; \frac{P(G^n = g' \;|\; \Theta, D^n)}
                           {\sum_{h : s \in S(h)} P(G^n = h \;|\; \Theta, D^n)}
                           \notag \\
=& P(G^n = g' \;|\; \Theta, D^n) \sum_{s \in S(g, g')} 
   P(S=s \;|\; G^n = g) \;
   \left[ \sum_{h : s \in S(h)} P(G^n = h \;|\; \Theta, D^n) \right]^{-1}
\end{align}
where the simplification in the second line is possible because $S$ is a
subgraph of $G^n$. 

If we choose subgraphs uniformly from the set $S(g)$, such that
$P(S=s \;|\; G^n=g) = 1/|S(g)|$, we can then write the acceptance
probability as
\begin{align}
 & A( g \to g' )  \notag \\
=& \min\left(1, 
   \frac{P(G^n = g'| \Theta, D^n)}{P(G^n = g | \Theta, D^n)} 
   \frac{P(G^n = g | \Theta, D^n) \sum_{s \in S(g, g')} 
    P(S=s | G^n = g') 
    \left[ \sum_{h : s \in S(h)} P(G^n = h | \Theta, D^n) \right]^{-1}}
   {P(G^n = g' | \Theta, D^n) \sum_{s \in S(g, g')} 
    P(S=s | G^n = g) 
    \left[ \sum_{h : s \in S(h)} P(G^n = h | \Theta, D^n) \right]^{-1}} 
   \right) \notag \\
=& \min\left(1, 
   \frac{\sum_{s \in S(g, g')} 
    |S(g')|^{-1}
    \left[ \sum_{h : s \in S(h)} P(G^n = h | \Theta, D^n) \right]^{-1}}
   {\sum_{s \in S(g, g')} 
    |S(g)|^{-1}
    \left[ \sum_{h : s \in S(h)} P(G^n = h | \Theta, D^n) \right]^{-1}} 
   \right) \notag \\
=& \min\left(1, 
   \frac{|S(g)| \sum_{s \in S(g, g')} 
    \left[ \sum_{h : s \in S(h)} P(G^n = h | \Theta, D^n) \right]^{-1}}
   {|S(g')| \sum_{s \in S(g, g')} 
    \left[ \sum_{h : s \in S(h)} P(G^n = h | \Theta, D^n) \right]^{-1}} 
   \right) \notag \\
=& \min\left(1, \frac{|S(g)|}{|S(g')|} \right).
\end{align}

For cases where $|S(g)| = |S(g')|$ the acceptance probability is
always 1 and the procedure is a valid a Gibbs sampler.  This is true
for the case when $S(g)$ is the set of subgraphs of $g$ where one
sequence is removed from the ARG.  If there are $n$ sequences then
$|S(g)| = n$.

For resampling internal branches, $|S(g)|$ is not as trivial to
calculate, but it can be calculated using dynamic programming
(see previous section).  However, $|S(g)|$ will not always be equal to
$|S(g')|$ and therefore there is a chance of rejection.

The rationale for using a proposal procedure based on conditioning on
the subgraph is that uses the data to drive the proposal.  Without
such a strategy, the acceptance probability would be driven by changes
in likelihood which can vary wildly when resampling large ARGs.

In order to establish that the stationary distribution of this Markov chain
equals the desired posterior distribution, we must show that the chain is
irreducible, aperiodic, and positive recurrent.  First, the chain is
irreducible because, given any two ARGs $g$ and $g'$, it is possible to
find a sequence of proposed moves that will transform $g$ to $g'$.  To see
that this is true, consider a subgraph $S(g, L)$ of $g$ that is defined by
removing threads from $g$ until only a set of leaves $L$ remains.  First
consider the base case of a single leaf, $L = \{l\}$.  In this case, we
trivially have $S(g, L) = S(g', L)$, because both subgraphs are simply
trunk genealogies. Now let us add one leaf at a time to $L$.  Each time we
add a leaf $l$ to the set $L$, we can ensure that $S(g, L) = S(g', L)$ by
removing the thread for $l$ from $g$ and then re-threading $l$ in such a
way that $S(g, L) = S(g', L)$. In this way, we can obtain any $g'$ from any
$g$ using the threading operation.

Next, to see that the chain is aperiodic and positive recurrent, note that
self-transitions $g \to g$ have nonzero probability.  In addition, every
transition in the Markov chain is reversible.  Specifically, for any
transition $g \to g'$, we choose a subgraph $s$ and then sample $g'$
conditional on $s$.  Notice that based on the design of our branch removal
procedure, if $g'$ was sampled conditioned on $s$, then $s$ can be obtained
by applying the branch removal procedure to $g'$.  Since $s$ is a subgraph
of $g$, $g$ can be sampled by the threading procedure conditioned on $s$.
Thus, the reverse transition $g' \to g$ must have non-zero probability.
Together, nonzero self transitions and reversible non-self transitions
guarantee that the chain is aperiodic.  The chain is positive recurrent
because it has a finite state space and is irreducible.

\section*{Supplementary Data Analysis}
\subsection*{Evaluation of Discretized Sequentially Markov Coalescent in
  Data Generation}
%{DSMC model closely approximates coalescent-with-recombination}
Following McVean
and Cardin 
\cite{McVean2005}, we compared data sets generated under 
the DSMC with ones generated under the sequentially Markov coalescent (SMC)
and the coalescent-with-recombination (CwR).  
First, we simulated 100 kb regions of 20 sequences using our standard
simulation parameters including four different $\mu/\rho$ ratios (see Methods).
%assuming an effective population size of $N=$ 10,000, a mutation 
%rate of $\mu=2.5\times 10^{-8}$ mutations/generation/site, and four 
%recombination rates ranging from $\rho=3.75\times 10^{-9}$ to 
%$\rho=1.5\times 10^{-8}$ recombination/generation/site.  
%For each 
%recombination rate, we simulated 100 replicates and counted the number 
%of recombination breakpoints produced during the simulation.
We carried out parallel 
simulations under the DSMC, the SMC, and the CwR, assuming various numbers
of time intervals ($K$) for the DSMC.
At all recombination rates, the DSMC, SMC, and CwR models
produced very similar distributions of recombination counts
(Supplementary Figure \ref{fig:sim-dsmc}A).  These distributions were essentially
indistinguishable at lower recombination rates, and the DSMC exhibited
only a slight excess of recombinations events at higher rates.
Interestingly, the DSMC appeared not to be highly sensitive to the
number of time intervals $K$, although the excess in recombination
events at high rates was most pronounced under the most coarse-grained
discretization scheme ($K=10$).

In a second simulation experiment, we assumed a single ratio of $\mu/\rho =
2$ and
considered four effective population sizes ($N$), ranging from 10,000 to
30,000 individuals.  In this
comparison, we used the number of segregating sites as the summary
statistic of interest.  As expected, this statistic increases approximately
linearly with $N$ under all models.  Once again, we found
that the CwR, SMC, and DSMC models produced nearly identical distributions
of counts, with only a minor inflation under the coarsest discretization
schemes (Supplementary Figure \ref{fig:sim-dsmc}B).  
%These simulated data
%sets could be additionally be compared in terms of decay of linkage
%disequilibrium, but it is unlikely that the discretization assumed in the
%DSMC would significantly degrade the performance of the SMC by this measure
%\cite{McVean2005}.  
Overall, these comparisons indicate that the
discretization scheme used by the DSMC has at most a minimal effect on
measurable patterns of mutation and recombination at realistic parameter
values for human populations, suggesting that the model will be adequate
for use in inference.

\end{document}